\documentclass[aps,prb,twocolumn,longbibliography,floatfix,preprintnumbers,superscriptaddress,amsmath,amssymb]{revtex4-2}

\usepackage{xcolor,bm,lineno,multirow,times}
\usepackage{physics}
\usepackage{amsmath,mathrsfs}
\usepackage{rotating}
\usepackage{verbatim}
\usepackage{graphicx,tabularx}
\usepackage[multiple]{footmisc}
\usepackage[sort&compress]{natbib}
\usepackage[T1]{fontenc}
\usepackage{soul}
\usepackage{dcolumn}
\usepackage{bm}
\usepackage[version=4]{mhchem}
\usepackage{booktabs}
\usepackage{gensymb}
\usepackage{color}
\usepackage[colorlinks=true,urlcolor=blue,linkcolor=blue,citecolor=blue]{hyperref}
\usepackage[title]{appendix}
\usepackage{orcidlink}
\usepackage{todonotes}
\usepackage{glossaries}
\newacronym{ED}{ED}{Exact diagonalization}
\newacronym{DMRG}{DMRG}{density matrix renormalization group} 
\newacronym{QMC}{QMC}{Quantum Monte Carlo} 
\newacronym{LSWT}{LSWT}{linear spin wave theory} 
\newacronym{GLSWT}{GLSWT}{generalized linear spin wave theory} 
\newacronym{LL}{LL}{Landau-Lifshitz} 
\newacronym{GLL}{GLL}{generalized Landau-Lifshitz} 
\newacronym{DSSF}{DSSF}{dynamical spin structure factor} 

\usepackage{floatrow}

\makeatother

\begin{document}
\title{Large-$N$ SU(4) Schwinger boson theory for coupled-dimer antiferromagnets}
\author{Shang-Shun~Zhang}
\affiliation{Department of Physics and Astronomy, The University of Tennessee, Knoxville, Tennessee 37996, USA}
\author{Yasuyuki Kato}
\affiliation{Department of Applied Physics, University of Fukui, Fukui 910-8507, Japan}
\author{E. A.~Ghioldi}
\affiliation{Department of Physics and Astronomy, The University of Tennessee, Knoxville, Tennessee 37996, USA}
\author{L. O.~Manuel}
\affiliation{Instituto de Física Rosario (CONICET) and Facultad de Ciencias Exactas, Ingeniería y Agrimensura, Universidad Nacional de Rosario, 2000 Rosario, Argentina}
\author{A. E.~Trumper}
\affiliation{Instituto de Física Rosario (CONICET) and Facultad de Ciencias Exactas, 
Ingeniería y Agrimensura, Universidad Nacional de Rosario, 2000 Rosario, Argentina}
\author{Cristian~D.~Batista}
\affiliation{Department of Physics and Astronomy, The University of Tennessee,
Knoxville, Tennessee 37996, USA}
\affiliation{Neutron Scattering Division and Shull-Wollan Center, Oak Ridge
National Laboratory, Oak Ridge, Tennessee 37831, USA}

\begin{abstract}
We develop a systematic large-$N$ expansion based on the Schwinger boson representation of SU(4) coherent states of dimers for the paradigmatic spin-$1/2$ bilayer square lattice Heisenberg antiferromagnet. This system exhibits a quantum phase transition between a quantum paramagnetic state and a Néel order state, driven by the coupling constant $g = J'/J$, which is defined as the ratio between the inter-dimer $J'$ and intra-dimer $J$ exchange interactions.  
We demonstrate that  this approach accurately describes static and  dynamic properties on both sides of the quantum phase transition. The critical coupling constant $g_c \approx 0.42$ and the dynamic spin structure factor reproduce quantum Monte Carlo results with high precision. Notably, the $1/N$ corrections reveal the longitudinal mode of the magnetically ordered phase along with the overdamping 
caused by its decay into the two-magnon continuum. 
The present large-$N$ SU(4) Schwinger boson theory can be extended to more general 
cases of quantum paramagnets that undergo a quantum phase transition into magnetically ordered states. 
\end{abstract}

\maketitle

\section{Introduction}

Networks of antiferromagnetic (AFM) dimers provide the simplest realization of quantum phase transitions between a quantum paramagnet (QPM) phase, adiabatically connected to a direct product of singlet states in each dimer, and a magnetically ordered phase~\cite{hester2019novel,jaime2004magnetic,ruegg2007multiple,ruegg2003bose,yamada2007magnetic,aczel2009bose,kofu2009magnetic,tsirlin2010microscopic,coldea2002direct,zheludev2007dynamics}. The elementary excitations of the QPM states are spin-one quasi-particles known as triplons. These arise from exciting one singlet 
dimer into a triplet state that propagates in a periodic dimer lattice with well-defined momentum. When the inter-dimer interaction $J'$ is much weaker 
than the intra-dimer exchange $J$, the single-triplon dispersion exhibits a spin gap of the order of $J$. As the ratio $g=J'/J$ increases, the gap reduces and often vanishes continuously at a critical value $g=g_c$, signaling the onset of long-range magnetic ordering via condensation of the soft triplon mode~\cite{zapf2014bose,giamarchi2008bose}. The magnetic ordering wave vector coincides with the wave vector of the triplon mode that becomes soft. While this simple and intuitive picture has been validated by multiple analytical, numerical, and experimental studies of quantum dimer magnets, an accurate and controlled theoretical description of these magnets has yet to be achieved.

A paradigmatic example is a spin-$1/2$ bilayer square lattice Heisenberg antiferromagnet~\cite{sachdev11}, where $J'$ connects nearest-neighbor spins on the same layer. 
This unfrustrated quantum model is free from the sign problem, allowing for accurate quantum Monte Carlo (QMC) simulations~\cite{Sandvik94,wang06,lohofer2015dynamical}. QMC studies reveal
a second order quantum phase transition (QPT) between a  QPM and a N\'eel ordered AFM  phase  at $g_c \approx 0.3965$. The QPT belongs to the 3D $O(3)$ universality class, where the spin gap closes as $\Delta = |g-g_c|^\nu$, with  a critical exponent $\nu \approx 0.71$~\cite{wang06}. 

Several other approaches have been employed to solve the bilayer antiferromagnet ~\cite{chubukov95,weihong97,kotov98,yu1999bond, rakhimov2011macroscopic,sandvik1995quantum,dodds2010theory,hida1992quantum,ganesh2011neel,lohofer2015dynamical,ng2017field,weber2022quantum,hering24}. 
From the magnetically ordered side, conventional linear spin-wave theory and its modified self-consistent versions, along with the Schwinger boson (SB) SU(2) mean-field approximation --that is, the large-$N$ SU(2) SB theory--, have incorrectly predicted a weak first-order transition at much smaller critical values of  $g_c$ (see Table ~\ref{table1}), in stark contrast with  QMC results. 
Chubukov and Morr~\cite{chubukov95} identified that the failure of these theories was due to the neglect of longitudinal fluctuations, specifically those along the local magnetization, which are expected to be significantly enhanced near the QPT. 

Approaching the problem from the quantum paramagnetic side, Chubukov~\cite{chubukov89}, and later Sachdev and Bhatt~\cite{sachdev1990bond}, introduced bond-operator theories (BOT), where spin operators within each dimer are expressed as bilinear forms of boson operators. These bond operators correspond to the four levels of a dimer: one singlet and three triplets. The BOT approach relies on correctly accounting for the degrees of freedom within each dimer, naturally incorporating intra-dimer entanglement, and has been applied to both the QPM and AFM phases. However, the BOT has a significant limitation due to the requirement of a local constraint on the number of bosons.

Broadly, three alternatives have been proposed to handle the constraint: (a) imposing it on average with a Lagrange multiplier within a mean-field approach where the singlet operator is replaced by a $c$-number~\cite{Sandvik94,yu1999bond}; (b) projecting out the double occupancy of triplets in each dimer~\cite{kotov98,joshi2015nonlinear,joshi2015nonlinear2}; and (c) using the constraint to express the singlet operator in terms of the triplet operators, akin to a Holstein-Primakoff transformation~\cite{matsumoto04}.

The mean-field treatment~\cite{yu1999bond} -case (a)— predicts a critical value $g_c$
more in line with the QMC result (see Table~\ref{table1}). However, a significant shortcoming of this approach is that the solution of the mean-field equations yields a spectrum for the Néel phase that violates the Goldstone theorem~\cite{zhang2013phase}. We attribute these problems to the absence of an expansion parameter.

This limitation of the bond-operator theory was addressed by introducing a systematic perturbation scheme~\cite{joshi2015nonlinear,joshi2015nonlinear2}-case (b)-, by reformulating the bond-operator approach as an expansion in $1/d$, where $d$ is the spatial dimension of the system. While this method preserves the Goldstone modes of the AFM phase order by order in $1/d$ and also predicts a continuous phase transition, the resulting value of $g_c$
is no longer close to the QMC result (see Table~\ref{table1}).

The four bond operators can be identified with the four Schwinger bosons associated with the fundamental irreducible representation (irrep) of SU(4)~\cite{sachdev1990bond,lecheminant2005phases,lecheminant2006competing}, as their bilinears faithfully represent the $\mathrm{SU}(4)$ generators of infinitesimal unitary transformations within the local four-dimensional Hilbert space of each dimer.  
This representation can be extended to any completely symmetric irrep of $\mathrm{SU}(4)$ by adjusting the local constraint from one  to $M$ bosons per dimer~\cite{dahlbom2024classical}, where $M$ is a positive integer that labels different irreps. Notably,  the classical limit of this theory is obtained by taking the large $M$ limit~\cite{Zhang21}. In this context, $M$ plays a role analogous to the spin value $S$ in the conventional SU(2) case.

A semi-classical (loop) expansion can be developed by implementing a Holstein-Primakoff transformation and expanding the bosonic propagators in powers of $1/M$, namely a generalized spin wave theory~\cite{muniz14,Zhang21} -case (c)-. However, to leading order, this large-$M$ expansion predicts a critical value $g_c \simeq 0.25$  which significantly deviates from the QMC result for the bilayer square lattice antiferromagnet. 

The SU(2) Schwinger boson theory can be refined by applying a Gutzwiller projection~\cite{miyazaki1996bilayer} to enforce the local constraint. This approach yields a more accurate critical value of $g_c$ compared to the mean-field result, although remains significantly lower than the QMC estimate (see Table~\ref{table1}). A more sophisticated variational scheme~\cite{liao2011variational} incorporating the Gutzwiller projection can further improve the accuracy of $g_c$; however, it fails to correctly capture the gapless Goldstone modes in the AFM phase. An alternative approach is to include $1/N$ corrections. However, both the Gutzwiller projection and the large-$N$ expansion originate from an SU(2) Schwinger boson representation of the spin-$S=1/2$ operators.

\begin{table}[h]
\begin{center}
\begin{tabular}{l| c|c |c|c} 
  \hline 
  \hline 
 Method & \hskip 0.5cm $g_c\;\;\;$ & Transition & Goldstone & Ref \\  
 \hline
 \hline
  QMC & 0.3965 & 2nd & yes & \cite{lohofer2015dynamical}\\ 
 \hline 
 LSWT & 0.0735 & 1st & yes & \cite{chubukov95} \\ 
 \hline
 MSWT  & 0.236 & 1st & yes & \cite{hida1990low}\\
 \hline
 SBMFT large-$N$ SU(2) SB & 0.2232 & 1st & yes & \cite{Millis93} \\
 \hline
 SB Gutzwiller & 0.285 & 2nd & - & \cite{miyazaki1996bilayer}\\
 \hline
 BOT - $\lambda$& 0.437 & 2nd & no & \cite{yu1999bond}\\
 \hline
 BOT - Brueckner & 0.389 & - & - & \cite{kotov98}\\
 \hline
  large-$d$ BOT  & 0.2968 &2nd  & yes & \cite{joshi2015nonlinear2}\\
 \hline 
 SE & 0.3942 & 2nd & yes &\cite{weihong97}\\
  \hline
  CST & 0.382  & 2nd & - & \cite{hering24}\\
  \hline
  large-$M$ SU(4) HP & 0.25 & 2nd  & yes  & \cite{lohofer2015dynamical}\\
    \hline 
 large-$N$ SU(4) SB & 0.42 & 2nd & yes & This work \\
 \hline
 \hline 
 \end{tabular}
\end{center}
  \caption{Representative numerical and analytical results for the bilayer Heisenberg quantum antiferromagnet. $g_c$ indicates the critical coupling, ``Transition'' the order of the phase transition, 
  "Goldstone`` indicates if the theory captures or not the Goldstone modes of the magnetically ordered phase.
  QMC stands for Quantum Monte Carlo, LSWT for linear spin wave theory, MSWT for modified spin wave theory, 
  large-$N$ SU(2) SB for the SU(2) Schwinger boson mean-field theory, SB Gutzwiller for SBMFT Gutzwiller projected, BOT-$\lambda$ for bond operator mean-field theory with the constraint imposed by a Lagrange multiplier, 
  BOT+Brueckner for bond-operator theory with a constraint of no triplet double-occupancy, large-$d$ BOT for the $1/d$ expansion of the BOT, 
  SE for dimer series expansions, CST for continuous similarity transformations, 
 large-$M$ SU(4) HP for the generalized Holstein-Primakoff spin wave theory (equivalent to large-$d$ BOT for $d=\infty$), and large-$N$ SU(4) SB corresponds to this work.}
  \label{table1}
  \end{table}

We emphasize the asymmetry between the aforementioned semi-classical  large-$M$ approach and the above-mentioned
large-$N$ method. While the former is implemented using SU(4) coherent states that capture intra-dimer entanglement at the classical level, the large-$N$ methods that have been implemented so far are based on SU(2) coherent states for each site of the dimer, requiring intra-dimer entanglement to be incorporated via fluctuations beyond the saddle-point (SP). One way to address this asymmetry  is to implement the latter using the same SU(4) Schwinger bosons that form the basis of the large-$M$ expansion. By capturing intra-dimer entanglement at the SP level of the SU(4) Schwinger boson theory, we can expect better agreement with numerical results.

In this work, we develop a systematic large-$N$ expansion based on the $\mathrm{SU}(4)$ coherent state description of a single dimer. This approach generalizes the original dimer problem with two antiferromagnetically coupled $\mathrm{SU}(2)$ spins on each site of the dimer to two interacting  $\mathrm{SU}(n)$ spins, with $n \geq 2$. Unlike the traditional SB approximation, which usually employs $\mathrm{SU}(n)$ SBs to represent the $\mathrm{SU}(n)$ spin components~\cite{arovas88}, we start from $\mathrm{SU}(N=n^2)$ SBs to include all quantum mechanical states  of the single-dimer problem in the manifold of $\mathrm{SU}(N)$ coherent states. 
This construction defines a family of SU($n$)-invariant models formulated in terms of Schwinger bosons transforming under SU($n^2$). Unlike conventional Schwinger boson theories on bipartite lattices~\cite{auerbach2012interacting}, where SU($n$) spins (i.e., generators of the SU($n$) group) interact exclusively through antiferromagnetic couplings, our model incorporates additional interaction terms beyond simple antiferromagnetic exchange. To emphasize this distinction, we refer to our approach as the ``large-$N$ SU(4) Schwinger boson theory,'' explicitly differentiating it from the traditional ``large-$N$ SU(2) Schwinger boson theory.''
Our theory gives rise to a controlled structure and a $1/N$ expansion of the BOT mean-field approximation, where the boson number constraint is imposed by a Lagrange multiplier.

We compute the dynamical susceptibility and take the limit $N=n^2 \to \infty$, using $1/N$ as the small expansion parameter.  
Remarkably, the leading-order  SP contribution already provides a very accurate description of both phases, QPM and AFM. 
The low energy spectrum of the QPM consist of a triplet of gapped triplon modes that become soft at the critical point, $g=g_c$.  The triplon condensation leads to the AFM state for $g>g_c$, where the transverse component of the dynamical spin structure factor (DSSF) is also accurately described by the SP approximation. Importantly, the DSSF exhibits magnon excitations with the expected transverse Goldstone modes 
and the value of $g_c \approx 0.42$ aligns well with QMC simulations. Moreover, the excitation spectrum and the spectral weight distribution of the dynamic spin structure factor (DSSF) also reproduce the QMC results with high precision.

Notably, the longitudinal mode of the magnetically ordered phase is also captured by the large-$N$ approach by including $1/N$ corrections beyond the SP level.
These corrections account for the overdamping caused by the decay of the longitudinal mode into the two-magnon continuum. These results, which also align well with existing numerical simulations, indicate that the large-$N$ 
approach effectively captures the essential physics of this system.

This paper is organized as follows. In Sec.~\ref{sec2}, we introduce the $\mathrm{SU}(4)$ coherent states of a single dimer. Sec.~\ref{sec3} presents the Schwinger boson theory for coupled dimer antiferromagnets based on these $\mathrm{SU}(4)$ coherent states. We generalize the theory 
by extending the $\mathrm{SU}(4)$ formulation to $\mathrm{SU}(N=n^2)$, corresponding to models of two antiferromagnetically coupled $\mathrm{SU}(n)$ spins on each dimer. A systematic expansion in powers of $1/N$ is then discussed within the path-integral formulation. In Sec.~\ref{sec4}, we explore the SP solution of the large-$N$ theory, presenting the continuous transition between the QPM and the N\'eel AFM order. In Sec.~\ref{sec:dssf}, we study the DSSF using the large-$N$ framework. Notably, we find that the SP approximation of the DSSF in the QPM phase and the transverse DSSF in the AFM phase agree very well with the QMC results (Sec.~\ref{sec:dssf1}). A correct description of the longitudinal DSSF requires the consideration of $1/N$ diagrams, which we discuss in Sec.~\ref{sec:dssf2}. Finally, the main conclusions of this study are summarized in Sec.~\ref{sec6}.

\vskip 1.cm


\section{$\mathrm {SU}(4)$ coherent-state description of a dimer}
\label{sec2}

\begin{figure}[!b]
\includegraphics[angle=0,width=\textwidth]{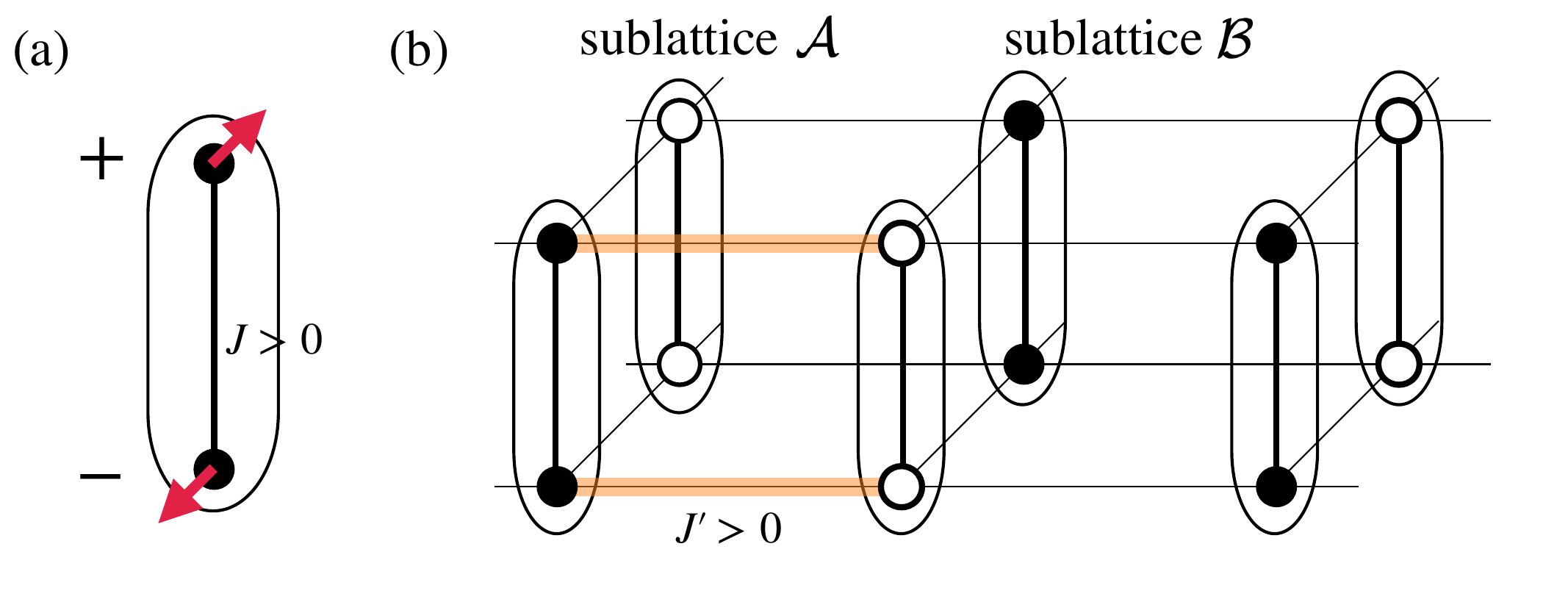}
\caption{(a) A single dimer formed by two $S=1/2$ spins coupled through antiferromagnetic interaction $J$. (b) Network of dimers forming a square lattice.
The splitting between sublattice ${\cal A}$ (white circles) and ${\cal B}$ (black circles) of the dimer lattice becomes relevant in presence of N\'eel AFM order.}
\label{Fig1}
\end{figure}

We consider a system of antiferromagnetically coupled dimers. Each dimer consists of two $S=1/2$ spins coupled through an antiferromagnetic Heisenberg interaction, described by the following Hamiltonian
\begin{equation}
\hat{\mathscr{H}}_0=J \sum_{j}\left(\hat{\boldsymbol{S}}_{j+} \cdot \hat{\boldsymbol{S}}_{j-}-\frac{1}{4}\right)
\end{equation}
The index $j$ labels the dimers, while $\pm$ refers to the two spins of each dimer (see Fig.~\ref{Fig1}(a)). The eigenstates of  a single dimer consist of a singlet ground state  and three degenerate triplet states, whose  wave functions and energy eigenvalues are
\begin{eqnarray}
| S=0 \rangle_j &=& \frac{1}{\sqrt{2}} (| \uparrow \downarrow \rangle - | \downarrow \uparrow \rangle),  \quad \quad \epsilon_s = -J,
\nonumber \\
| S=1, S^x=0 \rangle_j &=& \frac{1}{\sqrt{2}} (| \downarrow \downarrow \rangle -| \uparrow \uparrow \rangle) ,
\quad \quad 
\epsilon_{t,x} =  0, \nonumber \\
| S=1, S^y=0 \rangle_j &=& \frac{i}{\sqrt{2}} (| \uparrow \uparrow \rangle + | \downarrow \downarrow \rangle)  ,
\quad \quad 
\epsilon_{t,y} =  0,
\nonumber \\
| S=1, S^z=0 \rangle_j &=& \frac{1}{\sqrt{2}} (| \uparrow \downarrow \rangle + | \downarrow \uparrow \rangle) ,
\quad \quad \epsilon_{t,z} = 0.
\label{eq:singlet_triplet_basis}
\end{eqnarray}
In the absence of inter-dimer interaction, the ground state of the system is a product state of singlets on each dimer, and the excitations are local flips from a singlet to triplet (i.e., triplon excitation), with an energy gap $J$.
This solvable limit provides a qualitative picture for finite but weak inter-dimer interaction. The ground state is still a quantum paramagnet and the quantum mechanical state of each dimer retains a strong singlet character. In presence of translational invariance, the  elementary excitations are triplon modes with well-defined momentum that propagate through the lattice. 

The goal of this work is to understand the system's behavior upon increasing the inter-dimer interaction. Since the system may develop different types of instabilities due to the softening of the triplon modes, it is  important to introduce a formalism 
that can quantitatively describe these 
QPTs,
as well as the low-energy excitation spectrum on both sides of the  transition. 

A crucial consideration in choosing an adequate formalism is to note that each dimer is an ``entangled unit'' in the sense that the corresponding  wave function has a strong singlet character. It is then convenient to use a formalism in which the singlet state of each dimer is a coherent state of the Lie algebra associated with bosonic operators that are used to represent the spin operators. More specifically, a standard approach to the problem would start with a faithful representation of the spin operators at each site $j  \sigma$ ($\sigma=\pm$ denotes the layer) in terms of $\mathrm{SU}(2)$ SB
$\hat{b}^{\dagger}_{j \sigma, \mu}$ and 
$\hat{b}^{\;}_{j\sigma, \mu}$ ($\mu= \uparrow \downarrow$),
\begin{eqnarray}
\hat{S}^+_{j \sigma} &=& \hat{b}^{\dagger}_{j \sigma,\uparrow} \hat{b}^{\;}_{j \sigma,\downarrow}, \quad 
\hat{S}^-_{j \sigma} = \hat{b}^{\dagger}_{j \sigma,\downarrow} \hat{b}^{\;}_{j \sigma,\uparrow}, 
\nonumber \\
\hat{S}^z_{j \sigma} &=& \frac{1}{2} (\hat{b}^{\dagger}_{j \sigma,\uparrow} \hat{b}^{\;}_{j\sigma,\uparrow} - \hat{b}^{\dagger}_{j \sigma,\downarrow} \hat{b}^{\;}_{j \sigma,\downarrow}),
\label{eq:SU2gen}
\end{eqnarray}
that fulfill the constraint:
\begin{equation}
\hat{b}^{\dagger}_{j \sigma,\downarrow} \hat{b}^{\;}_{j \sigma,\downarrow}  + \hat{b}^{\dagger}_{j \sigma,\uparrow} \hat{b}^{\;}_{j \sigma,\uparrow} = 2S,
\end{equation}
where $S$  refers to the spin size, which labels the irrep of $\mathrm{SU}(2)$ that determines the matrix form of the generators given in Eq.~\eqref{eq:SU2gen}. Note that $S=1/2$ for the particular case of interest. However, as we mentioned in the Introduction, a large-$N$ approximation based on these SBs is not quantitatively accurate because the  path-integral formulation is parametrized in terms of direct products of  $\mathrm{SU}(2)$ coherent states. In other words, the intra-dimer entanglement must be built in via quantum fluctuations that favor linear combination of these product states. 

The above-mentioned problem can be avoided by introducing bosons with four flavours (instead of two), $\hat{b}^{\dagger}_{j, \mu}$ and $\hat{b}^{\;}_{j, \mu}$ ($\mu=0,1,2,3$),
that fulfill the constraint 
\begin{equation}
\sum_{\mu=0}^3 \hat{b}^{\dagger}_{j, \mu} \hat{b}^{\;}_{j, \mu} = M,
\label{eq:const}
\end{equation}
where $M$ is a positive integer ($M=1,2,3, ...$).
Bilinear forms in these bosons,
\begin{equation}
\hat{\cal S}^{\mu \nu}_j =    \hat{b}^{\dagger}_{j, \mu} \hat{b}_{j, \nu},
\end{equation}
with commutation relations
\begin{equation}
[ \hat{\cal S}^{\alpha \beta}_j, \hat{\cal S}^{\mu \nu}_j] = \delta_{\beta \mu} \hat{\cal S}^{\alpha \nu}_j - \delta_{\alpha \nu} \hat{\cal S}^{\mu \beta}_j,
\end{equation}
provide a faithful representation of generators of $\mathrm{SU}(4)$ in the completely symmetric irreps labelled by the integer $M$ (e.g. $M=1$ for the fundamental irrep of $\mathrm{SU}(4)$). 
Note that unlike the $\mathrm{SU}(2)$ SBs, these bosons have only a dimer index $j$ because they create all the possible quantum mechanical states of each dimer. In other words, 
the $\mathrm{SU}(4)$ SBs create coherent states of the $\mathrm{SU}(4)$ Lie algebra (for completely symmetric irreps) that span the CP$^3$ manifold of quantum mechanical states of a four-level quantum mechanical system. 

To make contact with our problem of interest, we can regard the single dimer $j$ as the four-level system, set $M=1$ (fundamental irrep of $\mathrm{SU}(4)$) and choose the basis of $\mathrm{SU}(4)$ SBs, such that the 
$\mu=0$ boson creates the singlet state:
\begin{equation}
| S=0 \rangle_j = \hat{b}^{\dagger}_{j,0} | 0\rangle,
\end{equation}
while the other three bosons $\mu=1,2,3$ create the three triplet states (corresponding to the index $\mu=x,y,z$ )
\begin{equation}
| S=1, S^\mu=0\rangle_j=\hat{b}^{\dagger}_{j,\mu} | 0\rangle.
\end{equation}

The important difference between the $\mathrm{SU}(2)$ and the $\mathrm{SU}(4)$ SBs is that coherent states of the latter include entangled intra-dimer singlet and triplet states, meaning that the intra-dimer entanglement is already built-in the $\mathrm{SU}(4)$ coherent states of the path-integral formulation and only the inter-dimer entanglement must be generated via the inclusion of quantum fluctuations. 
As expected from this simple observation, we will see in the next sections that the large-$N$ expansion of the $\mathrm{SU}(4)$ SB theory is indeed much more accurate than the large-$N$ expansion of the $\mathrm{SU}(2)$ SB theory for low-dimensional coupled-dimer systems. 

\section{$\mathrm {SU}(4)$ Schwinger boson theory}
\label{sec3}
To be specific, we shall consider a double-layer square lattice antiferromagnets, where two $S=1/2$ spins on the vertical inter-layer bonds connecting two layers form a antiferromagnetic dimer (see Fig.~\ref{Fig1} (b)). The spin Hamiltonian is the sum of the intra-dimer and inter-dimer spin Hamiltonians 
$\hat{\mathscr{H}}_{0}$ and $\hat{\mathscr{H}}^{\prime}$, respectively:
\begin{eqnarray}
\hat{\mathscr{H}} = \hat{\mathscr{H}}_{0} + \hat{\mathscr{H}}^{\prime},
\end{eqnarray}
where
\begin{equation}
\hat{\mathscr{H}}^{\prime}= J^{\prime} \sum_{j,
\delta} \sum_{\sigma=\pm} \hat{\boldsymbol{S}}_{j, \sigma} \cdot \hat{\boldsymbol{S}}_{j+\delta, \sigma},
\end{equation}
with $j$ being the dimer index, and $\delta=x, y$ running over the two inter-bond links associated with dimer $j$, and $j+\delta$ denoting the neighboring dimer connected to dimer $j$ through bond $\delta$.
$J^\prime$ is the inter-dimer spin-exchange illustrated in Fig.~\ref{Fig1}(b). Up to the overall energy scale $J$, this problem is solely characterized by the dimensionless coupling constant $g=J^\prime/J$.  


Since the $\mathrm{SU}(4)$ generators in the fundamental irrep ($M=1$), together with the identity, form a complete basis for the vector space of operators acting on the 4-dimensional Hilbert space of a single dimer,
the spin operators on each site of the dimer can be expressed as a linear combination of these generators
\begin{eqnarray}\label{eq:spin_SB}
    \hat{S}_{j \pm}^{\mu} &=& \pm  \frac{1}{2}\left( \hat{\cal S}_j^{0 \mu} + \hat{\cal S}_j^{\mu 0} \right) - \frac{i}{2} \sum_{\nu, \rho=1}^ 3 \epsilon^{\mu \nu \rho} \hat{\cal S}_j^{\nu \rho}.
\end{eqnarray}

In terms of the $\mathrm{SU}(4)$  SB, the intra-dimer Hamiltonian takes the diagonal form:
\begin{equation}\label{eq:h0}
\hat{\mathscr{H}}_0= - J \sum_{j}  \hat{\cal S}^{00}_j = - J \sum_{j}  \hat{b}^{\dagger}_{j,0} \hat{b}^{\;}_{j,0}. 
\end{equation}
Given the local constraint in Eq.~\eqref{eq:const} with $M=1$, the ground state $|\Psi_0\rangle$ of $\hat{\mathscr{H}}_0$ (direct product of singlet states) satisfies  $\hat{b}_{j,0}^\dagger \hat{b}_{j,0} |\Psi_0\rangle= |\Psi_0\rangle$ for $\forall j$. 


In the $\mathrm{SU}(4)$ SB theory, we rewrite the inter-dimer Hamiltonian in terms of $\mathrm{SU}(2)$-invariant ``link'' operators~\footnote{In the literature, these are usually called ``bond operators''. In this work we will call them ``link operators'' to distinguish them from the ``bond operators'' acting on each dimer unit.}
\begin{eqnarray}
\hat{\mathscr{H}}^{\prime} &=& \frac{J^\prime}{2} \sum_{j,\delta}  \left( \hat{S}_{j,\delta}^{\dagger} \hat{A}_{j,\delta} + \hat{T}_{j,\delta}^{\dagger} \hat{B}_{j,\delta} + \mathrm{h.c.}  \right)  \nonumber \\
&+& \frac{J^\prime}{2} \sum_{j,\delta}  \left( :\hat{B}_{j,\delta}^{\dagger} \hat{B}_{j,\delta}: - \hat{A}_{j,\delta}^{\dagger} \hat{A}_{j,\delta} \right),
\label{Eq:dec}
\end{eqnarray}
where $:\hat{B}_{j,\delta}^{\dagger} \hat{B}_{j,\delta}:$ indicates normal ordering of the operator $\hat{B}_{j,\delta}^{\dagger} \hat{B}_{j,\delta}$,
and the four $\mathrm{SU}(2)$ invariant link operators are
\begin{align}
\label{eq:bond_opr}
&\hat{S}_{j,\delta} = \hat{b}_{j,0}\hat{b}_{j+\delta,0}, \qquad \ 
\hat{T}_{j,\delta}^{\dagger} = \hat{b}_{j,0}^\dagger \hat{b}_{j+\delta,0}, \\
&\hat{A}_{j,\delta} = \sum_{\mu=1}^3 \hat{b}_{j,\mu}\hat{b}_{j+\delta,\mu}, \  \hat{B}_{j,\delta}^{\dagger} = \sum_{\mu=1}^3 \hat{b}_{j,\mu}^\dagger \hat{b}_{j+\delta,\mu}.
\end{align}
We note that the singlet boson $\hat{b}_{j,0}$ remains invariant (trivial irrep) under a global $\mathrm{SU}(2)$ rotation of the dimer $j$, while the triplet bosons $\hat{b}_{j,\mu}$ ($\mu=1,2,3$)
transform according to the $L=1$ (adjoint) irrep. The $\mathrm{SU}(2)$ invariance of $\hat{S}_{j,\delta}$ and $\hat{T}_{j,\delta}^{\dagger}$ follows directly from the singlet character of $\hat{b}^{\dagger}_{j,0}$ and $\hat{b}_{j,0}$. The other link operators, $\hat{A}_{j,\delta}$ and  $\hat{B}_{j,\delta}^{\dagger}$, are the two singlets  obtained by considering the two possible direct products of  bosons that transform like triplets, $1 \otimes 1 = 0 \oplus 1 \oplus 2 $, and projecting into the singlet $(0)$ component. 

$\hat{\mathscr{H}}^{\prime}$ in Eq.~(\ref{Eq:dec}) is expressed in a form that is manifestly SU(2) symmetric.
Physically, the different terms of $\hat{\mathscr{H}}^{\prime}$ can be interpreted in the following way: 
$\hat{A}^{\dagger} \hat{S}$ represents the process of destroying two neighboring singlets followed by the creation of two triplets, which together form a singlet (${\rm h.c.}$ corresponds to the inverse process); $\hat{T}^{\dagger} \hat{B}$ 
the exchange between singlet and triplet neighbours,  $:\hat{B}^{\dagger} \hat{B}:$ the exchange between neighboring triplets, and $\hat{A}^{\dagger} \hat{A}$ projects onto singlet formed between two neighboring triplets.

For strong enough inter-dimer interaction, the system undergoes a transition to a N\'eel AFM state, where the magnetic moments on 
nearest-neighbors within the same layer or within the same dimer are anti-aligned. Since this magnetic order breaks translation symmetry, we switch to a twisted reference frame  in which the ground state remains translationally invariant and the magnetic ordering is ferromagnetic in each layer, but antiferromagnetic between layers. To define the twisted reference frame it is convenient to introduce two interpenetrated  sublattices ${\cal A}$ and ${\cal B}$  of dimers (see Fig.~\ref{Fig1}). The twisted reference frame is obtained by applying a \(\pi\)-rotation about the spin-\(z\) axis on spins of the ${\cal B}$ sublattice: $S_{j\sigma}^{x/y}\to -S_{j\sigma}^{x/y}$, $S_{j\sigma}^{z}\to S_{j\sigma}^{z}$.
Under this transformation, the singlet and the three triplet states in Eq.~(\ref{eq:singlet_triplet_basis}) transform in the following way:
\begin{eqnarray}
| S=0 \rangle_j &\to& | S=0 \rangle_j,
\nonumber \\
| S=1, S^z=0 \rangle_j &\to& | S=1, S^z=0 \rangle_j,
\nonumber \\
| S=1, S^x=0 \rangle_j &\to& e^{i \bm{\pi} \cdot \bm{r}_j} | S=1, S^x=0 \rangle_j,
\nonumber \\
| S=1, S^y=0 \rangle_j  &\to& e^{i \bm{\pi} \cdot \bm{r}_j} | S=1, S^y=0 \rangle_j,
\end{eqnarray}
where ${\bm \pi}=(\pi,\pi)$, 
with the lattice constant taken as the length unit, 
and $\bm{r}_j$ is the position vector of the dimer $j$.
This gives rise to the following transformation of SBs,
\begin{eqnarray}
\hat{b}_{j,0} \rightarrow \hat{b}_{j,0}, \quad \hat{b}_{j,1} &\rightarrow & \hat{b}_{j,1} e^{i\bm{\pi}\cdot\bm{r}_j},
\nonumber \\
\hat{b}_{j,3} \rightarrow \hat{b}_{j,3}, \quad \hat{b}_{j,2} &\rightarrow & \hat{b}_{j,2} e^{i\bm{\pi}\cdot\bm{r}_j}.
\end{eqnarray}
In the twisted reference frame, the 
SB
representation of the spin operators is still given by Eq.~(\ref{eq:spin_SB}), while the four link operators transform to
\begin{align}\label{eq:bond_opr2}
\hat{S}_{j,\delta} &= \hat{b}_{j,0}\hat{b}_{j+\delta,0}, \ \
\hat{T}_{j,\delta}^{\dagger} = \hat{b}_{j,0}^\dagger \hat{b}_{j+\delta,0}, \\
\hat{A}_{j,\delta} &= \hat{b}_{j,3}\hat{b}_{j+\delta,3} - \hat{b}_{j,2}\hat{b}_{j+\delta,2} - \hat{b}_{j,1}\hat{b}_{j+\delta,1}, \\
\hat{B}_{j,\delta}^{\dagger} &= \hat{b}_{j,3}^\dagger \hat{b}_{j+\delta,3}  - \hat{b}_{j,2}^\dagger \hat{b}_{j+\delta,2} - \hat{b}_{j,1}^\dagger \hat{b}_{j+\delta,1}.
\end{align}
Unless otherwise specified, we will operate within this twisted reference frame.

\subsection{Large-$N$ generalization}

To study this interacting system of SBs, we employ the large-$N$ technique, extending the original $\mathrm{SU}(2)$-invariant model to a broader class of $\mathrm{SU}(n)$ Hamiltonians, expressed in terms SBs with  $N = n^2$ flavors. Notably, there are multiple ways to construct an $\mathrm{SU}(n)$-invariant interaction term using the generators of the $\mathrm{SU}(N)$ group. Our choice is guided by the following considerations:
\begin{itemize}
\item The generalized Hamiltonian is $SU(n)$-invariant.  
\item The form of the generalized Hamiltonian should be preserved  across different values of $n$. This ensures a unified approach to analyzing the system for arbitrary $n$. 
\end{itemize}
Regarding the second condition, note that if we were to generalize the model using $\mathrm{SU}(n)$-invariant forms in the $\mathrm{SU}(n)$ spins, the corresponding bond-operator representation would involve a varying number of bond operators depending on the value of $n$, introducing unnecessary complexity in taking the large-$N$ limit.

The Hamiltonian that satisfies both requirements is given by:
\begin{equation}
\hat{\mathscr{H}} = \hat{\mathscr{H}}_0 + \hat{\mathscr{H}}',
\end{equation}
where
\begin{equation}\label{eq:h0}
\hat{\mathscr{H}}_0 = - J \sum_{j} \hat{b}^{\dagger}_{j,0}\hat{b}_{j,0},
\end{equation}
and
\begin{eqnarray}\label{eq:dec2}
\hat{\mathscr{H}}'& = & \frac{2J'}{N} \sum_{j,\delta} \left(\hat{S}_{j,\delta}^{\dagger} \hat{A}_{j,\delta} + \hat{T}_{j,\delta}^{\dagger}\hat{B}_{j,\delta} + \mathrm{h.c.}\right) \nonumber \\
 &+& \frac{2J^\prime}{N}\sum_{j,\delta}\left(\hat{B}_{j,\delta}^{\dagger}\hat{B}_{j,\delta}-\hat{A}_{j,\delta}^{\dagger}\hat{A}_{j,\delta}\right).
\end{eqnarray}
The link operators introduced above are defined as:
\begin{align}
\hat{S}_{j,\delta}&=\hat{b}_{j,0}\hat{b}_{j+\delta,0}, & \hat{T}_{j,\delta}&=\hat{b}^{\dagger}_{j,0}\hat{b}_{j+\delta,0},\\
\hat{A}_{j,\delta}&=\sum_{\mu=1}^{N-1}\hat{b}_{j,\mu}\hat{b}_{j+\delta,\mu}, & \hat{B}_{j,\delta}&=\sum_{\mu=1}^{N-1}\hat{b}^{\dagger}_{j,\mu}\hat{b}_{j+\delta,\mu}.
\end{align}
Furthermore, we generalize the Schwinger boson number constraint \eqref{eq:const} to:
\begin{equation}\label{eq:constraint2}
\sum_{\mu=0}^{N-1}\hat{b}^{\dagger}_{j,\mu}\hat{b}_{j,\mu} = M,
\end{equation}
where $M$ is an arbitrary integer. For our original problem, we have $N=4$ and $M=1$.

The physical meaning of the generalized model introduced above becomes most transparent for the special case $M=1$. In this scenario, the intra-dimer Hamiltonian, $\hat{\mathscr{H}}_0$, describes dimers formed by two antiferromagnetically coupled $\mathrm{SU}(n)$ spins:
\begin{equation}
\hat{\mathscr{H}}_0 \propto \sum_j \sum_{\mu=0}^{N-1} \hat{O}_{j,+}^{\mu} \hat{\tilde{O}}_{j,-}^{\mu} + {\rm const.},
\end{equation}
where $\hat{O}^{\mu}_{j,+}$ represent the generators of the $\mathrm{SU}(n)$ group acting on the ``$+$'' site of the dimer in the fundamental irreducible representation, and $\hat{\tilde{O}}_{j,-}^{\mu}$ represent those on the ``$-$'' site in the conjugate representation.  Given the reducible tensor product of these representations, $n \otimes \bar{n} = 1 \oplus (n^2 - 1)$, the energy spectrum of $\hat{\mathscr{H}}_0$ consists of a singlet ground state and an $(N-1)$-fold degenerate multiplet transforming as the adjoint irrep of SU($n$), denoted by $\rvert \mu \rangle$. Analogously to the original problem, each energy level can be associated with Schwinger bosons as $\rvert \mu \rangle = \hat{b}_{j,\mu}^\dagger \rvert 0 \rangle$. In this framework, $\hat{b}_{j,0}$ describes an $\mathrm{SU}(N)$ singlet, while the remaining $N-1$ flavors, $\hat{b}_{j,\mu}$ (with $\mu = 1,2,\dots,N-1$), transform according to the self-conjugate adjoint representation of $\mathrm{SU}(N)$. With this identification, the $\mathrm{SU}(n)$ generators are expressed as:
\begin{equation}\label{eq:generator_largeN}
\hat{O}^{\mu}_{j,+} = \frac{1}{\sqrt{2n}}(\hat{b}_{j,0}^\dagger \hat{b}_{j,\mu} + \mathrm{h.c.}) + \sum_{\nu,\rho=1}^{N-1} \frac{d_{\mu \nu\rho} - i f_{\mu \nu\rho}}{2}\hat{b}_{j,\nu}^\dagger \hat{b}_{j,\rho},
\end{equation}
and
\begin{equation}\label{eq:generator_largeN2}
\hat{\tilde{O}}^{\mu}_{j,-} = \frac{-1}{\sqrt{2n}}(\hat{b}_{j,0}^\dagger \hat{b}_{j,\mu} + \mathrm{h.c.}) - \sum_{\nu,\rho=1}^{N-1} \frac{d_{\mu \nu\rho} + i f_{\mu \nu\rho}}{2}\hat{b}_{j,\nu}^\dagger \hat{b}_{j,\rho},
\end{equation}
where the indices satisfy $1 \leq \mu, \nu, \rho \leq N-1$. The coefficients $f_{\mu\nu\rho}$ are the structure constants of the $\mathfrak{su}(n)$ Lie algebra, which are fully antisymmetric under permutation of indices, and $d_{\mu\nu\rho}$ is a completely symmetric third-rank tensor defined through the anticommutator of $\mathrm{SU}(n)$ generators:
\begin{equation}
d_{\mu\nu\rho} = 2\operatorname{Tr}\left[\{T^\mu, T^\nu\}T^\rho\right].
\end{equation}
In the specific case of $\mathrm{SU}(2)$, $d_{\mu\nu\rho} \equiv 0$ and $f_{\mu\nu\rho} = \epsilon_{\mu\nu\rho}$, thereby recovering the familiar expression for spin operators given in Eq.~(\ref{eq:spin_SB}).

Translating the inter-dimer Hamiltonian into the language of ${\mathrm{SU}}(n)$ generators leads to a more intricate structure that inherently includes terms beyond simple inner products of the form $\sum_{\mu=0}^{N-1} \hat{O}_{j,\pm}^{\mu} \hat{\tilde{O}}_{k,\pm}^{\mu}$. Consequently, this generalized family of models departs significantly from conventional ${\mathrm{SU}}(n)$-invariant constructions based on the standard Schwinger boson framework, which relies on the ${\mathrm{SU}}(2)$ representation of spin-$1/2$ operators on bipartite lattices~\cite{auerbach2012interacting}. In fact, there are multiple ways to construct ${\mathrm{SU}}(n)$-invariant models using the $n^2 - 1$ generators of the ${\mathrm{SU}}(n)$ group. The particular approach adopted here introduces a unified representation of link operators that preserves the structure of the inter-dimer Hamiltonian for arbitrary $n$. This construction plays a key role in the broader theoretical framework developed in subsequent sections. Notably, all four link operators---$\hat{S}_{j,\delta}$, $\hat{T}_{j,\delta}$, $\hat{A}_{j,\delta}$, and $\hat{B}_{j,\delta}$---are explicitly ${\mathrm{SU}}(n)$-invariant, thereby ensuring that the generalized Hamiltonian $\hat{\mathscr{H}}$ respects ${\mathrm{SU}}(n)$ symmetry.

\begin{figure}
    \centering
    \includegraphics[width=\textwidth]{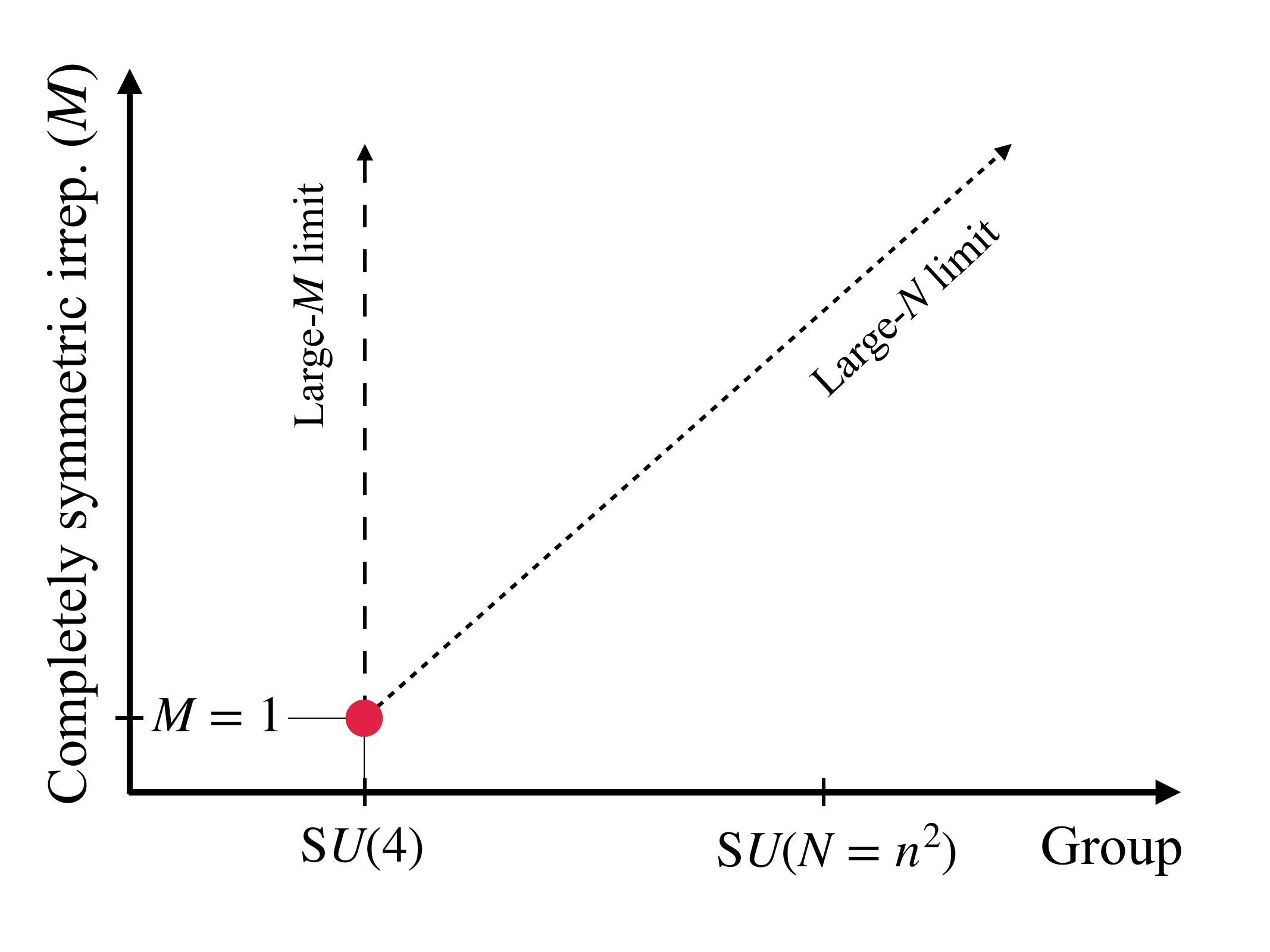}
    \caption{Large-$N$ versus large-$M$ limit of the extended model. Large-$M$ limit corresponds to $M\to \infty$ while fixing $N$ ($N=4$ for the case of interest). The large-$N$ (SB theory) corresponds to $N\to \infty$ under a fix ratio $\kappa=M/N$, i.e., along the dotted  line. }
    \label{fig:limits}
\end{figure}

For $M > 1$, both intra- and inter-dimer Hamiltonians include terms that go beyond simple ${\mathrm{SU}}(n)$-invariant inner products of operators in their higher-dimensional irreps. For example, the energy spectrum of the extended intra-dimer Hamiltonian, $\hat{\mathscr{H}}_0$, consists of equally spaced energy levels separated by $J$, with a non-degenerate singlet as the ground state. However, this spectrum does not match that of coupled ${\mathrm{SU}}(n)$ generators. The primary motivation for considering large values of $M$ is to enable theoretical approximations, such as semiclassical analyses via the large-$M$ approach (where $M \to \infty$) and the large-$N$ approach discussed herein. The former method captures the classical limit using ${\mathrm{SU}}(N)$ coherent states, with quantum corrections systematically included through a $1/M$ expansion~\cite{Zhang21,Dahlbom22,Dahlbom22b}. The latter, the large-$N$ approach, is especially useful for describing quantum states near critical points, such as the QPM-AFM transition explored in this paper. Importantly, $M$ must scale proportionally with $N$, maintaining a finite ratio $\kappa = M/N$, to ensure a nontrivial large-$N$ limit. If this condition is not met, the large-$N$ limit would lead to a vacuum of Schwinger bosons as the ground state, with local creation of Schwinger bosons as the excitations—states that are not of interest here. The distinction between the large-$M$ and large-$N$ approaches is despicted in Fig.~\ref{fig:limits}.

\subsection{Path-integral formulation}
\label{sec3b}
The path-integral formulation is implemented by introducing identities expressed in terms of the SU($N=n^2$) coherent states of each dimer, rather than the more conventional approach using $\mathrm{SU}(n)$ coherent states of each spin~\cite{arovas88,auerbach2012interacting}:
\begin{eqnarray}
    {\cal Z} &=& \int {\cal D}[\bar{b} b \lambda] \exp \left[ -\int_0^\beta d \tau \left( \sum_{j\mu} \bar{b}_{j,\mu} \partial_{\tau} b_{j,\mu} + \mathscr{H}(\bar{b},b) \right) \right. \nonumber \\
   && \left. + i \sum_j \lambda_j (\sum_{\mu} \bar{b}_{j,\mu} b_{j,\mu} - \kappa N) \right],
\end{eqnarray}
where $\tau$ denotes the imaginary time, $\beta$ the inverse of temperature, $\mathscr{H}(\bar{b},b)=\mathscr{H}_0(\bar{b},b)+\mathscr{H}^\prime(\bar{b},b)$ the Hamiltonian density, obtained from $\hat{\mathscr{H}}_0+\hat{\mathscr{H}}^\prime$ by replacing the operators $\hat{b}_{j,\mu}^\dagger,\hat{b}_{j,\mu}$ with complex fields $\bar{b}_{j,\mu},  b_{j,\mu} $ that parametrize the CP$^{N-1}$ manifold of $\mathrm{SU}(N=n^2)$ coherent states. The additional fields are the Lagrange multipliers $\lambda_j$, which enforce the local constraint given in Eq.~(\ref{eq:constraint2}).

To decouple the  terms that are quartic in the SBs, we first rewrite the link-operator factorization of ${\mathscr H}^\prime$ (see Eq.~(\ref{eq:dec2})) in the matrix form
\begin{eqnarray}
\mathscr{H}^{\prime} = {4\over N} \sum_{j,\delta}  \Psi_{j,\delta}^\dagger {\cal J}_{j,\delta} \Psi_{j,\delta},
\label{Eq:dec3}
\end{eqnarray} 
where $\Psi_{j,\delta} \equiv (S_{j,\delta},T_{j,\delta},A_{j,\delta},B_{j,\delta})^T$ denotes a 4-component link field~\footnote{These link-fields are usually called ``bond-fields'' in the literature. Here we use the name ``link-fields'' to distinguish them from the intra-dimer ``bond-fields''.}, ``$\dagger$'' denotes the conjugate transpose of a complex vector, and 
\begin{eqnarray}
    {\cal J}_{j,\delta} = {J^\prime \over 2} \left(
\begin{array}{cccc}
    0 &  0  &  1  & 0 \\
    0 &  0  &  0  &  1 \\
    1 &  0 &  -1 & 0 \\
    0 &  1  & 0   & 1 
\end{array}
\right),
\end{eqnarray}
is a positive-defined matrix. 
Next, a Hubbard-Stratonovich (HS) transformation is implemented by introducing the auxiliary complex field $W_{j,\delta} = (W_{j,\delta}^{(S)},W_{j,\delta}^{(T)},W_{j,\delta}^{(A)},W_{j,\delta}^{(B)})^T$,  that are conjugate  of the $\Psi$-fields
~\cite{ghioldi2018dynamical}: 
\begin{eqnarray}
    {\cal Z} = \int {\cal D}[\bar{W}W \bar{b} b \lambda] \exp \left[ - S(\bar{W},W,\bar{b},b,\lambda) \right],
\end{eqnarray}
where the action reads
\begin{widetext}
    \begin{eqnarray}
    S(\bar{W},W,\bar{b},b,\lambda) &=& \int_0^\beta d \tau \left[ \left( \sum_{j\mu} \bar{b}_{j,\mu} \partial_{\tau} b_{j,\mu} + {\mathscr{H}}_0(\bar{b},b) \right) + \sum_{j,\delta} \left( {N \over 4} \bar{W}_{j,\delta} {\cal J}_{j,\delta} W_{j,\delta} - \bar{W}_{j,\delta} {\cal J}_{j,\delta} \Psi_{j,\delta} + \Psi_{j,\delta}^\dagger {\cal J}_{j,\delta} W_{j,\delta} \right) \right. \nonumber \\
    &+& \left.  i \sum_j \lambda_j (\sum_{\mu} \bar{b}_{j,\mu} b_{j,\mu} - \kappa N) \right],
\end{eqnarray}
\end{widetext}
and $\bar{W}_{j,\delta}$ refers to the conjugate of the link field $W_{j,\delta}$. 
The phase fluctuations of the auxiliary fields represent the emergent gauge fluctuations of the SB theory~\cite{arovas88,auerbach2012interacting,ghioldi2018dynamical}. 


The new expression for the action is quadratic in the bosonic fields. 
In addition to the quadratic terms arising from ${\mathscr H}_0$ and from the on-site  Lagrange multiplier $\lambda_j$, there are complex hopping/pairing link terms arising from  inter-dimer interactions. 
The large-$N$ expansion is obtained after integrating out the bosonic variables,
\begin{eqnarray}
    e^{-{N\over 4} S_{\rm eff}(\bar{W},W,\lambda)} = \int {\cal D}[\bar{b} b] e^{-S(\bar{W},W,\bar{b},b,\lambda)},
\end{eqnarray}
and  expanding the resulting  effective action $S_{\rm eff}(\bar{W},W,\lambda)$ in the auxiliary fields $\bar{W}$, $W$, and $\lambda$ around the SP solution. The SP condition, 
\begin{equation} 
\left.\frac{\delta S_{\rm eff}}{\delta W_{j,\delta}} \right\rvert_{\rm sp} = 0, \ \ \left.\frac{\delta S_{\rm eff}}{\delta \bar{W}_{j,\delta}}\right\rvert_{\rm sp} = 0, \ \ \left.\frac{\delta S_{\rm eff}}{\delta \lambda_j} \right\rvert_{\rm sp} = 0,
\end{equation}
gives rise to the set of self-consistent equations
\begin{eqnarray}\label{eq:sp_cond}
W_{j,\delta}\rvert_{\rm sp} &=& {4\over N}\langle {\Psi}_{j,\delta}\rangle_{\rm sp},
\nonumber \\ 
\bar{W}_{j,\delta}\rvert_{\rm sp} &=& -{4\over N}\langle {\Psi}_{j,\delta}^\dagger \rangle_{\rm sp}, 
\nonumber \\
\kappa N &=& \sum_{\mu} \langle \bar{b}_{j,\mu} b_{j,\mu} \rangle_{\rm sp},
\end{eqnarray}
where the average is taken using the SP action $S_{\rm sp}(\bar{b}, b) \equiv S(\bar{W}_{\rm sp}, W_{\rm sp}, \bar{b}, b,\lambda_{\rm sp})$:
\begin{equation}\label{eq:sp_average}
\langle O \rangle_{\rm sp}    \equiv {\int {\cal D}[\bar{b} b] O e^{- S_{\rm sp}}\over \int {\cal D}[\bar{b} b] e^{- S_{\rm sp}}}.
\end{equation}
Since the SP bosonic Hamiltonian must be hermitian, the SP solution must satisfy 
$\bar{W}_{j,\delta}\rvert_{\rm sp} = - (W_{j,\delta}\rvert_{\rm sp})^*$ (here $*$ refers to complex conjugate)
and  $\lambda_j \rvert_{\rm sp}= -i \tilde{\lambda}_{j}\rvert_{\rm sp}$ with $\tilde{\lambda}_{j}\rvert_{\rm sp} \in \mathbb{R}$~\cite{ghioldi2018dynamical}. These SP equations correspond to the self-consistent equations of the mean-field SB theory in the canonical formalism.
One can verify that the  solutions $W_{j,\delta}\rvert_{\rm sp}$,  $\bar{W}_{j,\delta}\rvert_{\rm sp}$, and $\tilde{\lambda}_{j}\rvert_{\rm sp}$ scale as $N^0$, and consistently the energy spectrum of the bosons scales as $N^0$ as well.

To account for the fluctuations of  the auxiliary link fields, we introduce 
\begin{equation}
    \Phi_j \equiv \left( W_{j}^T-(W_{j}\rvert_{\rm sp})^T, \bar{W}_{j}^T-(\bar{W}_{j}\rvert_{\rm sp})^T, \lambda_j-\lambda_j\rvert_{\rm sp} \right)^T,
\end{equation}
where $W_{j} \equiv ((W_{j,x})^T, (W_{j,y})^T)^T$
and $\Phi_j^{\lambda} = \lambda_j-\lambda_j\rvert_{\rm sp}$ are the real fluctuations of the Lagrange multiplier fields. We denote the conjugate of $\Phi_j$ as
\begin{equation}
    \bar{\Phi}_j \equiv \left( \bar{W}_{j}^T-(\bar{W}_{j}\rvert_{\rm sp})^T, W_{j}^T-(W_{j}\rvert_{\rm sp})^T, \lambda_j-\lambda_j\rvert_{\rm sp} \right).
\end{equation}
For convenience, we also use the link variables $\Phi_{j,\delta} \equiv (W_{j,\delta}^T-(W_{j,\delta}\rvert_{\rm sp})^T)^T$ and $\bar{\Phi}_{j,\delta} \equiv (\bar{W}_{j,\delta}^T-(\bar{W}_{j,\delta}\rvert_{\rm sp})^T)$ to denote the fluctuation fields on bond $\delta$. Expanding the action around the SP yields: 
\begin{widetext}
    \begin{eqnarray} \label{eq:action}
    S(\bar{W},W,\bar{b},b,\lambda)&=& S_{\rm cl} + \int_0^\beta d \tau \left[ \left( \sum_{j\mu} \bar{b}_{j,\mu} \partial_{\tau} b_{j,\mu} + {\mathscr H}_{\rm sp}(\bar{b},b) \right) + 
    \sum_{j,\delta} \left( {N\over 4} \bar{\Phi}_{j,\delta} {\cal J}_{j,\delta} \Phi_{j,\delta} - \bar{\Phi}_{j,\delta} {\cal J}_{j,\delta} \Psi_{j,\delta} + \Psi_{j,\delta}^\dagger {\cal J}_{j,\delta} \Phi_{j,\delta} \right) \right. \nonumber \\
   &+& \left. i \sum_j \Phi_{j}^{\lambda}\sum_{\mu} \bar{b}_{j,\mu} b_{j,\mu}\right],
\end{eqnarray} 
\end{widetext}
where 
\begin{eqnarray}
S_{\rm cl} = \frac{N \beta}{4} 
\left[\sum_{j,\delta} 
(W_{j,\delta}\rvert_{\rm sp})^\dagger 
{\cal J}_{j,\delta}W_{j,\delta}\rvert_{\rm sp} + 4i\kappa \sum_j \lambda_j\rvert_{\rm sp} \right]
\nonumber
\end{eqnarray}
refers to the ``classical'' interaction energy for the auxiliary fields, which is proportional to  $N$, ${\mathscr H}_{\rm sp}(\bar{b},b)$ is the SP Hamiltonian density, which is also of order $N$:
\begin{eqnarray}\label{eq:hsp}
    {{\mathscr H}}_{\rm sp} &=& {{\mathscr H}}_{0} + \sum_{j,\delta} 
    (W_{j,\delta}\rvert_{\rm sp})^\dagger 
    {\cal J}_{j,\delta} {\Psi}_{j,\delta} + {\Psi}_{j,\delta}^\dagger {\cal J}_{j,\delta} W_{j,\delta}\rvert_{\rm sp}  \nonumber \\
    &+& \sum_j 
    \tilde{\lambda}_j 
    \rvert_{\rm sp} \sum_{\mu} \bar{b}_{j,\mu}^{} {b}_{j,\mu}.
\end{eqnarray}
The remaining components in $S(\bar{W},W,\bar{b},b,\lambda)$ represent the  fluctuations of the auxiliary fields and their interaction with the bosons.

\begin{figure}[t!]
\vspace{1cm}
    \centering
    \includegraphics[width=0.9\textwidth]{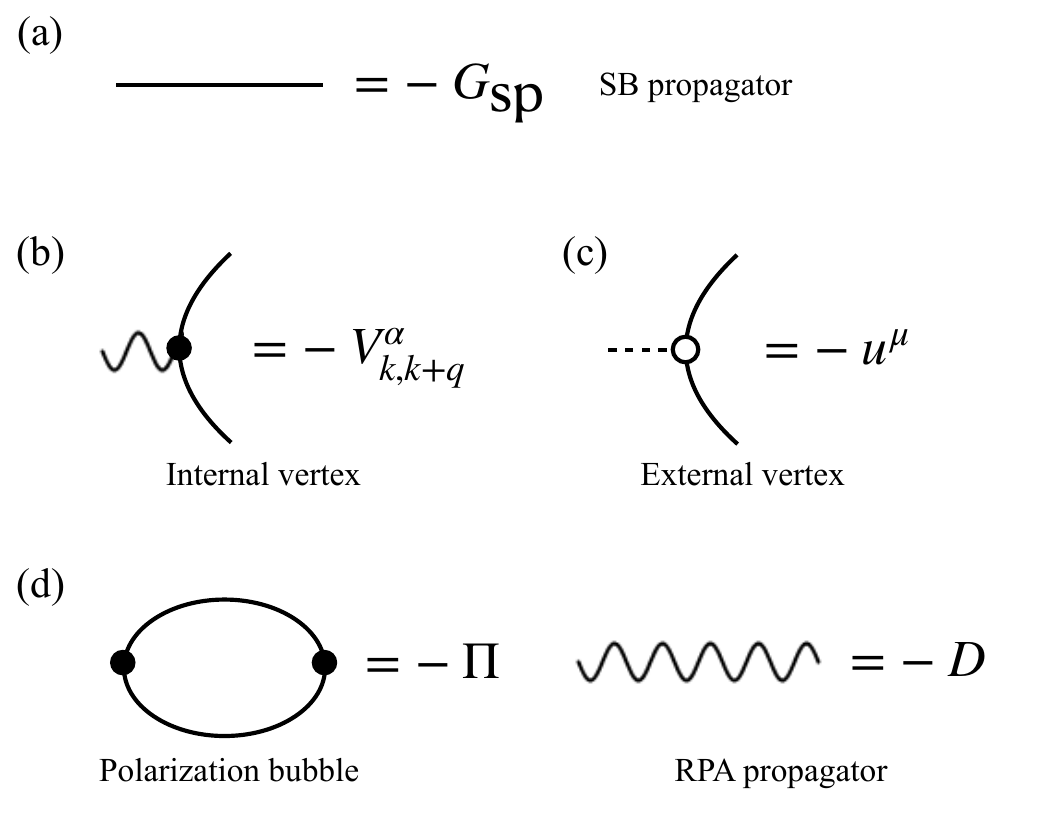}
    \caption{Building blocks for diagrammatic representation of the large-$N$ expansion. (a) Single-boson propagator at the SP level. (b) Internal vertex defined in
    Sec.~\ref{sec3b}. (c) External vertex that couples the external sources to the bosonic propagator. (d) Polarization bubble and corresponding RPA propagator of the auxiliary fields. }
    \label{fig:building_blocks}
\end{figure}

Since we are working in the twisted reference frame, the expected SP solution is translationally invariant and the quadratic action becomes diagonal in momentum space. It is then convenient to rewrite the action in momentum space:
\begin{equation}
    S=S_{\rm cl} + S_0(\bar{\Phi},\Phi) + S_1(\bar{\Phi},\Phi, \bar{\psi},\psi),
\end{equation}
where $S_{\rm cl}$ has been defined above. The second term reads 
\begin{eqnarray}
    S_0 = {N\over 8} \sum_{q,\alpha,\alpha^\prime} (\bar{\Phi})_{q}^{\alpha} {\Pi}_{0}^{\alpha \alpha^\prime} \Phi_{q}^{\alpha^\prime},
\end{eqnarray}
where $q\equiv(\bm q,i\omega_n)$, $\omega_n=2\pi n/\beta$ ($n \in {\mathbb Z}$) are the bosonic Matsubara frequencies,
\begin{eqnarray}
    \Phi_{ q} &=& \frac{1}{\sqrt{{\cal N}_D \beta}} \int_0^\beta d\tau \sum_j  \Phi_{j}(\tau) e^{-i{\bm q}\cdot {\bm r}_j + i \omega_n \tau}, \\
    \bar{\Phi}_{ q} &=& \frac{1}{\sqrt{{\cal N}_D \beta}} \int_0^\beta d\tau \sum_j  \bar{\Phi}_{j}(\tau) e^{i{\bm q}\cdot {\bm r}_j - i \omega_n \tau}
\nonumber \\
\end{eqnarray}
the Fourier transform of the fluctuation fields,  with ${\cal N}_D$ being the total number of dimers, and the block-diagonal constant matrix reads
\begin{eqnarray}
    {\Pi}_{0}=
    \left(
    \begin{array}{c|c|c|c|c}
       {\cal J}_{x}  &&&&  \\
       \hline
         &{\cal J}_{y}&&&  \\
         \hline
         &&{\cal J}_{x}^T&&  \\
         \hline
          &&&{\cal J}_{y}^T&  \\
         \hline
          &&&&0  \\
    \end{array}
    \right),
\end{eqnarray}
where ${\cal J}_{\delta} \equiv {\cal J}_{j,\delta}$ (independent to $j$) due to the translational invariance.
The third term in $S$ reads
\begin{equation}\label{eq:S1}
    S_1 = {1\over2\sqrt{{\cal N}_D \beta}} 
    \sum_{k q} \psi_{k}^\dagger
    \left( \delta_{q,0} G_{\rm sp}^{-1}(k)  +  
   2 \sum_\alpha \Phi^{\alpha}_{-q} V^{\alpha}_{k,k+q} \right) \psi_{k+q}.
\end{equation}
where $k\equiv(\bm k,i\nu_m)$ $\nu_m=2\pi m/\beta$ ($m \in {\mathbb Z}$) is the bosonic Matsubara frequency, $G_{\rm sp}^{-1}(k)$ is the inverse of the Green's function of the SB at the SP approximation, 
\begin{eqnarray}\label{eq:sp_green}
    G_{\rm sp}^{-1}(k) = -i\nu_m \sigma_z \otimes I_{N} + H_{\rm sp}({\bm k}).
\end{eqnarray}
with $I_{N}$ denoting the $N$-dimensional identity matrix and $H_{\rm sp}({\bm k})$ the SP Hamiltonian in momentum space, $\psi_k$ is the Nambu representation for the SB, defined by
\begin{eqnarray}
    \psi_j &=& (b_j^T, \bar{b}_j)^T, \quad b_j = (b_{j,0}, ...,b_{j,N-1})^T, \\
    \psi_k &=& {1\over\sqrt{{\cal N}_D \beta}}\int_0^\beta d\tau \sum_j \psi_j (\tau) e^{-i {\bm k}\cdot {\bm r}_j  + i \nu_m \tau }, \label{eq:numbu_sb}
\end{eqnarray}
and $V^{\alpha}_{k+q,k}$ is the ``internal'' interaction vertex between the fluctuation fields and the boson field, whose diagrammatic representation is illustrated in Fig~\ref{fig:building_blocks}~(b). Explicitly, by defining the bilinear form of the link fields under the Nambu basis of the SB,
\begin{eqnarray}
    {\Psi}_{j,\delta}^{a} = {\psi}_j^\dagger V^{a}_\delta {\psi}_{j+\delta},
\end{eqnarray}
where  $a=S,T,A,B$ refers to the four link operators introduced in Eq.~(\ref{eq:bond_opr}), internal vertex function takes the following form:
\begin{eqnarray}
    V^{a}_{k,k+q} 
    = {1\over 2} \sum_{a^\prime, \delta} {\cal J}^{a^\prime a}_{\delta} \left( (V_\delta^{a^\prime})^\dagger e^{-i {\bm k}\cdot{\bm \delta}} + P (V_\delta^{a^\prime})^*P e^{i({\bm k}+{\bm q})\cdot{\bm \delta}} \right),
\nonumber \\
\end{eqnarray}
where
$a$ denotes the component of fluctuation field $\Phi_{j,\delta}^{a}$ (see below for definition of matrix $P$), ${\bm \delta}$ is the bond vector of bond $\delta$,
\begin{eqnarray}
    V^{a}_{k,k+q} 
    = - {1\over 2} \sum_{a^\prime,\delta} {\cal J}^{a a^\prime}_\delta \left( V_\delta^{a^\prime} e^{i ({\bm k}+{\bm q})\cdot{\bm \delta}} + P (V_\delta^{a^\prime})^T P e^{-i{\bm k}\cdot{\bm \delta}} \right),
    \nonumber \\
\end{eqnarray}
where 
$a$ denotes the component of fluctuation field $\bar{\Phi}_{j,\delta}^{a}$, and 
\begin{eqnarray}
   V^{a=\lambda}_{k,k+q} 
    = {i\over 2} I_{2N}.
\end{eqnarray}
Note that, to simplify the diagrammatic calculation, we have symmetrized the internal vertex function by implementing the particle-hole symmetry of the Nambu field, $\psi_k=P (\psi_{-k}^{\dagger})^T$, with $P=(\sigma_x \otimes I_{N})$. 

Figure~\ref{fig:building_blocks}(a) and \ref{fig:building_blocks}(b) 
illustrate the building blocks for a diagrammatic representation based on the formulation described above.
We note that, since the Nambu fields $\Phi_{-q}$ and $\bar{\Phi}_{q}$ refer to the same physical degree of freedom, one can rewrite $S_1$ in an equivalent form:
\begin{eqnarray}\label{eq:S1b}
    S_1 \! = \!\! \frac{1}{2\sqrt{{\cal N}_D \beta}}
    \sum_{k q} \psi_{k}^\dagger 
    \left( \delta_{q,0} G_{\rm sp}^{-1}(k)
+  2 \sum_\alpha \bar{\Phi}^{\bar{\alpha}}_{q} V^{\alpha}_{k,k+q} \right) \psi_{k+q},
\nonumber  \\
\end{eqnarray}
where $\bar{\alpha}$ is defined such that $\bar{\Phi}^{\bar{\alpha}}_{- q} \equiv \Phi_{ q}^\alpha$. Diagrammatically, this corresponds to an internal vertex with an \textit{in-coming} wavy line (propagator of the fluctuation field, defined below), while that in Eq.~\eqref{eq:S1} to an internal vertex with an \textit{out-coming} wavy line.

\begin{figure}
\vspace{1cm}
    \centering
    \includegraphics[width=1\textwidth]{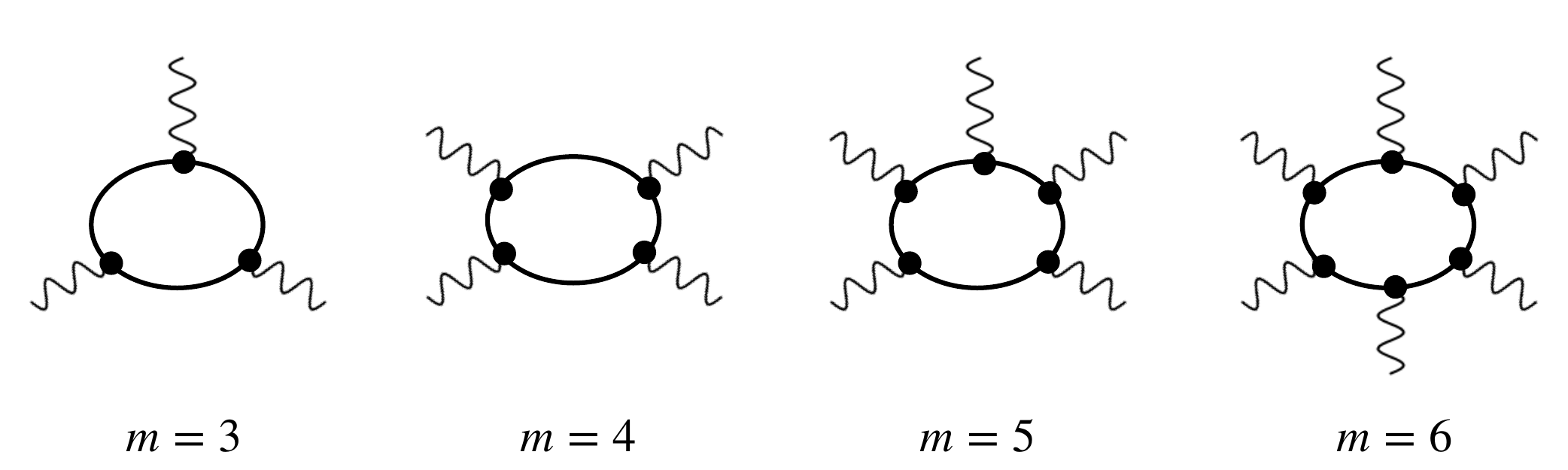}
    \caption{\textit{Internal loops} from the first four terms in $S_{\text{int}}$, Eq.~(\ref{eq:internal_loop}).}
    \label{fig:internal_loop}
\end{figure}

In  the last step, one can integrate out the boson fields, which yields
\begin{eqnarray}
    {\cal Z} = \int {\cal D}[\bar{\Phi}\Phi] \exp \left[ - {N \over 4} S_{\rm eff}(\bar{\Phi},\Phi) \right],
    \nonumber \\
\end{eqnarray}
where
\begin{eqnarray}
    S_{\rm eff}(\bar{\Phi},\Phi) = {1\over 2} \sum_{q,\alpha,\alpha^\prime} \bar{\Phi}_{q}^{\alpha} {\Pi}_{0}^{\alpha \alpha^\prime} \Phi_{q}^{\alpha^\prime} + {2\over N} \text{Tr}[\ln {\cal M}] + {4S_{\rm cl}\over N}.
\nonumber \\
\label{eq:Seff}
\end{eqnarray}
In the limit of $N\to \infty$, the fluctuations of the auxiliary field $\Phi$ are suppressed, and only the SP action contributes to the partition function (namely, the SP approximation). For  finite $N$, the typical amplitude of  fluctuations of the field $\Phi$ is of order ${\cal O}(1/\sqrt{N})$, which enables us to expand the second term of $S_{\rm eff}$ as follows:
\begin{equation}\label{eq:expand1}
{2\over N} \text{Tr}[\ln {\cal M}] = {2\over N} \text{Tr}\ln G_{\rm sp}^{-1} +  {2\over \sqrt{N}} \sum_{m=1}^{\infty} {(-1)^{m+1} \over m} \text{Tr} \left[ {\mathscr V}^m \right],
\end{equation}
where the trace ``${\rm Tr}$'' runs over $k$ and the Nambu indices, ${G^{-1}_{\rm sp}}$ is block diagonal in $k$, with each block given by $G^{-1}_{\rm sp} (k)$, while the matrix ${\mathscr{V}}$ is defined as
\begin{equation}
\mathscr{V}_{k,k+q} =     G_{\rm sp}(k) {2\over \sqrt{{\cal N}_D \beta}} \sum_{\alpha} \Phi^\alpha_{-q} V^{\alpha}_{k,k+q}.
\end{equation}
In this expansion, Eq.~\eqref{eq:expand1}, the first term is a constant and only contributes to static properties, like the ground state energy; the linear contribution in $\Phi$ vanishes for an expansion around a SP of the effective action; the quadratic contribution in $\Phi$ is combined with the first term of Eq.~\eqref{eq:Seff}
to obtain the Gaussian contribution, $S^{(2)}(\bar{\Phi},\Phi)$, to the expansion  of the effective action in powers of the fluctuations of the auxiliary fields.
$S^{(2)}(\bar{\Phi},\Phi)$ determines the ``bare'' propagator of the fluctuation fields,
which is usually known as ``random phase approximation'' (RPA) propagator. Thus, we reorganize the effective action as
\begin{eqnarray}
    S_{\rm eff}(\bar{\Phi},\Phi) &=& \! {4S_{\rm cl}\over N} \! + \! {2\over N} \text{Tr}\ln G_{\rm sp}^{-1} \! + \! S^{(2)}(\bar{\Phi},\Phi) \! + \! S_{\rm int}(\bar{\Phi},\Phi),
    \nonumber \\
    \label{eq:Seffexp}
\end{eqnarray}
where
\begin{eqnarray}\label{eq:srpa}
    S^{(2)}(\bar{\Phi},\Phi)={1\over 2}\sum_{q,\alpha,\beta} \bar{\Phi}_{q}^{\alpha} ({\Pi}_{0}^{\alpha \alpha^\prime} - \Pi^{\alpha \alpha^\prime} (q)) \Phi_{q}^{\alpha^\prime},
\end{eqnarray}
is the Gaussian action of the auxiliary fields and
 \begin{eqnarray}\label{eq:pi}
    \Pi^{\alpha \alpha^\prime} (q)={8\over N {\cal N}_D\beta} \sum_k  {\rm tr} \left( G_{\rm sp}(k)V^{\bar{\alpha}}_{k,k+q} G_{\rm sp}(k+q) V^{\alpha^\prime}_{k+q,k} \right),
    \nonumber \\
\end{eqnarray}
is  the polarization bubble illustrated in Fig.~\ref{fig:building_blocks}(d), where the trace ``${\rm tr}$''  runs over the Nambu indices. The polarization bubble determines the ``bare'' or RPA propagator of the auxiliary fields,
\begin{eqnarray}\label{eq:Drpa}
    D_{\alpha \alpha^\prime}(q)&=&{\int {\mathcal D}[\bar{\Phi},\Phi] e^{-{N\over 4}S^{(2)}(\bar{\Phi},\Phi)} \Phi_{\alpha}(q) \bar{\Phi}_{\alpha^\prime}(q)\over \int {\mathcal D}[\bar{\Phi},\Phi] e^{-{N\over 4}S^{(2)}(\bar{\Phi},\Phi)}}  \nonumber \\
    &=&{4\over N}\left[ (\Pi_0-\Pi(q))^{-1}\right]_{\alpha \alpha^\prime}.
\end{eqnarray}
In absence of a condensate, the RPA propagator scales as $1/N$ and  is represented by the wavy line shown in Fig.~\ref{fig:building_blocks}(d).
The last term of Eq.~\eqref{eq:Seffexp},
\begin{eqnarray}
\label{eq:internal_loop}
S_{\rm int}(\bar{\Phi},\Phi) = {2\over N} \sum_{m \geq 3}^{\infty} {(-1)^{m+1} \over m} 
\text{Tr} \left[ 
{\mathscr{V}}^m
\right],
\end{eqnarray}
describes the effective interaction between the fluctuation fields. As shown in Fig.~\ref{fig:internal_loop} for $ m \leq 6$, the  effective interaction vertex generated by each term of the expansion is diagrammatically represented as an ``internal loop''  of $m$ bosonic propagators and $m$ internal vertices.

\begin{figure}[t!]
\vspace{1cm}
    \centering
    \includegraphics[width=0.9\textwidth]{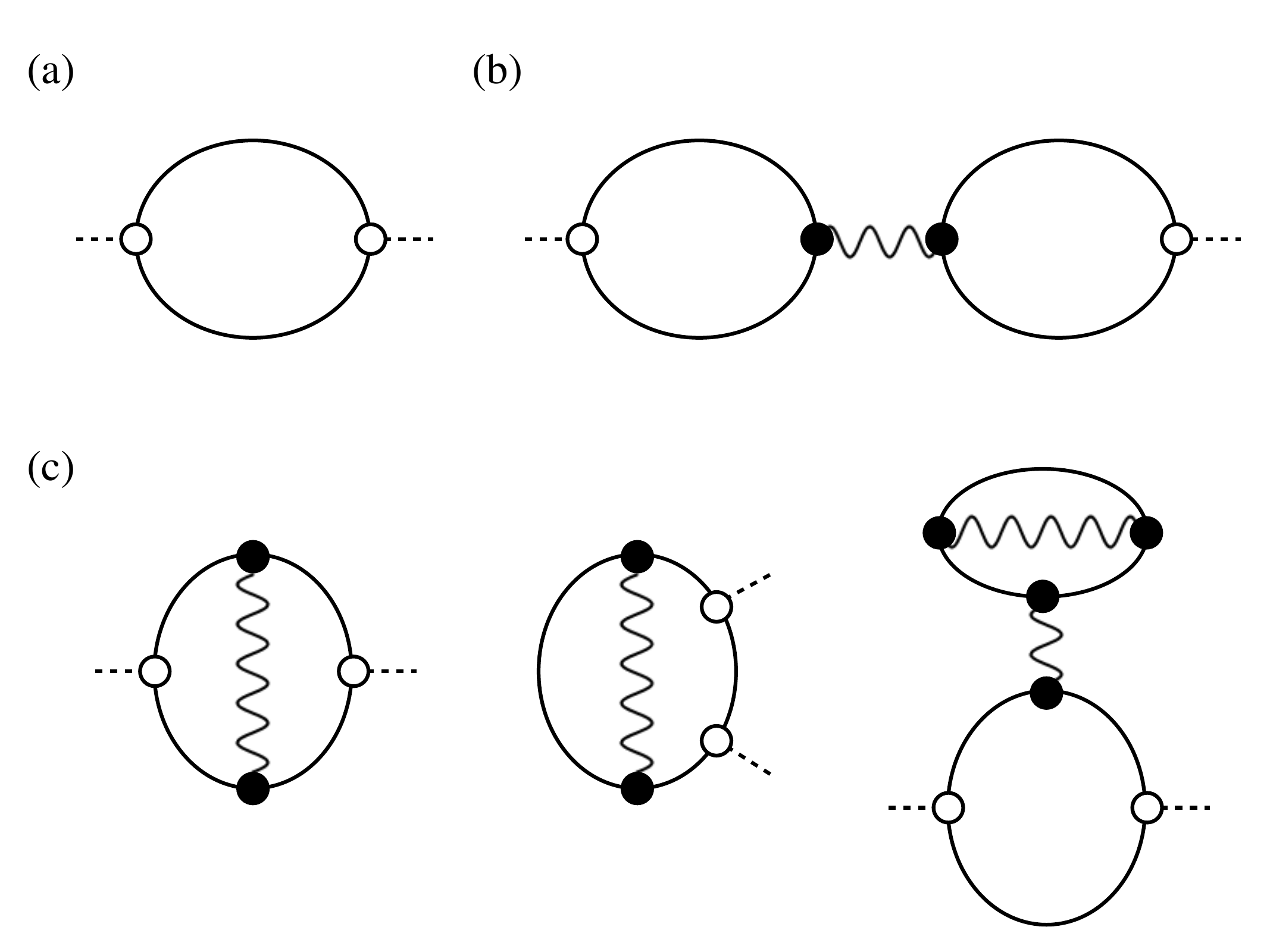}
    \caption{Large-$N$ expansion of the magnetic susceptibility: (a) leading order diagram, equivalent to SP approximation, (b,c) sub-leading order diagrams in $1/N$. Notation of propagators and vertex functions are explained in Fig.~\ref{fig:building_blocks}.  }
    \label{fig:large_N}
\end{figure}

In absence of a SB condensate, the power counting of each diagram, $1/N^\nu$, is determined by the difference between the number of  RPA propagators ($P$) and the number of internal loops ($L$): $\nu=P-L$.  As an example, Fig.~\ref{fig:large_N} shows  all the diagrams up to order $1/N$ of the magnetic susceptibility (see Sec.~\ref{sec:dssf} for definition and Appendix~\ref{app:largeN} for details of the large-$N$ expansion of the magnetic susceptibility). 
The leading order  diagram, $\nu=P-L=0$ because $P=L=0$,   is shown in Fig.~\ref{fig:large_N}~(a).
Fig.~\ref{fig:large_N}~(b) and (c) show the  diagrams of order $1/N$. However, when a fraction of the bosons condense, a subset of these diagrams violates this nominal power counting. Specifically, diagrams that are nominally of order $1/N$ can acquire singular contributions of order $1/N^0$, which are essential for producing physically sound results~\cite{ghioldi2018dynamical,ghioldi2022evidence,zhang2022schwinger}. 

In the following sections, we first describe the SP solution for both the QPM and AFM phases. We then analyze the $1/N$ corrections using the framework described in this section.

\section{Saddle point solution}

\label{sec4}
We consider the following ansatz for the SP solution of the  link fields:
\begin{eqnarray}
    \langle {S}_{j,\delta}\rangle\rvert_{\rm sp} &=& {\cal S}, \quad \langle {T}_{j,\delta}\rangle\rvert_{\rm sp} = {\cal T}, \nonumber \\
    \langle {A}_{j,\delta}\rangle\rvert_{\rm sp} &=& {\cal A}, \quad \langle {B}_{j,\delta}\rangle\rvert_{\rm sp} = {\cal B}, \quad \lambda_j\rvert_{\rm sp}=-i\tilde{\lambda}. \nonumber \\
    \label{eq:speqs}
\end{eqnarray}
which  is motivated by symmetry considerations. More specifically, the QPM phase 
results from the condensation of only the singlet boson, with the corresponding gapless singlet excitation at momentum ${\bm k} = {\bm 0} $. Note that in our $SU(4)$ theory the singlet boson has dynamics, in contrast to the conventional bond-operator theory. For a proper gauge choice of the condensate wavefunction, both $\langle {S}_{j,\delta}\rangle\rvert_{\rm sp}$ and $ \langle {T}_{j,\delta}\rangle\rvert_{\rm sp}$ are real numbers  and transform trivially under the space group. The expectation values of link fields $\langle {A}_{j,\delta}\rangle\rvert_{\rm sp}$ and $\langle {B}_{j,\delta}\rangle\rvert_{\rm sp}$ are determined by the coupling with fields ${S}_{j,\delta}$ and ${T}_{j,\delta}$, i.e., by Eq.~(\ref{eq:dec2}). Since the resulting SP Hamiltonian must retain the symmetries of the original  model, $\langle {A}_{j,\delta}\rangle\rvert_{\rm sp}$ and $\langle {B}_{j,\delta}\rangle\rvert_{\rm sp}$  must transform like $\langle {S}_{j,\delta}\rangle\rvert_{\rm sp}, \langle {T}_{j,\delta}\rangle\rvert_{\rm sp}$ under the symmetry group of $\hat{\mathscr{H}}$. 
The SP solution for the Lagrange multiplier, ${\tilde \lambda}_j\rvert_{\rm sp}={\tilde \lambda}$, is independent to site index $j$ because of  translation invariance.   

To determine the value of the above SP parameters, we need to solve the SP equations in Eq.~\eqref{eq:sp_cond}. On the right-hand side of the SP equations, the average of the link operators and the boson density depend on the Green's function of the SBs for a given SP configuration of the auxiliary fields, which is defined by the inverse of the dynamic matrix in Eq.~\eqref{eq:sp_green}. Essentially, this SP solution is equivalent to solving the eigenvalues and eigenstates of the following SP Hamiltonian in the canonical formulation of the problem
\begin{eqnarray}
    \hat{{\mathscr H}}_{\rm sp} = {1\over 2} \sum_{\bm k} \hat{\psi}_{\bm k} ^\dagger H_{\rm sp}({\bm k}) \hat{\psi}_{\bm k},\label{eq:spHk}
\end{eqnarray}
where $H_{\rm sp}({\bm k})$ has been defined in Eq.~\eqref{eq:sp_green} and $\hat{\psi}_{\bm k}$ are the Nambu operators of the SBs in canonical formulation (c.f. Eq.~\eqref{eq:numbu_sb}). 
Given the conservation of total spin, the SP Hamiltonian $H_{\rm sp}({\bm k})$ is block diagonal in the flavor of SBs, with each block given by
\begin{eqnarray}\label{eq:spHkblock}
     H_{{\rm sp},\mu}({\bm k}) &=& \left( 
    \begin{array}{cc}
        A_{{\bm k},\mu} &  B_{{\bm k},\mu}  \\
         B_{{\bm k},\mu}  & A_{{\bm k},\mu}
    \end{array}\right)
\end{eqnarray}
in the Nambu basis $(\hat{b}_{{\bm k},\mu},\hat{b}^\dagger_{-{\bm k},\mu})^T$. The matrix elements are:
\begin{eqnarray}
A_{{\bm k},0} &=& \tilde{\lambda}-J + J^{\prime} {\cal B} \gamma_{{\bm k}},
\quad B_{{\bm k},0} =     J^{\prime} {\cal A} \gamma_{{\bm k}}, \nonumber  \\
A_{{\bm k},1} &=& A_{{\bm k},2} = A_{{\bm k}+{\bm \pi},3} = \tilde{\lambda} - J^{\prime} ( {\cal T} +  {\cal B}) \gamma_{{\bm k}}, \nonumber \\
B_{{\bm k},1} &=& B_{{\bm k},2} = B_{{\bm k}+{\bm \pi},3} = - J^{\prime} ( {\cal S}- {\cal A}) \gamma_{{\bm k}},
\end{eqnarray}
and $$\gamma_{\bm k}=\cos{k_x}+\cos{k_y}.$$
The diagonal form of $H_{{\rm sp},\mu}({\bm k})$ is obtained by applying  the following Bogoliubov transformation for each flavor:
\begin{eqnarray}
    \left( 
    \begin{array}{c}
         \hat{b}_{{\bm k},\mu}  \\
         \hat{b}_{\bar{\bm k},\mu}^\dagger 
    \end{array}\right)
    =
    \left( 
    \begin{array}{cc}
         u_{{\bm k},\mu} & v_{{\bm k},\mu}  \\
         v_{{\bm k},\mu} & u_{{\bm k},\mu}
    \end{array}\right)
    \left( 
    \begin{array}{c}
         \hat{\beta}_{{\bm k},\mu}  \\
         \hat{\beta}_{\bar{\bm k},\mu}^\dagger 
    \end{array}\right),
\end{eqnarray}
where $\bar{\bm k} \equiv - {\bm k}$,
\begin{eqnarray}
u_{{\bm k},\mu}&=&\sqrt{{1\over 2}\left( {A_{{\bm k},\mu} \over \varepsilon_{{\bm k},\mu}}+1 \right)}, \nonumber \\
v_{{\bm k},\mu}&=&-\sqrt{{1\over 2}\left( {A_{{\bm k},\mu} \over \varepsilon_{{\bm k},\mu}}-1 \right)} \text{sgn}(B_{{\bm k}, \mu})
\end{eqnarray}
are coherent factors, and
\begin{equation}\label{eq:ek_singlet}
    \varepsilon_{{\bm k},\mu}=\sqrt{A^2_{{\bm k}, \mu} - B^2_{{\bm k},\mu}}
\end{equation}
is the SP energy spectrum. 

Using the above solution, the Green's function of the SBs can be computed straightforwardly by using Lehmann's representation, 
\begin{eqnarray}\label{eq:sp_green2}
    G_{\rm sp}(k) &=& \frac{1}{H_{\rm sp}({\bm k})-i\nu_m} \nonumber \\
    &=& \sum_{\varepsilon_{{\bm k},\mu} > 0} \frac{g_{{\bm k},\mu}}{\varepsilon_{{\bm k},\mu} - i \nu_m} + \frac{{\bar g}_{{\bm k},\mu}}{\varepsilon_{{\bm k},\mu} + i \nu_m},
\end{eqnarray}
where $g_{{\bm k},\mu}$ and ${\bar g}_{{\bm k},\mu}$ are $8 \times 8$ matrices with non-zero matrix elements only for the entries corresponding to Nambu spinor $(\hat{b}_{{\bm k},\mu},\hat{b}^\dagger_{-{\bm k},\mu})^T$, given by
\begin{eqnarray}
    g_{{\bm k},\mu} &=&  
    \left( 
    \begin{array}{c|cc|c}
        & & & \\
        \hline
         & u_{{\bm k},\mu}^2 &   v_{{\bm k},\mu}u_{{\bm k},\mu} & \\
         & v_{{\bm k},\mu}u_{{\bm k},\mu} & v_{{\bm k},\mu}^2 & \\
         \hline
         &&& 
    \end{array}
    \right), \\
    {\bar g}_{{\bm k},\mu} &=& 
    \left( 
    \begin{array}{c|cc|c}
         & & & \\
        \hline
         &v_{{\bm k},\mu}^2 &   v_{{\bm k},\mu}u_{{\bm k},\mu} & \\
         &v_{{\bm k},\mu}u_{{\bm k},\mu} & u_{{\bm k},\mu}^2   & \\
         \hline
         &&& 
    \end{array}
    \right).
\end{eqnarray}
It is important to note that, the above formula can be safely used for calculations when the system size is finite or when there is no condensation of bosons. A more general representation is described at the end of Sec.~\ref{sec4}. 

We can now compute the expectation values of the link operators and boson density in the SP equation~(\ref{eq:sp_cond}), which gives rise to the following self-consistent equations:
\begin{eqnarray}\label{eq:speq}
    {\cal S}&=&{1\over {\cal N}_D} \sum_{{\bm k}} u_{{\bm k},0} v_{{\bm k},0} \cos\left({{\bm k}\cdot {\bm \delta}}\right), \nonumber \\
    {\cal T}&=&{1\over {\cal N}_D} \sum_{{\bm k}}  v_{{\bm k},0}^2 \cos\left({{\bm k}\cdot {\bm \delta}}\right), \nonumber \\
    {\cal A}&=&{1\over {\cal N}_D} \sum_{{\bm k}}  \sum_{\mu=1}^3 \eta_\mu u_{{\bm k},\mu} v_{{\bm k},\mu}  \cos\left({{\bm k}\cdot {\bm \delta}}\right), \nonumber \\
    {\cal B}&=&{1\over {\cal N}_D} \sum_{{\bm k}} \sum_{\mu=1}^3 \eta_\mu v^2_{{\bm k},\mu}  \cos\left({{\bm k}\cdot {\bm \delta}}\right), \nonumber \\
    1&=&{1\over {\cal N}_D} \sum_{{\bm k}}  \sum_{\mu=0}^3 v^2_{{\bm k},\mu},
\end{eqnarray}
where $\eta_\mu=(-1,-1,1)$ for $\mu=1,2,3$ and ${\bm \delta} = (1,0)$ or $(0,1)$ refers to the two nearest neighbor bonds of the square lattice 
of dimers.
Since the system preserves the four-fold rotation symmetry, the above equations do not depend on the choice of ${\bm \delta}$.  

\begin{figure}
    \centering
    \includegraphics[width=\textwidth]{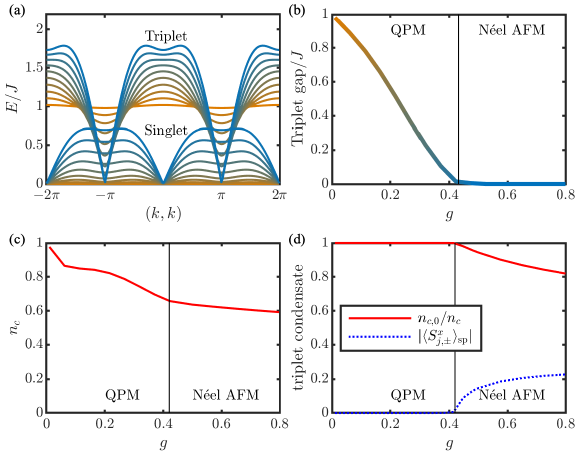}
    \caption{(a) SP spectrum of singlet 
    and triplet bosons 
    in the lab reference frame, where the triplet boson spectrum is three-fold degenerate. The color scale follows panel (b), indicating different values of $g$  between $g=0$ (brown) and $g=g_c\simeq 0.42$ (blue).
    In the twisted reference frame, the triplet boson dispersions for bosons $\hat{b}_{j,1}$ and $\hat{b}_{j,3}$ are shifted by ${\bm \pi}$, so they are gapless at the $\Gamma$ point. 
    (b) Minimal energy gap of triplet boson as a function of $g$. (c) Total condensate fraction $n_c$. (d) Fraction of singlet bosons in the  condensate, $n_{c,{\bm 0}}/n_c$, and SP magnetization, $\rvert \langle S_{j,\pm}^x \rangle_{\rm sp} \rvert=\sqrt{n_{c,\bf 0} n_{c\bf \pi}}$, as a function of $g$.
    Vertical black line in (b-d) marks the value of $g_c$ obtained from the SP solution.
    }
        \label{fig:sp_spectrum}
\end{figure}

Fig.~\ref{fig:sp_spectrum} (a) shows the single boson dispersion  for the singlet and triplet channels.
The dispersions of the three triplet bosons ($\mu=1,2,3$), presented in the laboratory reference frame, are identical because the global $\mathrm{SU}(2)$ symmetry of the Hamiltonian is preserved by the QPM phase. 
The singlet boson dispersion is gapless at the $\Gamma$ point, where the occupation number becomes macroscopic at zero temperature. 
The triplet-boson dispersion is gapped for small  $g$ values and it  has a minimum at ${\bm k}={\bm \pi}$. 
As illustrated in Fig.~\ref{fig:sp_spectrum}~(a) and \ref{fig:sp_spectrum}~(b),  this triplet mode softens as $|g-g_c|^\nu$ when $g$ approaches a critical point $g_c \approx 0.42$. This softening signals a phase transition to a state exhibiting magnetic long-range order. Notably, this critical value aligns closely with the value of $g_c \approx 0.3984$ obtained from QMC simulations~\cite{Sandvik94}. It is worth mentioning that a similar approach based on the more conventional $\mathrm{SU}(2)$ SB theory results in a critical value $g_c \approx 0.235$~\cite{Millis93,Millis94,miyazaki1996bilayer}, which significantly deviates from the QMC result. Interestingly, this critical value closely aligns with the value $g_c = 0.25$ that is  obtained from a semi-classical treatment based on $\mathrm{SU}(4)$ coherent states. 
 It should be stressed that the gap closes as $|g-g_c|^\nu$, with a critical exponent $\nu=1$. 
 This value, characteristic of a large-$N$ theory ~\cite{sachdev11}, differs from $\nu \simeq 0.71$ predicted by QMC~\cite{wang06}. A  more precise estimation of  $\nu$ requires the inclusion of higher order corrections in $1/N$.
 

Before delving into the SP solution for the broken-symmetry state, it is worthwhile to compare our SP solution for the QPM with the 'bond operator' approach introduced by Sachdev and Bhatt~\cite{sachdev1990bond}. The approximation proposed by Sachdev and Bhatt can be derived from our large-$N$ expansion  if we approximate the expectation values of the ${S}_{j,\delta}$ and ${T}_{j,\delta}$ link fields in Eq.~\eqref{eq:speqs} in the following way:
\begin{eqnarray}
\label{eq:bond_opr2}
&\langle S_{j,\delta} \rangle_{\rm sp} \approx \langle b_{j,0} \rangle_{\rm sp} \langle  b_{j+\delta,0}\rangle_{\rm sp}, \ \
&\langle T_{j,\delta}^{\;}\rangle_{\rm sp}  \approx \langle b^\dagger_{j+\delta,0}\rangle_{\rm sp} \langle b_{j,0} \rangle_{\rm sp} , 
\nonumber \\
\end{eqnarray}
where we have written the equations in the canonical formalism to make direct contact with Ref.~\cite{sachdev1990bond}. This approximation, which is characteristic of semi-classical treatments ($1/M$ expansion), removes the expansion parameter from the formalism (it is neither a $1/M$ nor a $1/N$ expansion). This is the basic reason why their approach does not capture correctly the Goldstone modes of the broken symmetry state. On the other hand, it leads to a triplon gap that closes as $|g -g_c|^\nu$ with $\nu =1$, characteristic of large-$N$ fixed  point~\cite{sachdev11,zhang2013phase}, in contrast to the large-$M$ (generalized spin wave theory) prediction  $\nu=1/2$.  This observation reveals the ``hidden'' large-$N$ nature of the original BOT mean-field approximation.

\begin{figure}[!b]
\includegraphics[angle=0,width=0.8\textwidth]{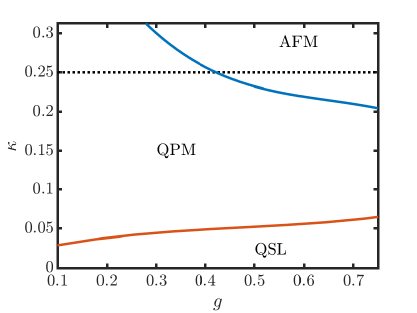}
\caption{Saddle-point phase diagram on the $g - \kappa$ plane.}
\label{fig:phase_diagram}
\end{figure}

For completeness, we computed the mean-field phase diagram in the \((g, \kappa)\) plane, which is shown in Fig.~\ref{fig:phase_diagram}. For the physical case, \(\kappa = 1/4\), the phase diagram features both quantum paramagnetic (QPM) and antiferromagnetic (AFM) phases. Reducing \(\kappa\) destabilizes both of these phases.

Within the QPM region, decreasing \(\kappa\) reduces the condensate fraction of the singlet boson, \(n_{c,0}\), eventually driving a transition into a gapped quantum spin liquid (QSL) phase once \(n_{c,0}\) vanishes. In the AFM phase, a similar reduction in \(\kappa\) leads to a sequential suppression of the boson condensates: the triplet condensate \(n_{c,\pi}\) vanishes at the AFM-to-QPM transition, followed by the disappearance of the singlet condensate \(n_{c,0}\) at the subsequent QPM-to-QSL transition.

In the QSL phase, the saddle-point solution is invariant under a staggered \({\mathrm U}(1)\) gauge transformation, defined by \(b_{j,\mu} \to b_{j,\mu}\) for sites \(j \in {\cal A}\) and \(b_{j,\mu} \to -b_{j,\mu}\) for sites \(j \in {\cal B}\). In accordance with this symmetry, the saddle-point expectation values of the link operators \(\hat{T}_{jk}\) and \(\hat{B}_{jk}\), which connect the \({\cal A}\) and \({\cal B}\) sublattices, vanish. This staggered gauge symmetry is spontaneously broken upon entering the QPM phase, as indicated by the condensation of the singlet boson \(b_{j,0}\).

\subsection{Broken symmetry phase}

The SP dispersion of the triplon modes remains gapless beyond the critical point. By evaluating the occupation number of the triplet bosons at the gapless momenta (${\bm \pi}$ in the laboratory frame), we find the density of  triplet bosons that are condensed. We label the condensate fraction of singlet and triplet bosons as $n_{c,{\bm 0}}$ and $n_{c,{\bm \pi}}$, respectively, and the total condensate fraction as $n_c=n_{c,{\bm 0}}+n_{c,{\bm \pi}}$. As shown in Fig.~\ref{fig:sp_spectrum}~(c), the total condensate density ($n_c$) decreases with increasing $g$. The fraction $n_{c,{\bm 0}}/n_c$ of singlet bosons in the condensate  is shown in Fig.~\ref{fig:sp_spectrum} (d).

When the triplet boson undergoes condensation, the system is generally expected to exhibit long-range Néel magnetic order. However, within the SP solution framework discussed above, the ground state expectation value of local spin operators remains zero because the SP solution retains the $\mathrm{SU}(2)$ invariance of the Hamiltonian. To accurately describe the condensate corresponding to a magnetically ordered state, it is necessary to introduce an infinitesimal symmetry-breaking field (SBF)~\cite{auerbach2012interacting}, which is sent to zero after taking the thermodynamic limit.

In our current scenario, the  SBF is  a staggered magnetic field $h$ linearly coupled to the local order parameter. In the twisted reference frame, the corresponding Zeeman term takes the form:
\begin{equation}\label{eq:sbfield}
    \hat{{\mathscr H}}_{\rm SBF}  =  -h \sum_j (S_{j,+}^x - S_{j,-}^x)  
     =  -h\sum_j (\hat{b}_{j,0}^\dagger \hat{b}_{j,1} + \hat{b}_{j,1}^\dagger \hat{b}_{j,0}).
\end{equation}
In the thermodynamic limit, the magnitude of this SBF is infinitesimally small, thereby leaving the SP spectrum unchanged. However, this field significantly alters the condensate wavefunction by controlling the relative population of bosons in the four gapless states. When solving the SP equation under the influence of this SBF, we find that all the bosons condense into a hybridized energy level formed by the singlet boson ($\mu=0$) and one flavor ($\mu=1$) of the triplet boson. The resulting condensate maintains the same fractions, $n_{c,{\bm 0}}$ and $n_{c,{\bm \pi}}$, as those obtained without the SBF, which is essential for satisfying the SP equation because the contribution from non-condensed bosons remains unchanged. Consequently, this condensate induces N\'eel magnetic order polarized along the $x$ direction, with the magnitude of the ordered moment determined by $\sqrt{n_{c,{\bm 0}}n_{c,{\bm \pi}}}$.

To understand how the presence of the SBF modifies the condensate, we consider the SP Hamiltonian for the condensed bosons under the SBF,
\begin{eqnarray}
    \hat{{\mathscr H}}_{\rm sp,c} = \frac{1}{2} \hat{\psi}_{01}^\dagger H_{01} \hat{\psi}_{01} + \frac{1}{2} \hat{\psi}_{2}^\dagger H_{2} \hat{\psi}_{2} 
     + \frac{1}{2} \hat{\psi}_{3}^\dagger H_{3} \hat{\psi}_{3},
     \nonumber \\
\end{eqnarray}
where $\hat{\psi}_{01}^\dagger \equiv (\hat b_{\bm 0,0}^\dagger, \hat b_{\bm 0,1}^\dagger, \hat b_{\bm 0,0}, \hat b_{\bm 0,1} ) $, $\hat{\psi}_{2}^\dagger \equiv (\hat{b}_{{\bm 0},2}^\dagger,\hat{b}_{{\bm 0},2})$, $\hat{\psi}_{3}^\dagger \equiv (\hat{b}_{{\bm \pi},3}^\dagger,\hat{b}_{{\bm \pi},3})$, and the Hamiltonian matrices are 
\begin{eqnarray}
    H_{01} &=& \left(
    \begin{array}{cccc}
        \Delta_0 +\delta_0 & -h  & \Delta_0    & 0   \\
        -h & \Delta_3 +\delta_3 &  0  & \Delta_3   \\
        \Delta_0 & 0 &  \Delta_0 +\delta_0  & -h  \\
        0   &  \Delta_3   &  -h  & \Delta_3 +\delta_3
    \end{array}
    \right), \nonumber \\
    H_{2}&=&H_{3} =\left(
    \begin{array}{cc}
        \Delta_3 +\delta_3 &\Delta_3  \\
       \Delta_3   & \Delta_3+\delta_3
    \end{array}
    \right),
\end{eqnarray}
with  $\Delta_0 = 2 J^\prime {\cal A}$ and $\Delta_3 = -2 J^\prime ( {\cal S} -  {\cal A})$.
The parameters $\delta_0$ and $\delta_3$ represent the corrections to the SP Hamiltonian due to the finite $h$.
In the thermodynamic limit, one can verify that the four energies reduce to zero when $\delta_0=\delta_3=0$, corresponding to $h=0$.
This implies that, in absence of the SBF $h$, the ground state is a  non-magnetic condensate 
as the three triplet states exhibit identical condensate densities.

For finite $h$, 
the spectrum of the hybridized singlet- ($\mu=1$) triplet bosons, determined by $H_{01}$, becomes
\begin{eqnarray}
    E_{0} = \sqrt{a-b}, \quad E_1 = \sqrt{a+b},
\end{eqnarray}
where 
\begin{eqnarray}
    a&=&\Delta_0 \delta_0 + \Delta_3\delta_3 + \frac{1}{2}(\delta_0^2 + \delta_3^2) + h^2, \\
    b&=&\frac{1}{2}\left[\left(2\Delta_0 \delta_0 + 2\Delta_3\delta_3 +\delta_0^2 + \delta_3^2 + h^2\right)^2  \right. \nonumber \\
    &-& \left. 4\left((2\Delta_0 + \delta_0 )(2\Delta_3 + \delta_3)-h^2\right)(\delta_0 \delta_3 -h^2)\right]^{1/2}.
    \nonumber \\
    \label{eq:hybH}
\end{eqnarray}
The spectrum of $H_2$ and $H_3$ reads
\begin{eqnarray}
    E_2=E_3 = \sqrt{2\Delta_3\delta_3 +\delta_3^2}.
    \label{eq:hybH2}
\end{eqnarray}
To determine the new ground state, we expand $\delta_0$ and $\delta_3$ in powers of $h$: $\delta_0=a_0 h+{\cal O}(h^2) $ and $\delta_3=a_3 h+{\cal O}(h^2) $.
By inserting these expressions in Eqs.~\eqref{eq:hybH} and \eqref{eq:hybH2}, we obtain the expressions for the squares of the four energies to linear order in $h$:
\begin{eqnarray}
    E_0^2 &\simeq&  \left[\Delta_0 a_0 + \Delta_3 a_3 - \sqrt{(\Delta_0 a_0 - \Delta_3 a_3)^2 + 4 \Delta_0 \Delta_3 }\right] h  , \nonumber \\
    E_1^2 &\simeq& \left[\Delta_0 a_0 + \Delta_3 a_3 + \sqrt{(\Delta_0 a_0 - \Delta_3 a_3)^2 + 4 \Delta_0 \Delta_3 } \right] h , \nonumber \\
    E_2^2 &=& E_3^2 \simeq 2\Delta_3 a_3 h .
\end{eqnarray}
The requirement of a SP solution with a finite condensate fraction necessitates that $E_0=0$, which in turn leads to the condition $a_0a_3=1$. This condition 
implies that the other three modes have excitation energies proportional to $\sqrt{h}$, so the bosons condense only in the single-particle state associated with the 
$E_0$ mode. 


The ratio $a_0/a_3$ is the determined from the condition that the ground state for infinitesimal field $h$ must preserve the ratio $n_{c,{\bm 0}}/n_{c,{\bm \pi}}$ between the condensate fractions 
$n_{c,{\bm 0}}$ and $n_{c,{\bm \pi}}$ in the singlet and triplet states.  

The wavefunction associated with the hybridized level $E_0$,
\begin{equation}\label{eq:Xc}
    X_c=\sqrt{{\cal N}_D}
    \left(\sqrt{n_{c,{\bm 0}}} ,\sqrt{n_{c,{\bm \pi}}},0,0,-\sqrt{n_{c,{\bm 0}}},-\sqrt{n_{c,{\bm \pi}}} ,0,0
    \right)^T,
\end{equation}
is given in the Nambu representation for SB, as described in  Eq.~\eqref{eq:numbu_sb} for $N=4$ specifically. A normalization factor $\sqrt{ {\cal N}_D }$ is introduced  to account for  finite size dimer lattices. 

This solution explicitly breaks the $\mathrm{SU}(2)$ spin rotation symmetry and thus gives rise to local magnetic moments in the ground state, which are obtained by replacing the boson operator $\hat{b}_{j,\mu}$ with  its SP expectation value: $\langle b_{j,\mu} \rangle_{\rm sp} = (\sqrt{n_{c,{\bm 0}}},\sqrt{n_{c,{\bm \pi}}},0,0)$. As expected, this condensate corresponds to an AFM state polarized along the $x$ direction, with magnitude of the ordered moment:
\begin{equation}
\langle S^x_{j,+} \rangle_{\rm sp} = - \langle S^x_{j,-} \rangle_{\rm sp} = \sqrt{n_{c,{\bm 0}}n_{c,{\bm \pi}}}.
\end{equation}

\begin{figure*}
    \centering
    \includegraphics[width=\textwidth]{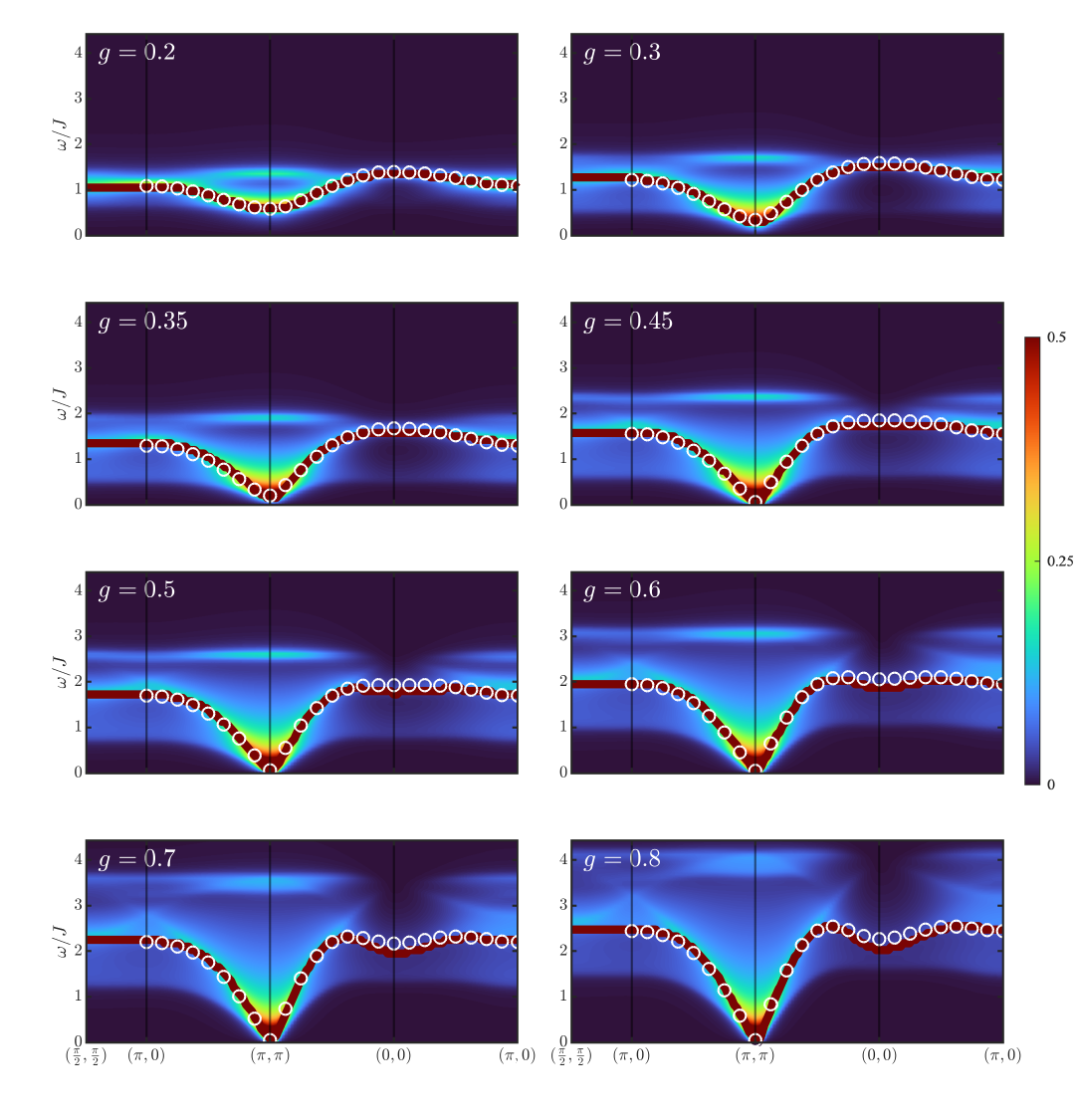}
    \caption{SP solution of the antisymmetric DSSF $S^{\mu \mu}_{{\rm sp},A}({\bm q},\omega)$ for different values of $g$, where $S^{xx}_{{\rm sp},A}({\bm q},\omega)=S^{yy}_{{\rm sp},A}({\bm q},\omega)=S^{zz}_{{\rm sp},A}({\bm q},\omega)$ is isotropic in QPM phase, and $S^{yy}_{{\rm sp},A}({\bm q},\omega)=S^{zz}_{{\rm sp},A}({\bm q},\omega)$ refers to the transverse spin components in AFM phase. For visual clarity, we represent the $\delta$-peak in the SB result with thick dark red lines. The white circles indicate the energy dispersion obtained from QMC simulations.}
    \label{fig:sqw_sp_odd}
\end{figure*}

For a finite condensate fraction, the SP Green's function becomes singular for momenta with a  gapless spectrum. This singularity can be regularized by considering a finite-size system, as is done when discussing the SP Green's function  in Eq.~\eqref{eq:sp_green2} and the SP equations in Eq.~\eqref{eq:speq}. Alternatively, one can isolate the singular part and then take the thermodynamic limit, which separates the SP Green's function into two components:
\begin{eqnarray}\label{eq:green_sp}
    G_{\rm sp}(k) = G_c(k) + G_{n}(k).
\end{eqnarray}
The first term on the right-hand side is the SP Green's function of the condensed SBs,
\begin{eqnarray}
    G_c(k) = g_{c}(2\pi)^3\delta({\bm k})\delta(\omega_m),
\end{eqnarray}
with
\begin{equation}
    g_c = \lim_{{\cal N}_D \to \infty} {1\over {\cal N}_D} X_c X_c^\dagger,
\end{equation}
and $X_c$ given by Eq.~(\ref{eq:Xc}). The second  term is the SP Green's function of the non-condensed SBs, which takes the same form as that in Eq.~\eqref{eq:sp_green2}. We note that, the above expression for the SP Green's function applies to both the QPM phase, for which $n_{c,{\bm \pi}} = 0$,  and the AFM phase. 

\section{Magnetic excitation spectrum}
\label{sec:dssf}

\begin{figure*}
    \centering
    \includegraphics[width=\textwidth]{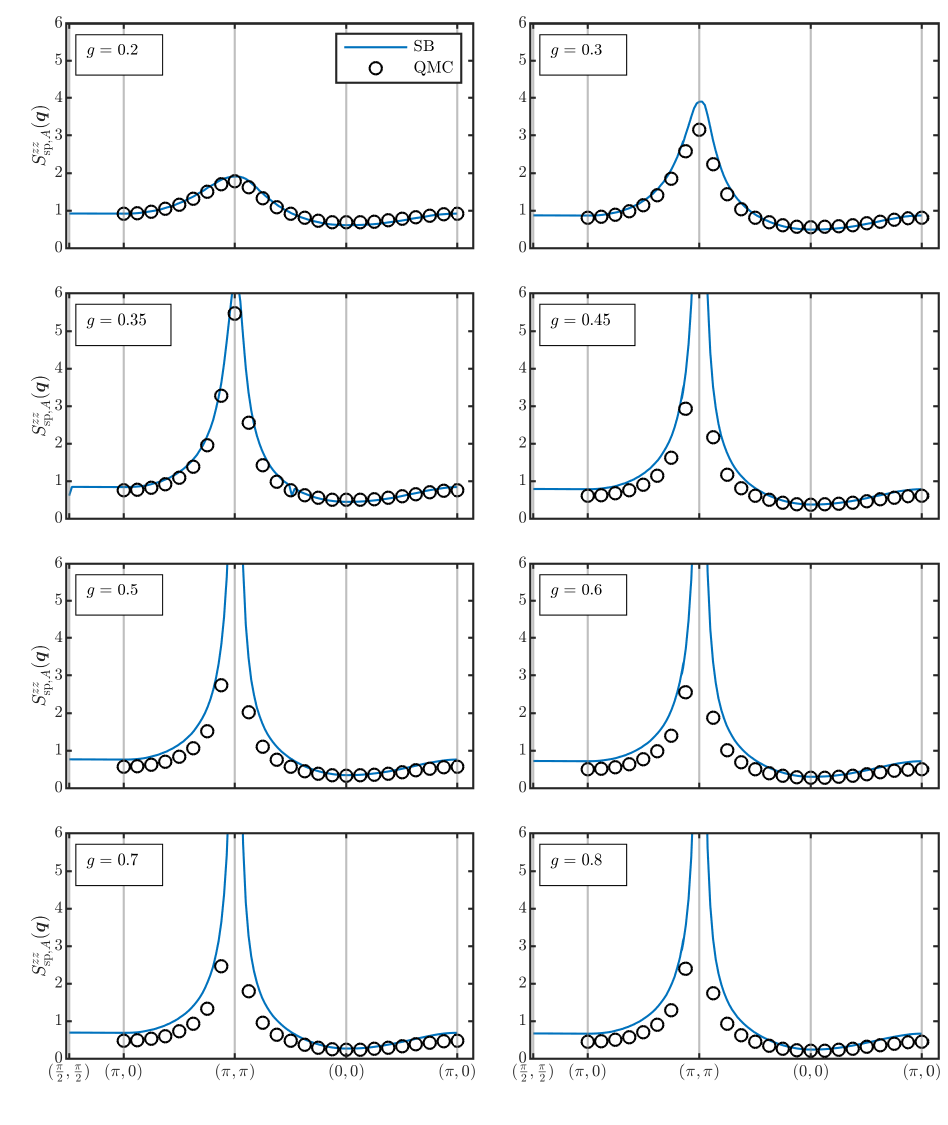}
    \caption{SP result of the static structure factor $S^{zz}_{{\rm sp},A}({\bm q})$ defined in Eq.~\eqref{eq:sq} for different values of $g$. The structure factor is independent of the $\mu$ component in the isotropic QPM phase.
    For the AFM phase, $S^{zz}_{{\rm sp},A}({\bm q})= S^{yy}_{{\rm sp},A}({\bm q})$ correspond to the two transverse spin components. The blue lines refer to the SP result, while black circles are obtained from the QMC simulations.}
    \label{fig:sq_sp_odd}
\end{figure*}

Having solved the SP equations, we are now equipped to study the magnetic excitations  described by the imaginary part of the magnetic susceptibility (fluctuation-dissipation theorem). Within the path-integral framework, the magnetic susceptibility is defined by the second derivative of the free energy with respect to an external time-space dependent magnetic field:
\begin{eqnarray}
    \chi^{\mu \nu}_{\sigma \sigma^\prime}({\bm q},i\omega_m) = \frac{\delta^2 \ln {\cal Z}(h)}{\delta h_{-q \sigma}^{\mu}\delta h_{q \sigma^\prime}^{\nu}},
\end{eqnarray}
where $\mu,\nu=x,y,z$ denotes the spin component indices, $\sigma,\sigma^\prime=\pm$ label the two sites within a dimer. ${\cal Z}(h)$ is the partition function in the presence of the external magnetic field $h_{j,\pm}^{\mu }$ coupled to the local moments via the action:
\begin{eqnarray}\label{eq:action_ext}
    {\cal S}_{\rm ext} = \frac{1}{2}\int_0^\beta d\tau \sum_j h^\mu_{j,\pm}(\tau) \psi_j^\dagger u^\mu_{\pm} \psi_j,
\end{eqnarray}
where $u^\mu_{\pm}$ is the matrix form of the spin operators in Nambu representation, providing the external vertex in Fig.~\ref{fig:building_blocks} (c), and $h_{q,\pm}^\mu = \frac{1}{\sqrt{{\cal N}_D\beta}}\sum_j \int_0^\beta d\tau h_{j,\pm}^\mu(\tau)e^{i\omega_n \tau - i {\bm q}\cdot {\bm r}_j}$ represents the Fourier transform of the external field. As detailed in Appendix~\ref{app:largeN}, the magnetic susceptibility can be expanded in powers of $1/N$. Through comparison with QMC results 
(for technical details of the QMC simulation, see Ref.~\cite{zhang2013phase}), we demonstrate that the SP approximation accurately describes the triplon excitations in the QPM phase and the two transverse magnon modes in the AFM phase.   However, capturing the longitudinal magnetic fluctuations requires the inclusion of essential $1/N$ corrections, which we discuss in the following section.

Given the invariance of the system under a mirror operation that exchanges the  two layers (up to a translation by one lattice space for the AFM phase), it is convenient to introduce  the symmetric (S) and antisymmetric (A) combinations of spin operators within a dimer:
\begin{equation}\label{eq:sa}
    S_{j,S}^{\mu} = S_{j,+}^{\mu} + S_{j,-}^{\mu}, \ \ S_{j,A}^{\mu} = S_{j,+}^{\mu} - S_{j,-}^{\mu}.
\end{equation}
along with the simplified notation for the magnetic susceptibility, $\chi_\alpha^{\mu \nu}$, where the subscript $\alpha=S,A$ denotes the symmetric or antisymmetric channels, respectively. 
Accordingly, we introduced the symmetric and antisymmetric external vertices, $u_{S}^{\mu} = u_{+}^{\mu} + u_{-}^{\mu}$ and $u_{A}^{\mu} = u_{+}^{\mu} - u_{-}^{\mu}.$
Due to the $\mathrm{SU}(2)$ rotation symmetry, or the residual $\mathrm{U}(1)$ symmetry about the local moments in the Néel order phase, the magnetic susceptibility is diagonal in the spin index. The diagonal components satisfy $\chi_\alpha^{xx}(q)=\chi_\alpha^{yy}(q)=\chi_\alpha^{zz}(q)$ in the QPM phase and $\chi_\alpha^{xx}(q)\neq \chi_\alpha^{yy}(q)=\chi_\alpha^{zz}(q)$ in the Néel ordered phase, assuming that the local moments align along the $x$ direction. 

In the remainder of this work, we focus on the antisymmetric channel, as it carries the dominant spectral weight within the parameter range of interest. For completeness, a parallel discussion of the symmetric channel is included in Appendix~\ref{app:sp_even}.

\subsection{Saddle-point approximation}
\label{sec:dssf1}

In the SP approximation, the magnetic susceptibility is given by the diagram shown in Fig.~\ref{fig:large_N}~(a):
\begin{eqnarray}\label{eq:chi_sp}
    \chi_{{\rm sp},\alpha}^{\mu \mu}(q) = \frac{1}{2{{\cal N}_D}\beta} \sum_{k} {\rm tr} \left[ G(k) u^{\mu}_{\alpha} G(k+q) u^{\mu}_{\alpha}  \right].
\end{eqnarray}
The susceptibility along the real-frequency axis is obtained through analytic continuation ($i\omega_n \to \omega + i 0^+$). According to the fluctuation-dissipation theorem, the dynamic spin structure factor is proportional to the imaginary part of the susceptibility:
\begin{eqnarray}
    S_{{\rm sp},\alpha}^{\mu \mu}({\bm q},\omega) = -\frac{1}{\pi} \text{Im}[\chi_{{\rm sp},\alpha}^{\mu \mu}({\bm q},\omega)].
\end{eqnarray}
As shown in Fig.~\ref{fig:sqw_sp_odd}, the DSSF in the antisymmetric channel ($\alpha=A$) consists of a $\delta$-peak and a continuum of excitations. To understand the origin of these two contributions we can split $\chi_{{\rm sp},A}^{\mu \mu}$ into two parts, $\chi_{{\rm sp},A}^{\mu\mu} = \chi_{{\rm sp},A}^{\mu\mu(1)} + \chi_{{\rm sp},A}^{\mu\mu(2)}$, where
\begin{eqnarray}
    \chi_{{\rm sp},A}^{\mu \mu(1)} ({\bm q},\omega) &=& \frac{1}{2}{\rm tr} \left[ g_c u^{\mu}_{A} G_n({\bm q},\omega + i 0^+) u^{\mu}_{A}  \right] \nonumber \\
    &+& \frac{1}{2}{\rm tr} \left[ G_n(-{\bm q},-\omega-i 0^+) u^{\mu}_{A} g_c u^{\mu}_{A}  \right]
\end{eqnarray}
involves one condensed SB. This contribution has a single-boson pole arising from  SP Green's function. The contribution from the second term, $\chi_{{\rm sp},A}^{\mu\mu(2)}$, becomes apparent after performing the Matsubara frequency summation,
\begin{widetext}
\begin{equation}
\label{eq:chi_sp2}
    \chi_{{\rm sp},A}^{\mu \mu(2)} ({\bm q},\omega) = \frac{1}{2 {{\cal N}_D}} \sum_{n,m=0}^3 {\sum_{\bm k}}^\prime
    \frac{{\rm tr} \left[ g_{{\bm k},n} u^{\mu}_{A} {\bar g}_{{\bm k}+{\bm q},m} u^{\mu}_{A}  \right]}{\varepsilon_{{\bm k},n} + \varepsilon_{{\bm k}+{\bm q},m} + \omega+ i0^+} + \frac{{\rm tr} \left[ {\bar g}_{{\bm k},n} u^{\mu}_{A} g_{{\bm k}+{\bm q},m} u^{\mu}_{A}  \right]}{\varepsilon_{{\bm k},n} + \varepsilon_{{\bm k}+{\bm q},m} - \omega-i0^+},
    \end{equation}
\end{widetext}
where the prime symbol indicates that the sum is restricted to momenta ${\bm k}$ such that $\varepsilon_{{\bm k},n} > 0, \varepsilon_{{\bm k}+{\bm q},m} > 0$.
This contribution gives rise to the two-particle continuum in the DSSF~\footnote{In the SP approximation, the two-particle continuum corresponds to the excitation of a singlet boson and a triplet boson (see Eq.~\eqref{eq:antisymm_spin}). However, this is a spurious continuum, as the singlet boson is not a physical mode. The physical continuum appears at higher energies and corresponds to a three-triplon excitations, which are only obtained after properly accounting for quantum fluctuations (this is a  generic feature of the large-$N$ theories). In contrast, the two-particle continuum in the symmetric channel of the DSSF is physical, as it consists of two triplon excitations.}.

In the QPM phase, the $\delta$-peak in the DSSF coincides with the SP spectrum of the three triplet bosons (triplon modes). Physically, these modes are excited by the antisymmetric spin operator:
\begin{equation}\label{eq:antisymm_spin}
    \hat{S}_{j A}^{\mu} = \hat{S}_{j +}^{\mu} - \hat{S}_{j -}^{\mu} = \hat{b}_{j,0}^\dagger \hat{b}_{j,\mu} + \hat{b}_{j,\mu}^\dagger \hat{b}_{j,0},
\end{equation}
because singlets and triplets have opposite parity under the mirror symmetry that  exchanges the two layers.
In the presence of a singlet-boson condensate, $\langle \hat{b}_{j,0}\rangle \simeq \sqrt{n_{c}}$ (here we fix the overall complex phase of the condensate to make this value real), we can approximate
\begin{equation}
    \hat{S}_{j A}^{\mu} \simeq \sqrt{n_c}\left( \hat{b}_{j,\mu} + \hat{b}_{j,\mu}^\dagger \right),
\end{equation}
which, upon acting on the ground state, creates a specific flavor of the triplet boson, corresponding to the $\delta$-peak in the DSSF. 

\begin{figure*}
    \centering
    \includegraphics[width=\textwidth]{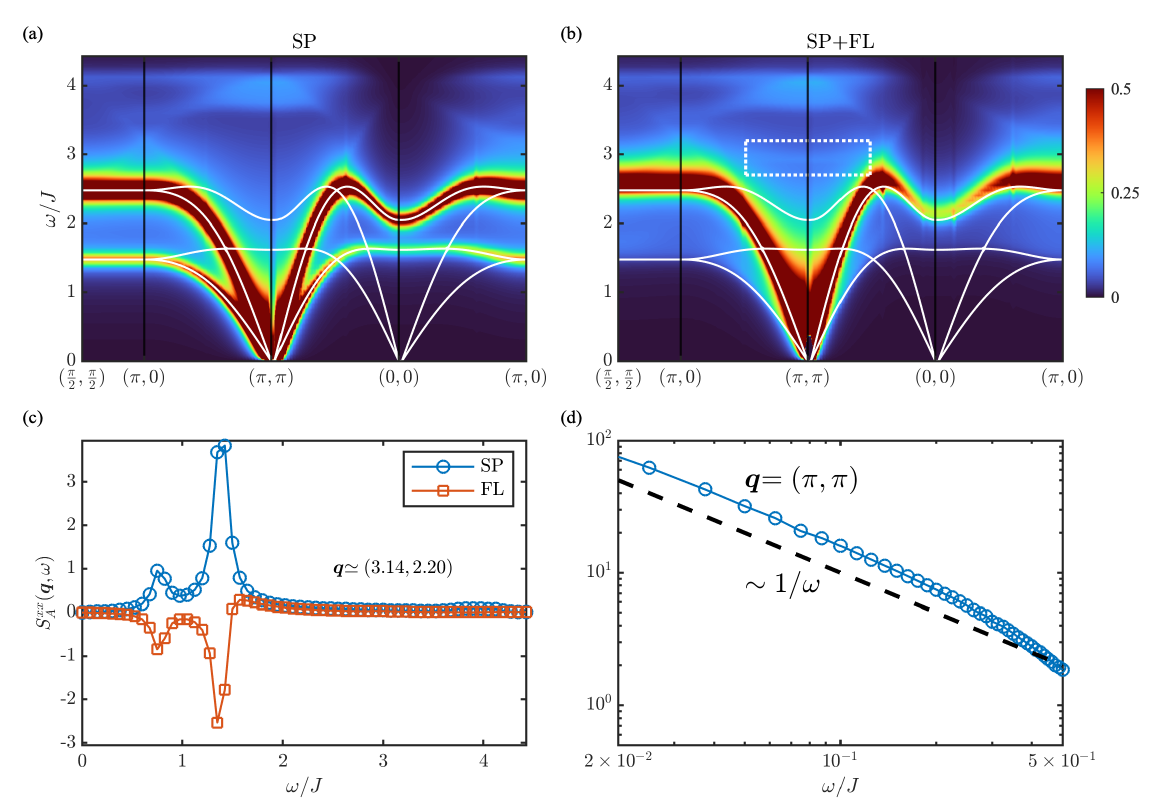}
    \caption{\textit{Longitudinal}  DSSF in the anti-symmetric channel for $g=0.8$. (a) SP result including two $\delta$-peaks corresponding to two unphysical modes. (b) Result after including the counter-diagram shown in Fig.~\ref{fig:diagram}~(b). The lines indicate the single-boson dispersion obtained from the SP Hamiltonian. The dotted square denotes the region where an extra low-intensity quasi-flat mode emerges from the fluctuations of the auxiliary fields. (c) Cancellation of the  $\delta$-peak of the SP susceptibility upon adding the counter-diagram shown in Fig.~\ref{fig:diagram}~(b). (d) Low-frequency scaling of the intensity of the continuum at the ordering wavevector $(\pi,\pi)$. The black dashed line indicates the slope associated with $1/\omega$ scaling.}
    \label{fig:sqw_fl}
\end{figure*}

In the AFM phase, both the singlet and the  triplet boson parallel to the ordered moment condense, $\langle \hat{b}_{j,0}\rangle \sim \sqrt{n_{c,{\bm 0}}}$ and $\langle \hat{b}_{j,1}\rangle \sim \sqrt{n_{c,\pi}}$ (up to a complex phase), while the other two \textit{transverse} triplet bosons remain uncondensed. Consequently, the two transverse spin operators can be approximated by :
\begin{equation}
    \hat{S}_{j A}^{\mu} \simeq \sqrt{n_{c,{\bm 0}}}\left( \hat{b}_{j,\mu} + \hat{b}_{j,\mu}^\dagger \right) \quad \text{for} \ \mu=y,z.
\end{equation}
This operator generates the corresponding flavor of triplet bosons upon acting on the ground state, corresponding to the magnon excitations of the system. In contrast, the longitudinal spin operator,
\begin{equation}
    \hat{S}_{j A}^{x} \simeq \sqrt{n_{c,{\bm 0}}}\left( \hat{b}_{j,1} + \hat{b}_{j,1}^\dagger \right)+\sqrt{n_{c,{\bm \pi}}}\left( \hat{b}_{j,0} + \hat{b}_{j,0}^\dagger \right),
\end{equation}
generates two types of $\delta$-peaks arising from the $\hat{b}_{j,0}$ and $\hat{b}_{j,1}$ bosons, as shown in Fig.~\ref{fig:sqw_fl} (a). In the next section we will demonstrate that the $\delta$-peak arising from  the $\hat{b}_{j,0}$ boson disappears upon adding the ``counter-diagram'' shown in Fig.~\ref{fig:large_N}~(b). Furthermore, the $\delta$-peak from  the $\hat{b}_{j,1}$ boson gets broadened.

Remarkably, the DSSF described by the SP approximation of the $\mathrm{SU}(4)$ SB approach shows very good agreement with QMC simulations. As demonstrated in Fig.~\ref{fig:sqw_sp_odd}, both the triplon dispersion in the QPM phase and the magnon dispersion in the N\'eel order phase match the QMC results with high precision. This agreement persists deep inside the magnetically ordered phase. Moreover, the spectral weight carried by these excitations  closely aligns with the QMC results. In Fig.~\ref{fig:sq_sp_odd}, we compare the total spectral weight at a given ${\bm q}$, obtained from the equal-time correlation function in the antisymmetric channel
\begin{eqnarray}\label{eq:sq}
S_A^{\mu \mu}({\bm q}) = \int_0^\infty \frac{d\omega}{2\pi} S_A^{\mu \mu}({\bm q},\omega).
\end{eqnarray}
We find that the ${\bm q}$ dependence of  $S_A^{\mu \mu}({\bm q})$ closely follows the QMC results, though the value obtained from the SB approach is systematically slightly higher than the QMC value. This overestimation of the spectral weight is a known feature of the mean field SB theory and it arises from violations of the local constraint. For instance, in the $\mathrm{SU}(2)$ SB approach, the integrated spectral weight is overestimated  by a factor of $3/2$~\cite{auerbach2012interacting}. As we will see in the next section, the inclusion a $1/N$ contribution reduces the integrated spectral weight and further improves the agreement between the SB approach and QMC.

\begin{figure}
    \centering
    \includegraphics[width=\textwidth]{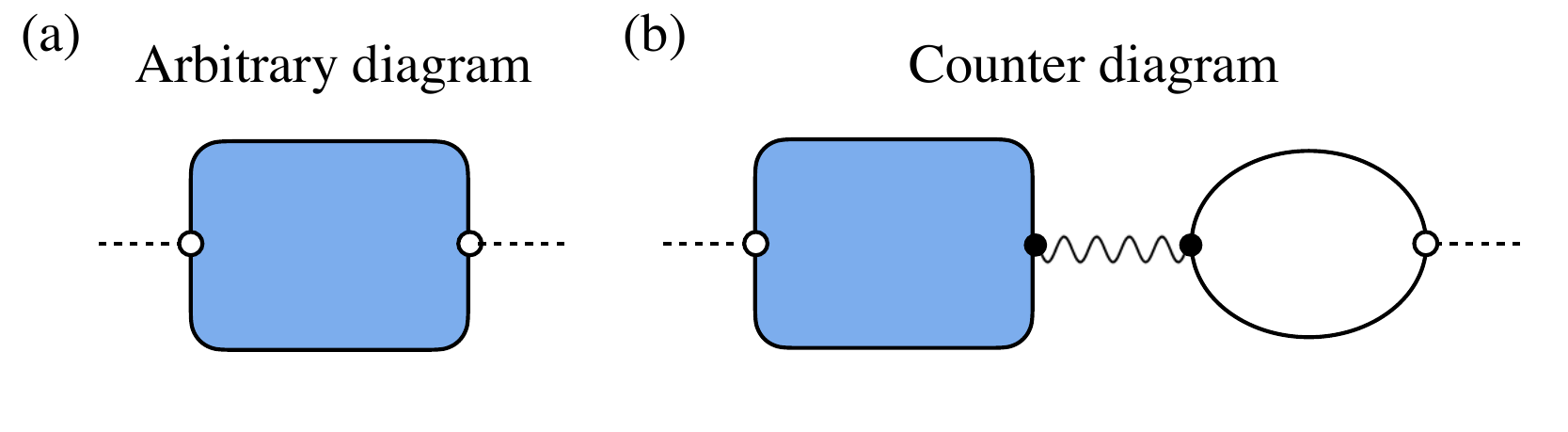}
    \caption{Each diagram shown in panel (a) must be accompanied by the counter diagram shown in panel (b). }
    \label{fig:recipe}
\end{figure}

\begin{figure}
    \centering
    \includegraphics[width=\textwidth]{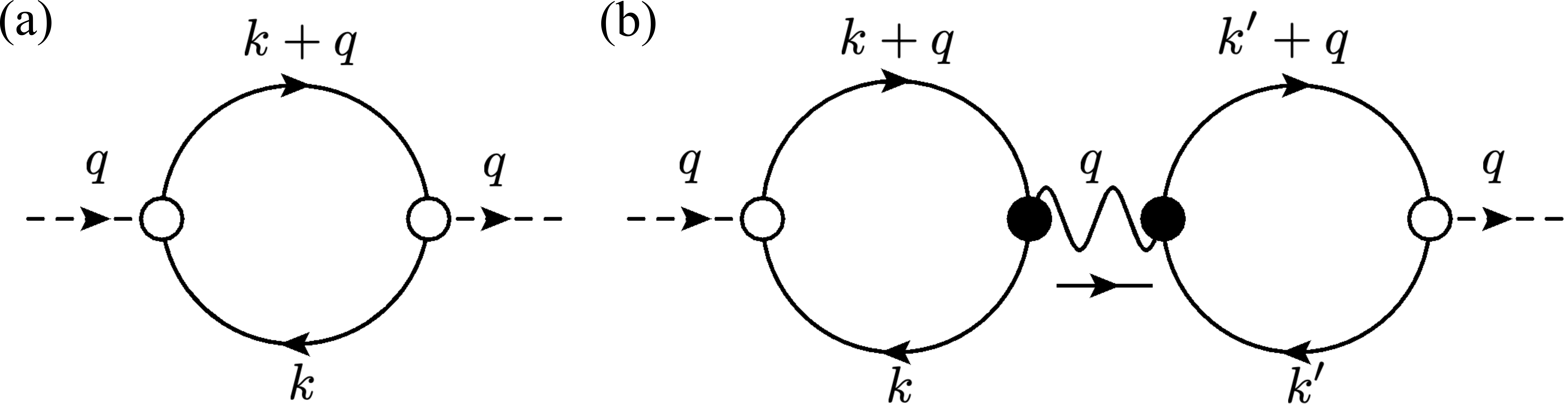}
    \caption{(a) SP diagram and (b) counter-diagram for the  magnetic susceptibility. The counter-diagram vanishes for all components of the magnetic susceptibility in the QPM phase,  and for the transverse components in the AFM phase. Arrows in the plot indicate the particle flow.}
    \label{fig:diagram}
\end{figure}

\subsection{Amplitude fluctuation in N\'eel AFM phase}
\label{sec:dssf2}

In this section, we examine the longitudinal spin fluctuations of the AFM phase. Although the longitudinal mode is expected to be gapped, it acquires an intrinsic broadening due to its decay into pairs of transverse Goldstone modes (i.e., magnons). In a two-dimensional system, the infrared divergence associated with this decay process raises the question of whether this mode can be distinguished from the two-magnon continuum. This issue has been investigated through various approaches in the literature, including field-theoretic analyses of the effective low-energy ${\mathrm O}(3)$ model describing long-wavelength fluctuations of local magnetic moments~\cite{podolsky2011visibility,podolsky2012spectral,gazit2013fate},  as well as QMC studies~\cite{lohofer2015dynamical,pollet2012higgs}. These studies conclude that the amplitude mode is barely visible in the dynamic spin structure factor for the 2D ${\mathrm O}(3)$ model. Here, we revisit this problem using the $\mathrm{SU}(4)$ SB approach and demonstrate that the same conclusion holds.

We first recall that, in the SP approximation, the longitudinal spin fluctuation $S_{{\rm sp},A}^{xx}({\bm q},\omega)$ consists of two $\delta$-peaks: one arising from the longitudinal triplet boson $\hat{b}_{{\bm q},1}$ and the other one arising from the singlet boson $\hat{b}_{{\bm q},0}$ [see Fig.~\ref{fig:sqw_fl} (a)].  This qualitatively incorrect result is characteristic of SP solutions with condensed bosons. As it has been explained at length for $\mathrm{SU}(2)$ SBs~\cite{ghioldi2018dynamical,ghioldi2022evidence,zhang2022schwinger}, in presence of a condensate, each diagram must be accompanied by a counter diagram where one external vertex is replaced by a ``tail'' consisting of the RPA propagator attached to a bubble with an internal and an external vertex (see Fig.~\ref{fig:recipe}). Since this bubble represents a cross susceptibility between the spin-components, which transform like a vector under $\mathrm{SU}(2)$ rotations and the auxiliary fields, which are scalars, it vanishes in the QPM phase along with the counter-diagram  of the SP diagram, shown in Fig.~\ref{fig:diagram} (b). This is the reason why the SP spectrum is qualitatively correct on the QPM side of the phase diagram. 

For the AFM phase, the residual symmetry group is $\mathrm{U}(1)$. Since the transverse spin-components still transform like a vector under this group, the counter-diagram still vanishes for the transverse components of the spin susceptibility. This is not true, however, for the longitudinal spin component that is a scalar of the residual symmetry group. The counter-diagram, whose nominal order is $1/N$, includes singular contributions of order $1/N^0$ that exactly cancel the residues of the two poles of $\chi_{{\rm sp},A}^{xx}({\bm q},\omega)$, eliminating the aforementioned pair of $\delta$-peaks.


\begin{figure}
    \centering
    \includegraphics[width=\textwidth]{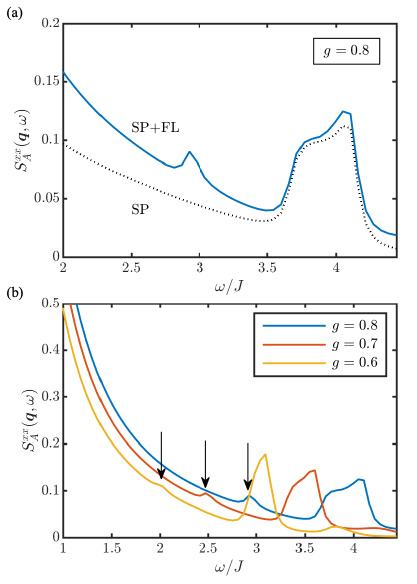}
    \caption{\textit{Longitudinal} antisymmetric DSSF at the ordering wave vector ${\bm \pi}$. (a) Extra peak induced by the inter-particle interaction. (b) Softening of the extra peak as reducing inter-dimer interaction.}
    \label{fig:sqw_fl_Q}
\end{figure}

As derived in Appendix~\ref{app:largeN}, the diagram in Fig.~\ref{fig:diagram} (b) is given by
\begin{eqnarray}\label{eq:chi_fl}
    \chi_{{\rm fl}, A}^{\mu \nu}(q) = \sum_{\alpha,\alpha^\prime} \Lambda^{\mu,\alpha}_A(q) D_{\alpha \alpha^\prime}(q) \Lambda_A^{\mu \bar{\alpha^\prime}}(-q),
\end{eqnarray}
where the subscript ``fl'' denotes ``correction due to fluctuations'' of the auxiliary fields,
\begin{eqnarray}
      \Lambda_A^{\mu,\alpha} ({q}) =  \frac{1}{{\cal N}_D\beta} \sum_{k} {\rm tr} \left[ G(k) u_A^{\mu} G({k+q}) v^{\alpha}_{k+q,k} \right]
      \nonumber \\
\end{eqnarray}
is the cross susceptibility between the spin-components and the auxiliary fields, and $D_{\alpha\alpha^\prime}(q)$ is the propagator of the fluctuation field, defined in Eq.~\eqref{eq:Drpa}.  Fig.~\ref{fig:sqw_fl} (b) presents the longitudinal antisymmetric DSSF in the AFM phase after including this contribution for  $g=0.8$. We observe that the two $\delta$-peaks in the SP result, 
shown in Fig.~\ref{fig:sqw_fl} (a), are removed, as explicitly shown in Fig.~\ref{fig:sqw_fl} (c)  for an arbitrary momentum ${\bm q}$. 

\begin{figure}
    \centering
    \includegraphics[width=\textwidth]{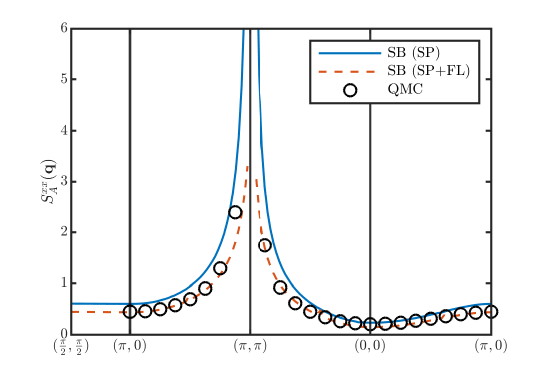}
    \caption{Equal-time \textit{longitudinal} spin structure factor $S^{xx}_{A}(\bm q)$ for $g=0.8$ in the SP approximation of SB approach and that including the proper $1/N$ corrections, and the QMC result.}
    \label{fig:sq_fl}
\end{figure}

With both $\delta$-peaks removed, the remaining spectral weight forms a structured continuum. Focusing on the low-frequency regime around the ordering wave vector, the intensity of the continuum scales as $1/\omega$ (see Fig.~\ref{fig:sqw_fl} (d)). This scaling is already evident at the SP level, as detailed in 
Appendix~\ref{app:sp_infrared}. It results from the product of the density of states, which scales as $\omega$ (linear dispersion) and the square of the matrix element that scales like $1/\omega^2$. This behavior is consistent with previous studies using different analytical and numerical methods~\cite{podolsky2011visibility,podolsky2012spectral,gazit2013fate,lohofer2015dynamical,pollet2012higgs}. 

Moving away from the ordering wave vector, the longitudinal DSSF displays a broadened peak at energy $v_L \rvert{\bm q}\rvert$ with an intrinsic broadening $\Gamma_{\bm q}$, where ${\bm q}$ is the momentum measured from the ordering wave vector ${\bm \pi}$. The velocity $v_L$ approximately equals that of the triplet bosons, as seen in Fig.~\ref{fig:sqw_fl} (b). The shape of this peak fits well with the formula $\text{Im}[Z_{\bm q}/\sqrt{(v_L \rvert {\bm q} \rvert)^2 -(\omega+i \Gamma_{\bm q})^2}]$, corresponding to a pole in the magnetic susceptibility in the lower half of the complex frequency plane, $z = v_L \rvert {\bm q} \rvert - i \Gamma_{\bm q}$. As ${\bm q} \to \bm 0$, both the real and imaginary parts of this pole reduce to zero, while $Z_{\bm q}$ remains finite, thereby recovering the $1/\omega$ form of the DSSF.

A closer examination of panels (a) and (b) of Fig.~\ref{fig:sqw_fl}  reveals a low intensity quasi-flat mode emerging from the continuum around energy $3J$. The feature, which is absent in the SP contribution is more clearly displayed by the constant ${\bm q}={\bm \pi}$ plot, shown in Fig.~\ref{fig:sqw_fl_Q} (a). The SP result shows a regular continuum over the energy range of interest, while the inclusion of $\chi_{{\rm fl},A}^{xx}(q)$ introduces a peak, signaling the emergence  of a resonance state due to fluctuations of the auxiliary fields. As the inter-dimer interaction decreases, this peak gradually shifts to lower energies and becomes less prominent near the critical point, where it is overshadowed by the high-intensity ($1/\omega$ scaling) of the low-frequency continuum (see Fig.~\ref{fig:sqw_fl_Q} (b)). Notably, the energy scale of this mode is comparable to the amplitude fluctuations observed in QMC simulations at $\omega \approx 1.5J$.
Additionally, the DSSF displays a broad peak around $\omega \sim 4J$, which is already present in the SP solution and undergoes only minor modifications due to quantum fluctuation effects. This broad structure also appears in QMC simulations at a similar energy scale~\cite{lohofer2015dynamical}.

Finally, in Fig.~\ref{fig:sq_fl} we compare the equal-time longitudinal structure factor, defined in Eq.~\eqref{eq:sq}, with the QMC result for $g=0.8$. The significant improvement over the SP result stems from a more accurate treatment of the local constraint in Eq.~\eqref{eq:constraint2}, leading to a better fulfillment of the sum rule.

\section{Concluding remarks}
\label{sec6}

As was observed by Perelomov in the seventies~\cite{Perelomov72},  coherent states of Lie algebras provide the natural link between quantum and classical mechanics. In fact, $N$-level quantum mechanical systems admit multiple classical limits corresponding to different choices of $N$-dimensional representations of distinct Lie algebras. Naturally, for each specific choice of Hamiltonian, there is one classical limit that better approximates   dynamics of the quantum system. 

Spin systems naturally provide examples of $N$-level units, where the unit may consist of a single spin $(N=2S+1)$ or a set of spins, such as the dimer units considered in this work. Recent studies have highlighted multiple instances of realistic spin Hamiltonians where the classical limit based on coherent states of the completely symmetric irreps of $\mathrm{SU}(N)$, labelled by the integer index $M$, offers a more accurate approximation of the exact dynamics than the classical limit based on $\mathrm{SU}(2)$ coherent states~\cite{Zhang21,Dahlbom22,Dahlbom22b,dahlbom2024classical}.

Since the classical theory becomes exact in the limit $M \to \infty$, quantum corrections for describing the spin dynamics for a finite $M$ irrep can be implemented via a $1/M$ expansion. This expansion is achieved by introducing a faithful representation of spin operators in terms of SU$(N)$ Schwinger bosons and accounting for the local constraint \eqref{eq:constraint2} via a generalization of the Holstein-Primakoff (HP) transformation. The $1/M$ expansion arises from the Taylor series expansion of the square root function inherent to the HP transformation. The resulting quadratic Hamiltonian in HP bosons corresponds to the generalized spin wave Hamiltonian, with the effects of non-quadratic terms systematically included order by order in $1/M$. The diagrammatic representation of this $1/M$ expansion leads to the so-called ``loop expansion'', where the order of each diagram in powers of $1/M$ is equal to the number of loops~\cite{Do2021,Bai2023}.

In contrast to the semi-classical large-$M$ expansions described above, one can also use the SB representation of spin operators as a basis for implementing a 
$1/N$ expansion, where the spin model is generalized to more general groups labeled by $N$. Just as standard semi-classical large-$S$
approaches rely on $\mathrm{SU}(2)$ coherent states, traditional large-$N$ approaches utilize $\mathrm{SU}(2)$ SBs to represent spin operators. However, since other Lie algebras may be more suitable for semi-classical expansions, it is natural to conjecture that these alternative algebras could also provide a better foundation for implementing a $1/N$ expansion. The results presented in this manuscript confirm this conjecture.  The two different strategies, large-$M$ and large-$N$, are schematically shown in Figure~\ref{fig:limits}.

To test the above-mentioned conjecture, in this work we developed a large-$N$ $\mathrm{SU}(4)$ Schwinger boson theory for the spin-$1/2$ bilayer square lattice Heisenberg antiferromagnet. The spin operators of each dimer are represented by four SBs, that create the coherent states that represent any quantum mechanical state of a given dimer. A key distinction from standard large-$N$ approaches based on $\mathrm{SU}(2)$ SBs~\cite{auerbach2012interacting} is that the $\mathrm{SU}(4)$ SBs fully capture the \emph{intra-dimer} entanglement, while the link fields and their fluctuations generate  \emph{inter-dimer} entanglement. To perform a large-$N$ expansion, the original dimer problem with two antiferromagnetically coupled $\mathrm{SU}(2)$ spins on each site of the dimer has been generalized to two antiferromagnetically coupled SU(n) spins, with $n \ge 2$. Then, $\mathrm{SU}(N=n^2)$ SBs have been used to include all quantum mechanical states of the generalized dimer in the manifold of $\mathrm{SU}(N)$ coherent states.
Both the static and dynamical properties are significantly better described by the large-$N$ expansion based on $\mathrm{SU}(4)$ 
SBs than by the large-$M$ $\mathrm{SU}$(4) limit.

Furthermore,  the theory introduces an expansion parameter  ($1/N$), that cures the shortcomings of previous mean-field approximations  based on bond operators~\cite{sachdev1990bond,matsushita1999bond,yu1999bond,ganesh2011neel}.  While alternative approaches based on expansions in the inverse system dimension~\cite{joshi2015nonlinear,joshi2015nonlinear2} address the same problem, they still produce values of $g_c$, which deviate significantly from the numerical (QMC) values. Similar limitations are also observed in 
large-$M$ expansions based on $\mathrm{SU}(4)$ coherent states or large-$N$ expansions based on $\mathrm{SU}(2)$ 
SBs~\cite{Millis93,Millis94}.

Achieving quantitative agreement with the exact value of $g_c$ is crucial for precisely describing real materials near the quantum critical point. 
Typically, the exact value of $g_c$ cannot be determined from QMC simulations due to the well-known sign problem that affects most 
frustrated Hamiltonians of interest. Without precise knowledge of $g_c$, it is challenging to overcome the quantitative limitations of the aforementioned methods by simply rescaling $g$.

Therefore, the remarkable accuracy of the large-$N$ method introduced in this work marks a significant advancement in modeling the static and dynamical properties of coupled antiferromagnetic dimers near continuous quantum phase transitions between magnetically ordered and paramagnetic states.\\

\begin{acknowledgments}
We thank Martin Mourigal for useful discussions. This work was supported by the U.S. Department of Energy, Office of Science, Basic Energy Sciences, Materials Sciences, and Engineering Division under award DE-SC-0018660. E.A.G. is supported by the Quantum Science Center (QSC), a National Quantum Information Science Research Center of the U.S. Department of Energy (DOE). This work was  performed in part at the Aspen Center for Physics, which is supported by National Science Foundation grant PHY-2210452. Y. K. is supported by Japan Society for the Promotion of Science (JSPS) KAKENHI Grant Numbers JP20H00122 and JP22K03509. A part of numerical calculations were performed using the facilities of the Supercomputer Center, The Institute for Solid State Physics, The University of Tokyo.
\end{acknowledgments}

\appendix

\section{Large-$N$ expansion of DSSF}
\label{app:largeN}

To compute the dynamic magnetic susceptibility within the path-integral framework, we introduce an external source term coupled to the local magnetic moment, as described by  Eq.~\eqref{eq:action_ext} in the main text. 
The Hubbard-Stratonovich transformation is then applied to the quartic interaction term, and the action is expanded around the saddle point (SP), similar to the procedure without the source term. It is important to note that the SP remains the same as in the absence of the source term. After integrating out the boson fields, the effective action in the presence of the source term is given by:
\begin{eqnarray}
    S_{\rm eff}(\bar{\Phi},\Phi,h) &=& \frac{1}{2} \sum_{q,\alpha,\alpha^\prime} (\bar{\Phi}^{\dagger})_{q}^{\alpha} {\Pi}_{0}^{\alpha \alpha^\prime} \Phi_{q}^{\alpha^\prime} + \frac{2}{N} \text{Tr}[\ln {\cal M}(h)] \nonumber \\
    &&+ \frac{4 S_{\rm cl}}{N}.
\end{eqnarray}
The external source term appears in the matrix ${\cal M}(h)$, whose matrix elements are defined by
\begin{eqnarray}
    {\cal M}_{k,k+q}=\delta_{q,0} G_{\rm sp}^{-1}(k) + 
   2 \Phi^{\alpha}_{-q} V^{\alpha}_{k,k+q} + \sum_{\sigma=\pm}h^{\mu}_{-q,\sigma} u_{\sigma}^{\mu},\nonumber \\
\end{eqnarray}
where $\sigma=\pm$ is the layer index, and $h^{\mu}_{q,\pm}$ is the Fourier transform of the external source $h_{j,\pm}(\tau)$ introduced in Eq.~\eqref{eq:action_ext} of the main text. The magnetic susceptibility is given by
\begin{eqnarray}
    \chi^{\mu\nu}_{\alpha}(q) = \frac{\delta^2 \ln {\cal Z}(h)}{\delta h^\mu_{-q,\alpha} \delta h^\nu_{q,\alpha}},
\end{eqnarray}
where the subscript $\alpha=S,A$ denotes the symmetric and antisymmetric combinations between the external source terms from the two layers (see Eq.~\eqref{eq:sa} in the main text). The susceptibility decomposes into two parts, $\chi_\alpha(q) = \chi^{{\rm (I)}}_\alpha(q) + \chi^{{\rm (II)}}_\alpha(q)$, with
\begin{eqnarray} \label{eq:chi_app}
    \chi^{{\rm (I)}\mu\nu}_\alpha(q) &=& \frac{1}{2{\cal N}_D \beta} \sum_k \int {\cal D}[\bar{\Phi},\Phi] \frac{e^{-S_{\rm eff}}}{\cal Z} \nonumber \\
    &&\times {\rm tr}\left[ {\cal M}^{-1}_{k+q,k} u_\alpha^\mu {\cal M}^{-1}_{k,k+q} u_\alpha^\nu \right], \nonumber \\
    \chi^{{\rm (II)}\mu\nu}_\alpha(q) &=& \frac{1}{4{\cal N}_D \beta} \sum_k \int {\cal D}[\bar{\Phi},\Phi] \frac{e^{-S_{\rm eff}}}{\cal Z} {\rm tr}\left[ {\cal M}^{-1}_{k+q,k} u_\alpha^\mu \right] \nonumber \\
    &&\times {\rm tr}\left[{\cal M}^{-1}_{k,k+q} u_\alpha^\nu \right] - {\cal N}_D \beta m^{\mu}_{\alpha,q} m^\nu_{\alpha,-q},
\end{eqnarray}
where
\begin{eqnarray}
    m^\mu_{\alpha,q} \!=\!  \frac{-1}{2 {\cal N}_D \beta } \sum_k \!\int\!\! {\cal D}[\bar{\Phi},\Phi] \frac{e^{-S_{\rm eff}}}{\cal Z} {\rm tr} \left[ {\cal M}^{-1}_{k+q,k} u_{\alpha}^\mu \right]
\end{eqnarray}
gives rise to the local magnetic moments. The last term in $\chi^{{\rm (II)}}(q)$ eliminates the disconnected diagrams generated from the first term. At this point, one can set $h^\mu_{q,\alpha}$ to zero.

The calculation of $\chi^{\mu\nu}_\alpha(q)$ can be carried out perturbatively in terms of $1/N$. For this purpose, one needs to expand ${\cal M}^{-1}$, which appears in the integrand of $\chi^{{\rm (I)}\mu\nu}_\alpha(q)$ and $\chi^{{\rm (II)}\mu\nu}_\alpha(q)$, and $\exp(-S_{\rm eff})$ with respect to the fluctuation field $\Phi$. We have
\begin{eqnarray}
    \left( {\cal M}\right)^{-1}_{k,k+q} &&= G_{\rm sp}(k)\delta_{q,0} \nonumber\\
    &&+\sum_{m=1}^\infty (-1)^m \left( 2 \Phi^\alpha G_{\rm sp} V^{\alpha} \right)^m_{k,k+q} G_{\rm sp}(k+q),
    \nonumber\\
\end{eqnarray}
where $\left( 2 \Phi^\alpha G_{\rm sp}V^{\alpha}\right)_{k,k+q} \equiv 2 \Phi^\alpha_{-q} G_{\rm sp}(k) V^{\alpha}_{k,k+q}$. According to Eq.~\eqref{eq:Seffexp},
\begin{eqnarray}
    e^{-S_{\rm eff}} = e^{-\frac{4 S_{\rm cl}}{N}-\frac{2}{N}\text{Tr}[\ln G_{\rm sp}^{-1}]-S^{(2)}(\bar{\Phi},\Phi)}\sum_{p=0}^{\infty} \frac{(-1)^p}{p!} S^p_{\rm int},
    \nonumber \\
\end{eqnarray}
where $S_{\rm int}$ is given by Eq.~\eqref{eq:internal_loop}. The functional integral over the fluctuation fields is then carried out using Wick's theorem. 

Note that disconnected diagrams generated from this integration are completely canceled out by the perturbative expansion of the partition function ${\cal Z}$. The resulting series expansion can be conveniently represented by Feynman diagrams. The nominal order ${\cal O}(1/N^{L-P})$ of each diagram is determined by the number of internal loops ($L$, defined in Eq.~\eqref{eq:internal_loop}) and the number of RPA propagators.

In Fig.~\ref{fig:large_N}, we present the large-$N$ expansion of the magnetic susceptibility up to nominal ${\cal O}(1/N)$. The leading order in Fig.~\ref{fig:large_N} (a), i.e., the SP approximation, corresponds to the leading term in $\chi_\alpha^{\rm (I)}$, obtained by taking ${\cal M}^{-1} = G_{\rm sp}$. The $1/N$ diagrams in Fig.~\ref{fig:large_N} (c) correspond to higher-order terms in $\chi_\alpha^{\rm (I)}$. The other $1/N$ diagram in Fig.~\ref{fig:large_N} (b) corresponds to the leading term in $\chi_\alpha^{\rm (II)}$. As emphasized in the main text, when certain flavors of SBs condense, the $1/N$ diagrams contain singular contributions of ${\cal O}(1)$, which must be considered on an equal footing with the SP approximation.

\begin{figure*}
    \centering
    \includegraphics[width=\textwidth]{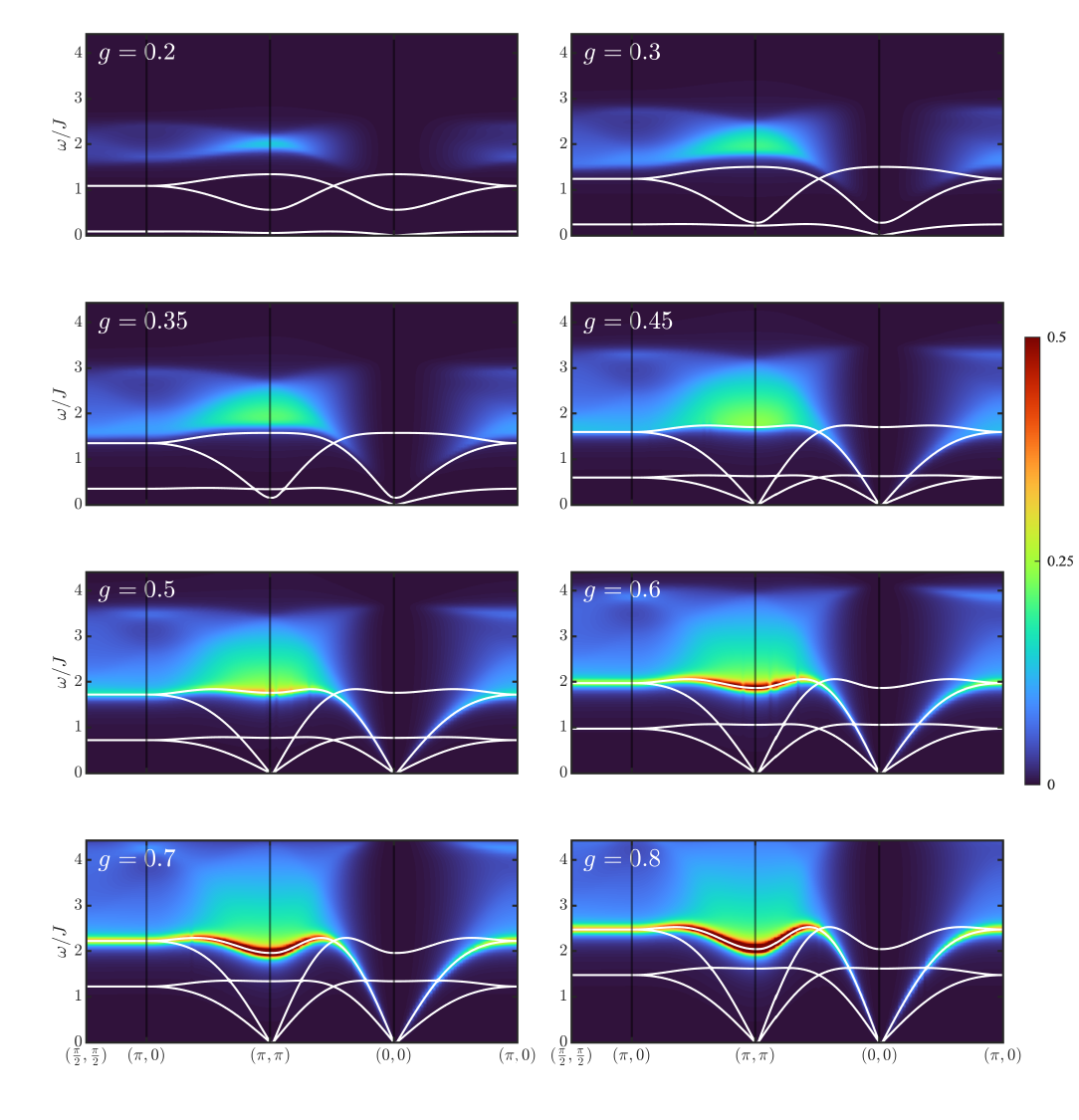}
    \caption{SP solution of the symmetric DSSF $S^{\mu \mu}_{{\rm sp},S}({\bm q},\omega)$ for different values of $g$, where $S^{xx}_{{\rm sp},S}({\bm q},\omega)=S^{yy}_{{\rm sp},S}({\bm q},\omega)=S^{zz}_{{\rm sp},S}({\bm q},\omega)$ is isotropic in QPM phase, and $S^{yy}_{{\rm sp},S}({\bm q},\omega)=S^{zz}_{{\rm sp},S}({\bm q},\omega)$ refers to the transverse spin components in AFM phase. White lines denote $\varepsilon_{{\rm q},\mu}$ in the QPM phase; extra lines corresponding to $\varepsilon_{{\bm q}+{\bm \pi},\mu}$ are shown in the N\'eel AFM phase.}
    \label{fig:sqw_sp_even_transverse}
\end{figure*}

\section{Longitudinal DSSF at low frequency}
\label{app:sp_infrared}

Here we analyze the low frequency behavior of the antisymmetric longitudinal DSSF for the ordering wave vector ${\bm q} = {\bm \pi}$. 

In the SP approximation, the longitudinal DSSF in the antisymmetric channel reads

\begin{eqnarray}
\label{eq:sxx_sp_app}
    S^{xx}_{{\rm sp},A}({\bm \pi},\omega) = && \Theta(\omega) \int \frac{d^2{\bm k}}{(2\pi)^2} (v_{{\bm k},0}u_{{\bm k},1} + u_{{\bm k},0} v_{{\bm k},1})^2 \nonumber\\
    &&\delta(\varepsilon_{{\bm k},0}+\varepsilon_{{\bm k},1}-\omega).
\end{eqnarray}
For small $\omega$, the singlet and triplet boson dispersions are linear near ${\bm q} = {\bm 0}$ and ${\bm q} = {\bm \pi}$ with velocities $v_s$ and $v_t$, respectively. Since the spin operator creates a singlet and a triplet boson with momenta ${\bm k}$ and ${\bm \pi} - {\bm k}$, only $\rvert{\bm k}\rvert \sim \omega/(v_s+v_t)$ contributes to the integral. Thus, in the long-wavelength limit we have:
\begin{eqnarray}
&& \varepsilon_{{\bm k},0} = v_s \rvert{\bm k}\rvert  (1 + {\cal O}(\rvert{\bm k}\rvert^2)), \\
&& \varepsilon_{{\bm k}+{\bm \pi},1} = v_t \rvert{\bm k}\rvert  (1 + {\cal O}(\rvert{\bm k}\rvert^2)), \\
    &&u_{{\bm k},0} \approx - v_{{\bm k},0} = \sqrt{\frac{\Delta_s}{2v_s \rvert{\bm k}\rvert}} (1 + {\cal O}(\rvert{\bm k}\rvert^2)), \\
    &&u_{{\bm k}+{\bm \pi},1} \approx - v_{{\bm k}+{\bm \pi},1} = \sqrt{\frac{\Delta_t}{2v_t \rvert{\bm k}\rvert}}  (1 + {\cal O}(\rvert{\bm k}\rvert^2)),
\end{eqnarray}
where $\Delta_s = 2J^\prime {\cal A}$, $\Delta_t = 2 J^\prime ( {\cal A} -  {\cal S})$, $v_s = \sqrt{\Delta_s(\tilde{\lambda}-J)/2}$, $v_t = \sqrt{\Delta_t \tilde{\lambda}/2}$. 
For $g = 0.8$
considered in the main text, we have $\Delta_s \approx 1.0336 J$, $\Delta_t \approx 2.0542 J$, $v_s \approx 0.8737 J$, and $v_t \approx 1.5951 J$. By substituting these expressions to Eq.~\eqref{eq:sxx_sp_app}, we find
\begin{eqnarray}
    S^{xx}_{{\rm sp},A}({\bm \pi},\omega) = {\Theta(\omega)\over 2\pi} {\Delta_s \Delta_t \over v_s v_t} {1\over \omega}.
\end{eqnarray}
The numerical solution presented in the main text matches this form very well in the low-frequency regime. This low frequency behavior is preserved after including  the contribution from the counter-diagram shown in Fig.~\ref{fig:diagram}~(b). The correction only modifies the prefactor of $1/\omega$.

\section{DSSF in the symmetric channel}
\label{app:sp_even}

\begin{figure}
    \centering
    \includegraphics[width=\textwidth]{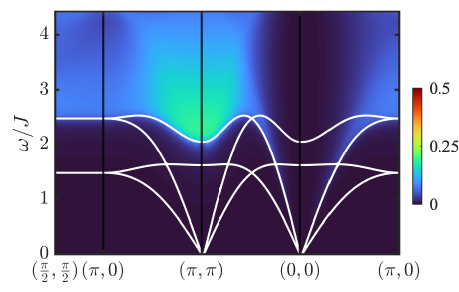}
    \caption{SP solution of the longitudinal symmetric DSSF $S^{xx}_{{\rm sp},S}({\rm q},\omega)$ for a representative $g=0.8$. White lines denote both $\varepsilon_{{\rm q},\mu}$ and $\varepsilon_{{\bm q}+{\bm \pi},\mu}$.}
    \label{fig:sqw_sp_even_longitudinal}
\end{figure}

Given the exchange symmetry of the two layers (up to a translation in the AFM phase), the symmetric (S) and antisymmetric (A) channels of the magnetic susceptibility, $\chi^{\mu\nu}_{S}(q)$ and $\chi^{\mu\nu}_{A}(q)$, are decoupled. In the main text, we focus on the antisymmetric channel, which exhibits triplon modes in the QPM phase, where their softening leads to the continuous phase transition to the N\'eel AFM phase, as well as magnon modes and amplitude fluctuation modes in the AFM phase. For completeness, in this Appendix, we present the results for the symmetric channel. Fig.~\ref{fig:sqw_sp_even_transverse} and Fig.~\ref{fig:sqw_sp_even_longitudinal} show the SP results. Notably, the counter diagram for the symmetric channel does not have singular contributions of ${\cal O}(N^0)$ and, thus, it is not necessary to correct the leading-order (i.e., the SP) approximation.

The excited states revealed by $\chi^{\mu\nu}_{S}(q)$ are generated by the symmetric spin operator,
\begin{equation}
    \hat{S}_{j S}^{\mu} = \hat{S}_{j +}^{\mu} + \hat{S}_{j -}^{\mu} = -i \sum_{\nu,\rho=1}^3 \epsilon^{\mu\nu\rho} \hat{b}_{j,\nu}^\dagger \hat{b}_{j,\rho}.
\end{equation}
This operator is bilinear in two triplet bosons, as it needs to be invariant under the exchange of layers (a bilinear form in singlet bosons is not possible because it would be a scalar under $\mathrm{SU}(2)$ spin rotation). This contrasts with the antisymmetric spin operator, which is bilinear in a singlet and a triplet boson and is therefore odd under the exchange of layers.

In the QPM phase, none of the triplet bosons condense. Thus, each of the two-triplon excitations generated by the symmetric spin operator $\hat{S}_{j S}^{\mu}$ carries spectral weight $\propto 1/{\cal N}_D$, forming a continuum of excitations in the DSSF without singular $\delta$ peaks. This results in the DSSF shown in Fig.~\ref{fig:sqw_sp_even_transverse}. In contrast, the antisymmetric DSSF shown in Fig.~\ref{fig:sqw_sp_odd} consists of a $\delta$ peak, arising from the creation of a singlet boson in the condensate and a gapped triplet boson, whose spectral weight is of ${\cal O}(1)$ due to the finite condensate fraction of the singlet bosons. It also exhibits a continuum formed by one singlet boson and one triplet boson, which differs from the continuum in the symmetric DSSF.

In the N\'eel AFM phase, the condensation of the $\mu=1$ triplet boson, $\langle \hat{b}_{j,1} \rangle = \sqrt{n_{c,{\bm \pi}}}$, gives rise to local magnetic moments along the $\hat{x}$ direction. For convenience, we denote the longitudinal Schwinger boson (LSB) by ${\hat b}_{j,1}$ and the transverse Schwinger bosons (TSBs) by ${\hat b}_{j,2/3}$ . The transverse symmetric spin operator, $\hat{S}_{j S}^{y/z}$, is bilinear in the LSB and TSB. Since the LSB has a finite condensate fraction, one can replace the LSB operator by $\langle \hat{b}_{j,1} \rangle = \sqrt{n_{c,{\bm \pi}}}$, leading to the approximation:
\begin{align}
    \hat{S}_{j S}^{y} &\simeq i \sqrt{n_{c,{\bm \pi}}} (\hat{b}_{j,3} - \hat{b}_{j,3}^\dagger),\\
    \hat{S}_{j S}^{z} &\simeq i \sqrt{n_{c,{\bm \pi}}} (\hat{b}_{j,2}^\dagger - \hat{b}_{j,2}).
\end{align}
These transverse symmetric spin operators generate quasiparticles (magnons) described by the TSBs, $\hat{b}_{j,2}$ and $\hat{b}_{j,3}$, forming $\delta$ peaks in the symmetric DSSF (see Fig.~\ref{fig:sqw_sp_even_transverse}). In the laboratory reference frame, the LSB condenses at the momentum ${\bm \pi}$. By momentum conservation, the dispersion of the $\delta$ peak is given by $\varepsilon_{{\bm \pi}+{\bm q},\mu}$, with $\mu=2, 3$,  for an external momentum ${\bm q}$ in the DSSF, which becomes gapless at the $\Gamma$ point. The spectral weight associated with these $\delta$ peaks is proportional to $n_{c,{\bm \pi}}$, which vanishes as the quantum critical point is approached. Fluctuations of the LSB condensate generate a continuum in the symmetric DSSF, arising from a two-particle excitation formed by LSB and TSB, differing again from the antisymmetric channel.

The longitudinal symmetric spin operator, however, is bilinear in two uncondensed TSBs:
\begin{equation}
    \hat{S}_{j S}^{x} = -i (\hat{b}_{j,2}^\dagger \hat{b}_{j,3} - \hat{b}_{j,3}^\dagger \hat{b}_{j,2}).
\end{equation}
It merely excites a two-particle continuum formed by two TSBs, corresponding to a two-magnon continuum since each TSB describes a magnon excitation. This results in the longitudinal DSSF in the symmetric channel shown in Fig.~\ref{fig:sqw_sp_even_longitudinal}. In contrast to the antisymmetric channel shown in Fig.~\ref{fig:sqw_fl}, the symmetric channel does not exhibit spurious modes. In other words, the SP description of the longitudinal DSSF in the symmetric channel is qualitatively correct.

As we already mentioned, the counter diagram shown in Fig.~\ref{fig:diagram}(b) vanishes for the symmetric channel. Specifically, the cross susceptibility between the external symmetric spin operators and the auxiliary field operators must vanish. For the QPM phase and the transverse spin components in the N\'eel AFM phase, the same symmetry argument provided in the main text for the antisymmetric channel still applies. For the longitudinal spin components in the N\'eel AFM phase, while both the longitudinal spin operator $\hat{S}_{j S}^{x}$ and the auxiliary field operators are invariant under the residual $\mathrm{U}(1)$ spin rotation, their cross susceptibility still vanishes because $\hat{S}_{j S}^{x}$ creates two uncondensed TSBs that preserve all symmetries of the model. Note that, for the cross susceptibility to be finite, the intermediate two-particle state must break certain symmetries of the system. As in the antisymmetric channel, the longitudinal spin operator $\hat{S}_{j A}^{x}$ creates a singlet boson and an LSB, both of which condense in a hybridized energy level that results in the formation of local magnetic moments.


\begin{thebibliography}{62}%
\makeatletter
\providecommand \@ifxundefined [1]{%
 \@ifx{#1\undefined}
}%
\providecommand \@ifnum [1]{%
 \ifnum #1\expandafter \@firstoftwo
 \else \expandafter \@secondoftwo
 \fi
}%
\providecommand \@ifx [1]{%
 \ifx #1\expandafter \@firstoftwo
 \else \expandafter \@secondoftwo
 \fi
}%
\providecommand \natexlab [1]{#1}%
\providecommand \enquote  [1]{``#1''}%
\providecommand \bibnamefont  [1]{#1}%
\providecommand \bibfnamefont [1]{#1}%
\providecommand \citenamefont [1]{#1}%
\providecommand \href@noop [0]{\@secondoftwo}%
\providecommand \href [0]{\begingroup \@sanitize@url \@href}%
\providecommand \@href[1]{\@@startlink{#1}\@@href}%
\providecommand \@@href[1]{\endgroup#1\@@endlink}%
\providecommand \@sanitize@url [0]{\catcode `\\12\catcode `\$12\catcode
  `\&12\catcode `\#12\catcode `\^12\catcode `\_12\catcode `\%12\relax}%
\providecommand \@@startlink[1]{}%
\providecommand \@@endlink[0]{}%
\providecommand \url  [0]{\begingroup\@sanitize@url \@url }%
\providecommand \@url [1]{\endgroup\@href {#1}{\urlprefix }}%
\providecommand \urlprefix  [0]{URL }%
\providecommand \Eprint [0]{\href }%
\providecommand \doibase [0]{https://doi.org/}%
\providecommand \selectlanguage [0]{\@gobble}%
\providecommand \bibinfo  [0]{\@secondoftwo}%
\providecommand \bibfield  [0]{\@secondoftwo}%
\providecommand \translation [1]{[#1]}%
\providecommand \BibitemOpen [0]{}%
\providecommand \bibitemStop [0]{}%
\providecommand \bibitemNoStop [0]{.\EOS\space}%
\providecommand \EOS [0]{\spacefactor3000\relax}%
\providecommand \BibitemShut  [1]{\csname bibitem#1\endcsname}%
\let\auto@bib@innerbib\@empty
\bibitem [{\citenamefont {Hester}\ \emph {et~al.}(2019)\citenamefont {Hester},
  \citenamefont {Nair}, \citenamefont {Reeder}, \citenamefont {Yahne},
  \citenamefont {DeLazzer}, \citenamefont {Berges}, \citenamefont {Ziat},
  \citenamefont {Neilson}, \citenamefont {Aczel}, \citenamefont {Sala},
  \citenamefont {Quilliam},\ and\ \citenamefont {Ross}}]{hester2019novel}%
  \BibitemOpen
  \bibfield  {author} {\bibinfo {author} {\bibfnamefont {G.}~\bibnamefont
  {Hester}}, \bibinfo {author} {\bibfnamefont {H.~S.}\ \bibnamefont {Nair}},
  \bibinfo {author} {\bibfnamefont {T.}~\bibnamefont {Reeder}}, \bibinfo
  {author} {\bibfnamefont {D.~R.}\ \bibnamefont {Yahne}}, \bibinfo {author}
  {\bibfnamefont {T.~N.}\ \bibnamefont {DeLazzer}}, \bibinfo {author}
  {\bibfnamefont {L.}~\bibnamefont {Berges}}, \bibinfo {author} {\bibfnamefont
  {D.}~\bibnamefont {Ziat}}, \bibinfo {author} {\bibfnamefont {J.~R.}\
  \bibnamefont {Neilson}}, \bibinfo {author} {\bibfnamefont {A.~A.}\
  \bibnamefont {Aczel}}, \bibinfo {author} {\bibfnamefont {G.}~\bibnamefont
  {Sala}}, \bibinfo {author} {\bibfnamefont {J.~A.}\ \bibnamefont {Quilliam}},\
  and\ \bibinfo {author} {\bibfnamefont {K.~A.}\ \bibnamefont {Ross}},\
  }\bibfield  {title} {\bibinfo {title} {Novel strongly spin-orbit coupled
  quantum dimer magnet: {Yb}$_{2}${S}i$_{2}${O}$_{7}$},\ }\href
  {https://doi.org/10.1103/PhysRevLett.123.027201} {\bibfield  {journal}
  {\bibinfo  {journal} {Phys. Rev. Lett.}\ }\textbf {\bibinfo {volume} {123}},\
  \bibinfo {pages} {027201} (\bibinfo {year} {2019})}\BibitemShut {NoStop}%
\bibitem [{\citenamefont {Jaime}\ \emph {et~al.}(2004)\citenamefont {Jaime},
  \citenamefont {Correa}, \citenamefont {Harrison}, \citenamefont {Batista},
  \citenamefont {Kawashima}, \citenamefont {Kazuma}, \citenamefont {Jorge},
  \citenamefont {Stern}, \citenamefont {Heinmaa}, \citenamefont {Zvyagin} \emph
  {et~al.}}]{jaime2004magnetic}%
  \BibitemOpen
  \bibfield  {author} {\bibinfo {author} {\bibfnamefont {M.}~\bibnamefont
  {Jaime}}, \bibinfo {author} {\bibfnamefont {V.}~\bibnamefont {Correa}},
  \bibinfo {author} {\bibfnamefont {N.}~\bibnamefont {Harrison}}, \bibinfo
  {author} {\bibfnamefont {C.}~\bibnamefont {Batista}}, \bibinfo {author}
  {\bibfnamefont {N.}~\bibnamefont {Kawashima}}, \bibinfo {author}
  {\bibfnamefont {Y.}~\bibnamefont {Kazuma}}, \bibinfo {author} {\bibfnamefont
  {G.}~\bibnamefont {Jorge}}, \bibinfo {author} {\bibfnamefont
  {R.}~\bibnamefont {Stern}}, \bibinfo {author} {\bibfnamefont
  {I.}~\bibnamefont {Heinmaa}}, \bibinfo {author} {\bibfnamefont
  {S.}~\bibnamefont {Zvyagin}}, \emph {et~al.},\ }\bibfield  {title} {\bibinfo
  {title} {Magnetic-field-induced condensation of triplons in han purple
  pigment {B}a{C}u{S}i$_2${O}$_6$},\ }\href@noop {} {\bibfield  {journal}
  {\bibinfo  {journal} {Physical Review Letters}\ }\textbf {\bibinfo {volume}
  {93}},\ \bibinfo {pages} {087203} (\bibinfo {year} {2004})}\BibitemShut
  {NoStop}%
\bibitem [{\citenamefont {R{\"u}egg}\ \emph {et~al.}(2007)\citenamefont
  {R{\"u}egg}, \citenamefont {McMorrow}, \citenamefont {Normand}, \citenamefont
  {R{\o}nnow}, \citenamefont {Sebastian}, \citenamefont {Fisher}, \citenamefont
  {Batista}, \citenamefont {Gvasaliya}, \citenamefont {Niedermayer},\ and\
  \citenamefont {Stahn}}]{ruegg2007multiple}%
  \BibitemOpen
  \bibfield  {author} {\bibinfo {author} {\bibfnamefont {C.}~\bibnamefont
  {R{\"u}egg}}, \bibinfo {author} {\bibfnamefont {D.}~\bibnamefont {McMorrow}},
  \bibinfo {author} {\bibfnamefont {B.}~\bibnamefont {Normand}}, \bibinfo
  {author} {\bibfnamefont {H.~M.}\ \bibnamefont {R{\o}nnow}}, \bibinfo {author}
  {\bibfnamefont {S.}~\bibnamefont {Sebastian}}, \bibinfo {author}
  {\bibfnamefont {I.}~\bibnamefont {Fisher}}, \bibinfo {author} {\bibfnamefont
  {C.}~\bibnamefont {Batista}}, \bibinfo {author} {\bibfnamefont
  {S.}~\bibnamefont {Gvasaliya}}, \bibinfo {author} {\bibfnamefont
  {C.}~\bibnamefont {Niedermayer}},\ and\ \bibinfo {author} {\bibfnamefont
  {J.}~\bibnamefont {Stahn}},\ }\bibfield  {title} {\bibinfo {title} {Multiple
  magnon modes and consequences for the bose-einstein condensed phase in
  {B}a{C}u{S}i$_2${O}$_6$},\ }\href@noop {} {\bibfield  {journal} {\bibinfo
  {journal} {Physical review letters}\ }\textbf {\bibinfo {volume} {98}},\
  \bibinfo {pages} {017202} (\bibinfo {year} {2007})}\BibitemShut {NoStop}%
\bibitem [{\citenamefont {R{\"u}egg}\ \emph {et~al.}(2003)\citenamefont
  {R{\"u}egg}, \citenamefont {Cavadini}, \citenamefont {Furrer}, \citenamefont
  {G{\"u}del}, \citenamefont {Kr{\"a}mer}, \citenamefont {Mutka}, \citenamefont
  {Wildes}, \citenamefont {Habicht},\ and\ \citenamefont
  {Vorderwisch}}]{ruegg2003bose}%
  \BibitemOpen
  \bibfield  {author} {\bibinfo {author} {\bibfnamefont {C.}~\bibnamefont
  {R{\"u}egg}}, \bibinfo {author} {\bibfnamefont {N.}~\bibnamefont {Cavadini}},
  \bibinfo {author} {\bibfnamefont {A.}~\bibnamefont {Furrer}}, \bibinfo
  {author} {\bibfnamefont {H.-U.}\ \bibnamefont {G{\"u}del}}, \bibinfo {author}
  {\bibfnamefont {K.}~\bibnamefont {Kr{\"a}mer}}, \bibinfo {author}
  {\bibfnamefont {H.}~\bibnamefont {Mutka}}, \bibinfo {author} {\bibfnamefont
  {A.}~\bibnamefont {Wildes}}, \bibinfo {author} {\bibfnamefont
  {K.}~\bibnamefont {Habicht}},\ and\ \bibinfo {author} {\bibfnamefont
  {P.}~\bibnamefont {Vorderwisch}},\ }\bibfield  {title} {\bibinfo {title}
  {Bose--einstein condensation of the triplet states in the magnetic insulator
  {T}l{C}u{C}l$_3$},\ }\href@noop {} {\bibfield  {journal} {\bibinfo  {journal}
  {Nature}\ }\textbf {\bibinfo {volume} {423}},\ \bibinfo {pages} {62}
  (\bibinfo {year} {2003})}\BibitemShut {NoStop}%
\bibitem [{\citenamefont {Yamada}\ \emph {et~al.}(2007)\citenamefont {Yamada},
  \citenamefont {Ono}, \citenamefont {Fujisawa}, \citenamefont {Tanaka},\ and\
  \citenamefont {Sakakibara}}]{yamada2007magnetic}%
  \BibitemOpen
  \bibfield  {author} {\bibinfo {author} {\bibfnamefont {F.}~\bibnamefont
  {Yamada}}, \bibinfo {author} {\bibfnamefont {T.}~\bibnamefont {Ono}},
  \bibinfo {author} {\bibfnamefont {M.}~\bibnamefont {Fujisawa}}, \bibinfo
  {author} {\bibfnamefont {H.}~\bibnamefont {Tanaka}},\ and\ \bibinfo {author}
  {\bibfnamefont {T.}~\bibnamefont {Sakakibara}},\ }\bibfield  {title}
  {\bibinfo {title} {Magnetic-field induced quantum phase transition and
  critical behavior in a gapped spin system {T}l{C}u{C}l$_3$},\ }\href@noop {}
  {\bibfield  {journal} {\bibinfo  {journal} {Journal of magnetism and magnetic
  materials}\ }\textbf {\bibinfo {volume} {310}},\ \bibinfo {pages} {1352}
  (\bibinfo {year} {2007})}\BibitemShut {NoStop}%
\bibitem [{\citenamefont {Aczel}\ \emph {et~al.}(2009)\citenamefont {Aczel},
  \citenamefont {Kohama}, \citenamefont {Jaime}, \citenamefont {Ninios},
  \citenamefont {Chan}, \citenamefont {Balicas}, \citenamefont {Dabkowska},\
  and\ \citenamefont {Luke}}]{aczel2009bose}%
  \BibitemOpen
  \bibfield  {author} {\bibinfo {author} {\bibfnamefont {A.}~\bibnamefont
  {Aczel}}, \bibinfo {author} {\bibfnamefont {Y.}~\bibnamefont {Kohama}},
  \bibinfo {author} {\bibfnamefont {M.}~\bibnamefont {Jaime}}, \bibinfo
  {author} {\bibfnamefont {K.}~\bibnamefont {Ninios}}, \bibinfo {author}
  {\bibfnamefont {H.~B.}\ \bibnamefont {Chan}}, \bibinfo {author}
  {\bibfnamefont {L.}~\bibnamefont {Balicas}}, \bibinfo {author} {\bibfnamefont
  {H.}~\bibnamefont {Dabkowska}},\ and\ \bibinfo {author} {\bibfnamefont
  {G.}~\bibnamefont {Luke}},\ }\bibfield  {title} {\bibinfo {title}
  {Bose-einstein condensation of triplons in {B}a$_3${C}r$_2${O}$_8$},\
  }\href@noop {} {\bibfield  {journal} {\bibinfo  {journal} {Physical Review
  B—Condensed Matter and Materials Physics}\ }\textbf {\bibinfo {volume}
  {79}},\ \bibinfo {pages} {100409} (\bibinfo {year} {2009})}\BibitemShut
  {NoStop}%
\bibitem [{\citenamefont {Kofu}\ \emph {et~al.}(2009)\citenamefont {Kofu},
  \citenamefont {Ueda}, \citenamefont {Nojiri}, \citenamefont {Oshima},
  \citenamefont {Zenmoto}, \citenamefont {Rule}, \citenamefont {Gerischer},
  \citenamefont {Lake}, \citenamefont {Batista}, \citenamefont {Ueda},\ and\
  \citenamefont {Lee}}]{kofu2009magnetic}%
  \BibitemOpen
  \bibfield  {author} {\bibinfo {author} {\bibfnamefont {M.}~\bibnamefont
  {Kofu}}, \bibinfo {author} {\bibfnamefont {H.}~\bibnamefont {Ueda}}, \bibinfo
  {author} {\bibfnamefont {H.}~\bibnamefont {Nojiri}}, \bibinfo {author}
  {\bibfnamefont {Y.}~\bibnamefont {Oshima}}, \bibinfo {author} {\bibfnamefont
  {T.}~\bibnamefont {Zenmoto}}, \bibinfo {author} {\bibfnamefont
  {K.}~\bibnamefont {Rule}}, \bibinfo {author} {\bibfnamefont {S.}~\bibnamefont
  {Gerischer}}, \bibinfo {author} {\bibfnamefont {B.}~\bibnamefont {Lake}},
  \bibinfo {author} {\bibfnamefont {C.}~\bibnamefont {Batista}}, \bibinfo
  {author} {\bibfnamefont {Y.}~\bibnamefont {Ueda}},\ and\ \bibinfo {author}
  {\bibfnamefont {S.-H.}\ \bibnamefont {Lee}},\ }\bibfield  {title} {\bibinfo
  {title} {Magnetic-field induced phase transitions in a weakly coupled $s=
  1/2$ quantum spin dimer system {B}a$_3${C}r$_2${O}$_8$},\ }\href@noop {}
  {\bibfield  {journal} {\bibinfo  {journal} {Physical review letters}\
  }\textbf {\bibinfo {volume} {102}},\ \bibinfo {pages} {177204} (\bibinfo
  {year} {2009})}\BibitemShut {NoStop}%
\bibitem [{\citenamefont {Tsirlin}\ and\ \citenamefont
  {Rosner}(2010)}]{tsirlin2010microscopic}%
  \BibitemOpen
  \bibfield  {author} {\bibinfo {author} {\bibfnamefont {A.~A.}\ \bibnamefont
  {Tsirlin}}\ and\ \bibinfo {author} {\bibfnamefont {H.}~\bibnamefont
  {Rosner}},\ }\bibfield  {title} {\bibinfo {title} {Microscopic model of
  ({C}u{C}l){L}a{N}b$_2${O}$_7$: Coupled spin dimers replace a frustrated
  square lattice},\ }\href@noop {} {\bibfield  {journal} {\bibinfo  {journal}
  {Physical Review B—Condensed Matter and Materials Physics}\ }\textbf
  {\bibinfo {volume} {82}},\ \bibinfo {pages} {060409} (\bibinfo {year}
  {2010})}\BibitemShut {NoStop}%
\bibitem [{\citenamefont {Coldea}\ \emph {et~al.}(2002)\citenamefont {Coldea},
  \citenamefont {Tennant}, \citenamefont {Habicht}, \citenamefont {Smeibidl},
  \citenamefont {Wolters},\ and\ \citenamefont
  {Tylczynski}}]{coldea2002direct}%
  \BibitemOpen
  \bibfield  {author} {\bibinfo {author} {\bibfnamefont {R.}~\bibnamefont
  {Coldea}}, \bibinfo {author} {\bibfnamefont {D.}~\bibnamefont {Tennant}},
  \bibinfo {author} {\bibfnamefont {K.}~\bibnamefont {Habicht}}, \bibinfo
  {author} {\bibfnamefont {P.}~\bibnamefont {Smeibidl}}, \bibinfo {author}
  {\bibfnamefont {C.}~\bibnamefont {Wolters}},\ and\ \bibinfo {author}
  {\bibfnamefont {Z.}~\bibnamefont {Tylczynski}},\ }\bibfield  {title}
  {\bibinfo {title} {Direct measurement of the spin hamiltonian and observation
  of condensation of magnons in the 2d frustrated quantum magnet
  {C}s$_2${C}u{C}l$_4$},\ }\href@noop {} {\bibfield  {journal} {\bibinfo
  {journal} {Physical review letters}\ }\textbf {\bibinfo {volume} {88}},\
  \bibinfo {pages} {137203} (\bibinfo {year} {2002})}\BibitemShut {NoStop}%
\bibitem [{\citenamefont {Zheludev}\ \emph {et~al.}(2007)\citenamefont
  {Zheludev}, \citenamefont {Garlea}, \citenamefont {Masuda}, \citenamefont
  {Manaka}, \citenamefont {Regnault}, \citenamefont {Ressouche}, \citenamefont
  {Grenier}, \citenamefont {Chung}, \citenamefont {Qiu}, \citenamefont
  {Habicht} \emph {et~al.}}]{zheludev2007dynamics}%
  \BibitemOpen
  \bibfield  {author} {\bibinfo {author} {\bibfnamefont {A.}~\bibnamefont
  {Zheludev}}, \bibinfo {author} {\bibfnamefont {V.~O.}\ \bibnamefont
  {Garlea}}, \bibinfo {author} {\bibfnamefont {T.}~\bibnamefont {Masuda}},
  \bibinfo {author} {\bibfnamefont {H.}~\bibnamefont {Manaka}}, \bibinfo
  {author} {\bibfnamefont {L.-P.}\ \bibnamefont {Regnault}}, \bibinfo {author}
  {\bibfnamefont {E.}~\bibnamefont {Ressouche}}, \bibinfo {author}
  {\bibfnamefont {B.}~\bibnamefont {Grenier}}, \bibinfo {author} {\bibfnamefont
  {J.-H.}\ \bibnamefont {Chung}}, \bibinfo {author} {\bibfnamefont
  {Y.}~\bibnamefont {Qiu}}, \bibinfo {author} {\bibfnamefont {K.}~\bibnamefont
  {Habicht}}, \emph {et~al.},\ }\bibfield  {title} {\bibinfo {title} {Dynamics
  of quantum spin liquid and spin solid phases in {IPA}-{C}u{C}l$_3$ under an
  applied magnetic field studied with neutron scattering},\ }\href@noop {}
  {\bibfield  {journal} {\bibinfo  {journal} {Physical Review B—Condensed
  Matter and Materials Physics}\ }\textbf {\bibinfo {volume} {76}},\ \bibinfo
  {pages} {054450} (\bibinfo {year} {2007})}\BibitemShut {NoStop}%
\bibitem [{\citenamefont {Zapf}\ \emph {et~al.}(2014)\citenamefont {Zapf},
  \citenamefont {Jaime},\ and\ \citenamefont {Batista}}]{zapf2014bose}%
  \BibitemOpen
  \bibfield  {author} {\bibinfo {author} {\bibfnamefont {V.}~\bibnamefont
  {Zapf}}, \bibinfo {author} {\bibfnamefont {M.}~\bibnamefont {Jaime}},\ and\
  \bibinfo {author} {\bibfnamefont {C.}~\bibnamefont {Batista}},\ }\bibfield
  {title} {\bibinfo {title} {Bose-einstein condensation in quantum magnets},\
  }\href@noop {} {\bibfield  {journal} {\bibinfo  {journal} {Reviews of Modern
  Physics}\ }\textbf {\bibinfo {volume} {86}},\ \bibinfo {pages} {563}
  (\bibinfo {year} {2014})}\BibitemShut {NoStop}%
\bibitem [{\citenamefont {Giamarchi}\ \emph {et~al.}(2008)\citenamefont
  {Giamarchi}, \citenamefont {R{\"u}egg},\ and\ \citenamefont
  {Tchernyshyov}}]{giamarchi2008bose}%
  \BibitemOpen
  \bibfield  {author} {\bibinfo {author} {\bibfnamefont {T.}~\bibnamefont
  {Giamarchi}}, \bibinfo {author} {\bibfnamefont {C.}~\bibnamefont
  {R{\"u}egg}},\ and\ \bibinfo {author} {\bibfnamefont {O.}~\bibnamefont
  {Tchernyshyov}},\ }\bibfield  {title} {\bibinfo {title} {Bose--einstein
  condensation in magnetic insulators},\ }\href@noop {} {\bibfield  {journal}
  {\bibinfo  {journal} {Nature Physics}\ }\textbf {\bibinfo {volume} {4}},\
  \bibinfo {pages} {198} (\bibinfo {year} {2008})}\BibitemShut {NoStop}%
\bibitem [{\citenamefont {Sachdev}(2011)}]{sachdev11}%
  \BibitemOpen
  \bibfield  {author} {\bibinfo {author} {\bibfnamefont {S.}~\bibnamefont
  {Sachdev}},\ }\href@noop {} {\emph {\bibinfo {title} {Quantum Phase
  Transitions}}}\ (\bibinfo  {publisher} {Cambridge University Press},\
  \bibinfo {year} {2011})\BibitemShut {NoStop}%
\bibitem [{\citenamefont {Sandvik}\ and\ \citenamefont
  {Scalapino}(1994)}]{Sandvik94}%
  \BibitemOpen
  \bibfield  {author} {\bibinfo {author} {\bibfnamefont {A.~W.}\ \bibnamefont
  {Sandvik}}\ and\ \bibinfo {author} {\bibfnamefont {D.~J.}\ \bibnamefont
  {Scalapino}},\ }\bibfield  {title} {\bibinfo {title} {Order-disorder
  transition in a two-layer quantum antiferromagnet},\ }\href
  {https://doi.org/10.1103/PhysRevLett.72.2777} {\bibfield  {journal} {\bibinfo
   {journal} {Phys. Rev. Lett.}\ }\textbf {\bibinfo {volume} {72}},\ \bibinfo
  {pages} {2777} (\bibinfo {year} {1994})}\BibitemShut {NoStop}%
\bibitem [{\citenamefont {Wang}\ \emph {et~al.}(2006)\citenamefont {Wang},
  \citenamefont {Beach},\ and\ \citenamefont {Sandvik}}]{wang06}%
  \BibitemOpen
  \bibfield  {author} {\bibinfo {author} {\bibfnamefont {L.}~\bibnamefont
  {Wang}}, \bibinfo {author} {\bibfnamefont {K.~S.~D.}\ \bibnamefont {Beach}},\
  and\ \bibinfo {author} {\bibfnamefont {A.~W.}\ \bibnamefont {Sandvik}},\
  }\bibfield  {title} {\bibinfo {title} {High-precision finite-size scaling
  analysis of the quantum-critical point of s=1/2 heisenberg antiferromagnetic
  bilayers},\ }\href {https://doi.org/10.1103/PhysRevB.73.014431} {\bibfield
  {journal} {\bibinfo  {journal} {Phys. Rev. B}\ }\textbf {\bibinfo {volume}
  {73}},\ \bibinfo {pages} {014431} (\bibinfo {year} {2006})}\BibitemShut
  {NoStop}%
\bibitem [{\citenamefont {Loh{\"o}fer}\ \emph {et~al.}(2015)\citenamefont
  {Loh{\"o}fer}, \citenamefont {Coletta}, \citenamefont {Joshi}, \citenamefont
  {Assaad}, \citenamefont {Vojta}, \citenamefont {Wessel},\ and\ \citenamefont
  {Mila}}]{lohofer2015dynamical}%
  \BibitemOpen
  \bibfield  {author} {\bibinfo {author} {\bibfnamefont {M.}~\bibnamefont
  {Loh{\"o}fer}}, \bibinfo {author} {\bibfnamefont {T.}~\bibnamefont
  {Coletta}}, \bibinfo {author} {\bibfnamefont {D.}~\bibnamefont {Joshi}},
  \bibinfo {author} {\bibfnamefont {F.}~\bibnamefont {Assaad}}, \bibinfo
  {author} {\bibfnamefont {M.}~\bibnamefont {Vojta}}, \bibinfo {author}
  {\bibfnamefont {S.}~\bibnamefont {Wessel}},\ and\ \bibinfo {author}
  {\bibfnamefont {F.}~\bibnamefont {Mila}},\ }\bibfield  {title} {\bibinfo
  {title} {Dynamical structure factors and excitation modes of the bilayer
  heisenberg model},\ }\href@noop {} {\bibfield  {journal} {\bibinfo  {journal}
  {Physical Review B}\ }\textbf {\bibinfo {volume} {92}},\ \bibinfo {pages}
  {245137} (\bibinfo {year} {2015})}\BibitemShut {NoStop}%
\bibitem [{\citenamefont {Chubukov}\ and\ \citenamefont
  {Morr}(1995)}]{chubukov95}%
  \BibitemOpen
  \bibfield  {author} {\bibinfo {author} {\bibfnamefont {A.~V.}\ \bibnamefont
  {Chubukov}}\ and\ \bibinfo {author} {\bibfnamefont {D.~K.}\ \bibnamefont
  {Morr}},\ }\bibfield  {title} {\bibinfo {title} {Phase transition,
  longitudinal spin fluctuations, and scaling in a two-layer antiferromagnet},\
  }\href {https://doi.org/10.1103/PhysRevB.52.3521} {\bibfield  {journal}
  {\bibinfo  {journal} {Phys. Rev. B}\ }\textbf {\bibinfo {volume} {52}},\
  \bibinfo {pages} {3521} (\bibinfo {year} {1995})}\BibitemShut {NoStop}%
\bibitem [{\citenamefont {Weihong}(1997)}]{weihong97}%
  \BibitemOpen
  \bibfield  {author} {\bibinfo {author} {\bibfnamefont {Z.}~\bibnamefont
  {Weihong}},\ }\bibfield  {title} {\bibinfo {title} {Various series expansions
  for the bilayer s= heisenberg antiferromagnet},\ }\href
  {https://doi.org/10.1103/PhysRevB.55.12267} {\bibfield  {journal} {\bibinfo
  {journal} {Phys. Rev. B}\ }\textbf {\bibinfo {volume} {55}},\ \bibinfo
  {pages} {12267} (\bibinfo {year} {1997})}\BibitemShut {NoStop}%
\bibitem [{\citenamefont {Kotov}\ \emph {et~al.}(1998)\citenamefont {Kotov},
  \citenamefont {Sushkov}, \citenamefont {Weihong},\ and\ \citenamefont
  {Oitmaa}}]{kotov98}%
  \BibitemOpen
  \bibfield  {author} {\bibinfo {author} {\bibfnamefont {V.~N.}\ \bibnamefont
  {Kotov}}, \bibinfo {author} {\bibfnamefont {O.}~\bibnamefont {Sushkov}},
  \bibinfo {author} {\bibfnamefont {Z.}~\bibnamefont {Weihong}},\ and\ \bibinfo
  {author} {\bibfnamefont {J.}~\bibnamefont {Oitmaa}},\ }\bibfield  {title}
  {\bibinfo {title} {Novel approach to description of spin-liquid phases in
  low-dimensional quantum antiferromagnets},\ }\href
  {https://doi.org/10.1103/PhysRevLett.80.5790} {\bibfield  {journal} {\bibinfo
   {journal} {Phys. Rev. Lett.}\ }\textbf {\bibinfo {volume} {80}},\ \bibinfo
  {pages} {5790} (\bibinfo {year} {1998})}\BibitemShut {NoStop}%
\bibitem [{\citenamefont {Yu}\ \emph {et~al.}(1999)\citenamefont {Yu},
  \citenamefont {Gu}, \citenamefont {Wang},\ and\ \citenamefont
  {Shen}}]{yu1999bond}%
  \BibitemOpen
  \bibfield  {author} {\bibinfo {author} {\bibfnamefont {D.-K.}\ \bibnamefont
  {Yu}}, \bibinfo {author} {\bibfnamefont {Q.}~\bibnamefont {Gu}}, \bibinfo
  {author} {\bibfnamefont {H.-T.}\ \bibnamefont {Wang}},\ and\ \bibinfo
  {author} {\bibfnamefont {J.-L.}\ \bibnamefont {Shen}},\ }\bibfield  {title}
  {\bibinfo {title} {Bond-operator approach to the bilayer heisenberg
  antiferromagnet},\ }\href@noop {} {\bibfield  {journal} {\bibinfo  {journal}
  {Physical Review B}\ }\textbf {\bibinfo {volume} {59}},\ \bibinfo {pages}
  {111} (\bibinfo {year} {1999})}\BibitemShut {NoStop}%
\bibitem [{\citenamefont {Rakhimov}\ \emph {et~al.}(2011)\citenamefont
  {Rakhimov}, \citenamefont {Mardonov},\ and\ \citenamefont
  {Sherman}}]{rakhimov2011macroscopic}%
  \BibitemOpen
  \bibfield  {author} {\bibinfo {author} {\bibfnamefont {A.}~\bibnamefont
  {Rakhimov}}, \bibinfo {author} {\bibfnamefont {S.}~\bibnamefont {Mardonov}},\
  and\ \bibinfo {author} {\bibfnamefont {E.~Y.}\ \bibnamefont {Sherman}},\
  }\bibfield  {title} {\bibinfo {title} {Macroscopic properties of triplon
  bose--einstein condensates},\ }\href@noop {} {\bibfield  {journal} {\bibinfo
  {journal} {Annals of Physics}\ }\textbf {\bibinfo {volume} {326}},\ \bibinfo
  {pages} {2499} (\bibinfo {year} {2011})}\BibitemShut {NoStop}%
\bibitem [{\citenamefont {Sandvik}\ \emph {et~al.}(1995)\citenamefont
  {Sandvik}, \citenamefont {Chubukov},\ and\ \citenamefont
  {Sachdev}}]{sandvik1995quantum}%
  \BibitemOpen
  \bibfield  {author} {\bibinfo {author} {\bibfnamefont {A.~W.}\ \bibnamefont
  {Sandvik}}, \bibinfo {author} {\bibfnamefont {A.~V.}\ \bibnamefont
  {Chubukov}},\ and\ \bibinfo {author} {\bibfnamefont {S.}~\bibnamefont
  {Sachdev}},\ }\bibfield  {title} {\bibinfo {title} {Quantum critical behavior
  in a two-layer antiferromagnet},\ }\href@noop {} {\bibfield  {journal}
  {\bibinfo  {journal} {Physical Review B}\ }\textbf {\bibinfo {volume} {51}},\
  \bibinfo {pages} {16483} (\bibinfo {year} {1995})}\BibitemShut {NoStop}%
\bibitem [{\citenamefont {Dodds}\ \emph {et~al.}(2010)\citenamefont {Dodds},
  \citenamefont {Yang},\ and\ \citenamefont {Kim}}]{dodds2010theory}%
  \BibitemOpen
  \bibfield  {author} {\bibinfo {author} {\bibfnamefont {T.}~\bibnamefont
  {Dodds}}, \bibinfo {author} {\bibfnamefont {B.-J.}\ \bibnamefont {Yang}},\
  and\ \bibinfo {author} {\bibfnamefont {Y.~B.}\ \bibnamefont {Kim}},\
  }\bibfield  {title} {\bibinfo {title} {Theory of magnetic-field-induced
  bose-einstein condensation of triplons in {B}a$_3${C}r$_2${O}$_8$},\
  }\href@noop {} {\bibfield  {journal} {\bibinfo  {journal} {Physical Review
  B—Condensed Matter and Materials Physics}\ }\textbf {\bibinfo {volume}
  {81}},\ \bibinfo {pages} {054412} (\bibinfo {year} {2010})}\BibitemShut
  {NoStop}%
\bibitem [{\citenamefont {Hida}(1992)}]{hida1992quantum}%
  \BibitemOpen
  \bibfield  {author} {\bibinfo {author} {\bibfnamefont {K.}~\bibnamefont
  {Hida}},\ }\bibfield  {title} {\bibinfo {title} {Quantum disordered state
  without frustration in the double layer heisenberg antiferromagnet—dimer
  expansion and projector monte carlo study—},\ }\href@noop {} {\bibfield
  {journal} {\bibinfo  {journal} {Journal of the Physical Society of Japan}\
  }\textbf {\bibinfo {volume} {61}},\ \bibinfo {pages} {1013} (\bibinfo {year}
  {1992})}\BibitemShut {NoStop}%
\bibitem [{\citenamefont {Ganesh}\ \emph {et~al.}(2011)\citenamefont {Ganesh},
  \citenamefont {Isakov},\ and\ \citenamefont {Paramekanti}}]{ganesh2011neel}%
  \BibitemOpen
  \bibfield  {author} {\bibinfo {author} {\bibfnamefont {R.}~\bibnamefont
  {Ganesh}}, \bibinfo {author} {\bibfnamefont {S.~V.}\ \bibnamefont {Isakov}},\
  and\ \bibinfo {author} {\bibfnamefont {A.}~\bibnamefont {Paramekanti}},\
  }\bibfield  {title} {\bibinfo {title} {Neel to dimer transition in spin-s
  antiferromagnets: Comparing bond operator theory with quantum monte carlo
  simulations for bilayer heisenberg models},\ }\href@noop {} {\bibfield
  {journal} {\bibinfo  {journal} {Physical Review B}\ }\textbf {\bibinfo
  {volume} {84}},\ \bibinfo {pages} {214412} (\bibinfo {year}
  {2011})}\BibitemShut {NoStop}%
\bibitem [{\citenamefont {Ng}\ and\ \citenamefont {Yang}(2017)}]{ng2017field}%
  \BibitemOpen
  \bibfield  {author} {\bibinfo {author} {\bibfnamefont {K.-K.}\ \bibnamefont
  {Ng}}\ and\ \bibinfo {author} {\bibfnamefont {M.-F.}\ \bibnamefont {Yang}},\
  }\bibfield  {title} {\bibinfo {title} {Field-induced quantum phases in a
  frustrated spin-dimer model: A sign-problem-free quantum monte carlo study},\
  }\href@noop {} {\bibfield  {journal} {\bibinfo  {journal} {Physical Review
  B}\ }\textbf {\bibinfo {volume} {95}},\ \bibinfo {pages} {064431} (\bibinfo
  {year} {2017})}\BibitemShut {NoStop}%
\bibitem [{\citenamefont {Weber}\ \emph {et~al.}(2022)\citenamefont {Weber},
  \citenamefont {Honecker}, \citenamefont {Normand}, \citenamefont {Corboz},
  \citenamefont {Mila},\ and\ \citenamefont {Wessel}}]{weber2022quantum}%
  \BibitemOpen
  \bibfield  {author} {\bibinfo {author} {\bibfnamefont {L.}~\bibnamefont
  {Weber}}, \bibinfo {author} {\bibfnamefont {A.}~\bibnamefont {Honecker}},
  \bibinfo {author} {\bibfnamefont {B.}~\bibnamefont {Normand}}, \bibinfo
  {author} {\bibfnamefont {P.}~\bibnamefont {Corboz}}, \bibinfo {author}
  {\bibfnamefont {F.}~\bibnamefont {Mila}},\ and\ \bibinfo {author}
  {\bibfnamefont {S.}~\bibnamefont {Wessel}},\ }\bibfield  {title} {\bibinfo
  {title} {Quantum monte carlo simulations in the trimer basis: first-order
  transitions and thermal critical points in frustrated trilayer magnets},\
  }\href@noop {} {\bibfield  {journal} {\bibinfo  {journal} {SciPost Physics}\
  }\textbf {\bibinfo {volume} {12}},\ \bibinfo {pages} {054} (\bibinfo {year}
  {2022})}\BibitemShut {NoStop}%
\bibitem [{\citenamefont {Hering}\ \emph {et~al.}(2024)\citenamefont {Hering},
  \citenamefont {Walther}, \citenamefont {Schmidt},\ and\ \citenamefont
  {Uhrig}}]{hering24}%
  \BibitemOpen
  \bibfield  {author} {\bibinfo {author} {\bibfnamefont {D.-B.}\ \bibnamefont
  {Hering}}, \bibinfo {author} {\bibfnamefont {M.~R.}\ \bibnamefont {Walther}},
  \bibinfo {author} {\bibfnamefont {K.~P.}\ \bibnamefont {Schmidt}},\ and\
  \bibinfo {author} {\bibfnamefont {G.~S.}\ \bibnamefont {Uhrig}},\ }\bibfield
  {title} {\bibinfo {title} {Quantum melting of long-range ordered quantum
  antiferromagnets investigated by momentum-space continuous similarity
  transformations},\ }\href {https://doi.org/10.1103/PhysRevB.110.085115}
  {\bibfield  {journal} {\bibinfo  {journal} {Phys. Rev. B}\ }\textbf {\bibinfo
  {volume} {110}},\ \bibinfo {pages} {085115} (\bibinfo {year}
  {2024})}\BibitemShut {NoStop}%
\bibitem [{\citenamefont {Chubukov}(1989)}]{chubukov89}%
  \BibitemOpen
  \bibfield  {author} {\bibinfo {author} {\bibfnamefont {A.~V.}\ \bibnamefont
  {Chubukov}},\ }\bibfield  {title} {\bibinfo {title} {A difference in the
  properties of one-dimensional antiferromagnets with integer and half-integer
  spins},\ }\href {http://jetpletters.ru/ps/0/article_16848.shtml} {\bibfield
  {journal} {\bibinfo  {journal} {Pis'ma Zh. Eksp. Teor. Fiz.}\ }\textbf
  {\bibinfo {volume} {49}},\ \bibinfo {pages} {108} (\bibinfo {year}
  {1989})}\BibitemShut {NoStop}%
\bibitem [{\citenamefont {Sachdev}\ and\ \citenamefont
  {Bhatt}(1990)}]{sachdev1990bond}%
  \BibitemOpen
  \bibfield  {author} {\bibinfo {author} {\bibfnamefont {S.}~\bibnamefont
  {Sachdev}}\ and\ \bibinfo {author} {\bibfnamefont {R.}~\bibnamefont
  {Bhatt}},\ }\bibfield  {title} {\bibinfo {title} {Bond-operator
  representation of quantum spins: Mean-field theory of frustrated quantum
  heisenberg antiferromagnets},\ }\href@noop {} {\bibfield  {journal} {\bibinfo
   {journal} {Physical Review B}\ }\textbf {\bibinfo {volume} {41}},\ \bibinfo
  {pages} {9323} (\bibinfo {year} {1990})}\BibitemShut {NoStop}%
\bibitem [{\citenamefont {Joshi}\ \emph {et~al.}(2015)\citenamefont {Joshi},
  \citenamefont {Coester}, \citenamefont {Schmidt},\ and\ \citenamefont
  {Vojta}}]{joshi2015nonlinear}%
  \BibitemOpen
  \bibfield  {author} {\bibinfo {author} {\bibfnamefont {D.~G.}\ \bibnamefont
  {Joshi}}, \bibinfo {author} {\bibfnamefont {K.}~\bibnamefont {Coester}},
  \bibinfo {author} {\bibfnamefont {K.~P.}\ \bibnamefont {Schmidt}},\ and\
  \bibinfo {author} {\bibfnamefont {M.}~\bibnamefont {Vojta}},\ }\bibfield
  {title} {\bibinfo {title} {Nonlinear bond-operator theory and 1/d expansion
  for coupled-dimer magnets. i. paramagnetic phase},\ }\href@noop {} {\bibfield
   {journal} {\bibinfo  {journal} {Physical Review B}\ }\textbf {\bibinfo
  {volume} {91}},\ \bibinfo {pages} {094404} (\bibinfo {year}
  {2015})}\BibitemShut {NoStop}%
\bibitem [{\citenamefont {Joshi}\ and\ \citenamefont
  {Vojta}(2015)}]{joshi2015nonlinear2}%
  \BibitemOpen
  \bibfield  {author} {\bibinfo {author} {\bibfnamefont {D.~G.}\ \bibnamefont
  {Joshi}}\ and\ \bibinfo {author} {\bibfnamefont {M.}~\bibnamefont {Vojta}},\
  }\bibfield  {title} {\bibinfo {title} {Nonlinear bond-operator theory and 1/d
  expansion for coupled-dimer magnets. ii. antiferromagnetic phase and quantum
  phase transition},\ }\href@noop {} {\bibfield  {journal} {\bibinfo  {journal}
  {Physical Review B}\ }\textbf {\bibinfo {volume} {91}},\ \bibinfo {pages}
  {094405} (\bibinfo {year} {2015})}\BibitemShut {NoStop}%
\bibitem [{\citenamefont {Matsumoto}\ \emph {et~al.}(2004)\citenamefont
  {Matsumoto}, \citenamefont {Normand}, \citenamefont {Rice},\ and\
  \citenamefont {Sigrist}}]{matsumoto04}%
  \BibitemOpen
  \bibfield  {author} {\bibinfo {author} {\bibfnamefont {M.}~\bibnamefont
  {Matsumoto}}, \bibinfo {author} {\bibfnamefont {B.}~\bibnamefont {Normand}},
  \bibinfo {author} {\bibfnamefont {T.~M.}\ \bibnamefont {Rice}},\ and\
  \bibinfo {author} {\bibfnamefont {M.}~\bibnamefont {Sigrist}},\ }\bibfield
  {title} {\bibinfo {title} {Field- and pressure-induced magnetic quantum phase
  transitions in ${\mathrm{tlcucl}}_{3}$},\ }\href
  {https://doi.org/10.1103/PhysRevB.69.054423} {\bibfield  {journal} {\bibinfo
  {journal} {Phys. Rev. B}\ }\textbf {\bibinfo {volume} {69}},\ \bibinfo
  {pages} {054423} (\bibinfo {year} {2004})}\BibitemShut {NoStop}%
\bibitem [{\citenamefont {Zhang}\ \emph {et~al.}(2013)\citenamefont {Zhang},
  \citenamefont {Wierschem}, \citenamefont {Yap}, \citenamefont {Kato},
  \citenamefont {Batista},\ and\ \citenamefont {Sengupta}}]{zhang2013phase}%
  \BibitemOpen
  \bibfield  {author} {\bibinfo {author} {\bibfnamefont {Z.}~\bibnamefont
  {Zhang}}, \bibinfo {author} {\bibfnamefont {K.}~\bibnamefont {Wierschem}},
  \bibinfo {author} {\bibfnamefont {I.}~\bibnamefont {Yap}}, \bibinfo {author}
  {\bibfnamefont {Y.}~\bibnamefont {Kato}}, \bibinfo {author} {\bibfnamefont
  {C.~D.}\ \bibnamefont {Batista}},\ and\ \bibinfo {author} {\bibfnamefont
  {P.}~\bibnamefont {Sengupta}},\ }\bibfield  {title} {\bibinfo {title} {Phase
  diagram and magnetic excitations of anisotropic spin-one magnets},\
  }\href@noop {} {\bibfield  {journal} {\bibinfo  {journal} {Physical Review
  B}\ }\textbf {\bibinfo {volume} {87}},\ \bibinfo {pages} {174405} (\bibinfo
  {year} {2013})}\BibitemShut {NoStop}%
\bibitem [{\citenamefont {Lecheminant}\ and\ \citenamefont
  {Totsuka}(2005)}]{lecheminant2005phases}%
  \BibitemOpen
  \bibfield  {author} {\bibinfo {author} {\bibfnamefont {P.}~\bibnamefont
  {Lecheminant}}\ and\ \bibinfo {author} {\bibfnamefont {K.}~\bibnamefont
  {Totsuka}},\ }\bibfield  {title} {\bibinfo {title} {Phases of the generalized
  two-leg spin ladder: A view from the su (4) symmetry},\ }\href@noop {}
  {\bibfield  {journal} {\bibinfo  {journal} {Physical Review B—Condensed
  Matter and Materials Physics}\ }\textbf {\bibinfo {volume} {71}},\ \bibinfo
  {pages} {020407} (\bibinfo {year} {2005})}\BibitemShut {NoStop}%
\bibitem [{\citenamefont {Lecheminant}\ and\ \citenamefont
  {Totsuka}(2006)}]{lecheminant2006competing}%
  \BibitemOpen
  \bibfield  {author} {\bibinfo {author} {\bibfnamefont {P.}~\bibnamefont
  {Lecheminant}}\ and\ \bibinfo {author} {\bibfnamefont {K.}~\bibnamefont
  {Totsuka}},\ }\bibfield  {title} {\bibinfo {title} {Competing orders and
  hidden duality symmetries in two-leg spin ladder systems},\ }\href@noop {}
  {\bibfield  {journal} {\bibinfo  {journal} {Physical Review B—Condensed
  Matter and Materials Physics}\ }\textbf {\bibinfo {volume} {74}},\ \bibinfo
  {pages} {224426} (\bibinfo {year} {2006})}\BibitemShut {NoStop}%
\bibitem [{\citenamefont {Dahlbom}\ \emph {et~al.}(2024)\citenamefont
  {Dahlbom}, \citenamefont {Thomas}, \citenamefont {Johnston}, \citenamefont
  {Barros},\ and\ \citenamefont {Batista}}]{dahlbom2024classical}%
  \BibitemOpen
  \bibfield  {author} {\bibinfo {author} {\bibfnamefont {D.~A.}\ \bibnamefont
  {Dahlbom}}, \bibinfo {author} {\bibfnamefont {J.}~\bibnamefont {Thomas}},
  \bibinfo {author} {\bibfnamefont {S.}~\bibnamefont {Johnston}}, \bibinfo
  {author} {\bibfnamefont {K.}~\bibnamefont {Barros}},\ and\ \bibinfo {author}
  {\bibfnamefont {C.~D.}\ \bibnamefont {Batista}},\ }\bibfield  {title}
  {\bibinfo {title} {Classical dynamics of the antiferromagnetic heisenberg $s=
  1/2$ spin ladder},\ }\href@noop {} {\bibfield  {journal} {\bibinfo  {journal}
  {arXiv preprint arXiv:2405.16315}\ } (\bibinfo {year} {2024})}\BibitemShut
  {NoStop}%
\bibitem [{\citenamefont {Zhang}\ and\ \citenamefont
  {Batista}(2021)}]{Zhang21}%
  \BibitemOpen
  \bibfield  {author} {\bibinfo {author} {\bibfnamefont {H.}~\bibnamefont
  {Zhang}}\ and\ \bibinfo {author} {\bibfnamefont {C.~D.}\ \bibnamefont
  {Batista}},\ }\bibfield  {title} {\bibinfo {title} {{Classical spin dynamics
  based on {\normalfont SU$({N})$} coherent states}},\ }\href
  {https://doi.org/10.1103/PhysRevB.104.104409} {\bibfield  {journal} {\bibinfo
   {journal} {Phys. Rev. B}\ }\textbf {\bibinfo {volume} {104}},\ \bibinfo
  {pages} {104409} (\bibinfo {year} {2021})}\BibitemShut {NoStop}%
\bibitem [{\citenamefont {Muniz}\ \emph {et~al.}(2014)\citenamefont {Muniz},
  \citenamefont {Kato},\ and\ \citenamefont {Batista}}]{muniz14}%
  \BibitemOpen
  \bibfield  {author} {\bibinfo {author} {\bibfnamefont {R.~A.}\ \bibnamefont
  {Muniz}}, \bibinfo {author} {\bibfnamefont {Y.}~\bibnamefont {Kato}},\ and\
  \bibinfo {author} {\bibfnamefont {C.~D.}\ \bibnamefont {Batista}},\
  }\bibfield  {title} {\bibinfo {title} {{Generalized spin-wave theory:
  Application to the bilinear–biquadratic model}},\ }\href
  {https://doi.org/10.1093/ptep/ptu109} {\bibfield  {journal} {\bibinfo
  {journal} {Progress of Theoretical and Experimental Physics}\ }\textbf
  {\bibinfo {volume} {2014}},\ \bibinfo {pages} {083I01} (\bibinfo {year}
  {2014})},\ \Eprint
  {https://arxiv.org/abs/https://academic.oup.com/ptep/article-pdf/2014/8/083I01/4321660/ptu109.pdf}
  {https://academic.oup.com/ptep/article-pdf/2014/8/083I01/4321660/ptu109.pdf}
  \BibitemShut {NoStop}%
\bibitem [{\citenamefont {Miyazaki}\ \emph {et~al.}(1996)\citenamefont
  {Miyazaki}, \citenamefont {Nakamura},\ and\ \citenamefont
  {Yoshioka}}]{miyazaki1996bilayer}%
  \BibitemOpen
  \bibfield  {author} {\bibinfo {author} {\bibfnamefont {T.}~\bibnamefont
  {Miyazaki}}, \bibinfo {author} {\bibfnamefont {I.}~\bibnamefont {Nakamura}},\
  and\ \bibinfo {author} {\bibfnamefont {D.}~\bibnamefont {Yoshioka}},\
  }\bibfield  {title} {\bibinfo {title} {Bilayer heisenberg model studied by
  the schwinger-boson gutzwiller-projection method},\ }\href@noop {} {\bibfield
   {journal} {\bibinfo  {journal} {Physical Review B}\ }\textbf {\bibinfo
  {volume} {53}},\ \bibinfo {pages} {12206} (\bibinfo {year}
  {1996})}\BibitemShut {NoStop}%
\bibitem [{\citenamefont {Liao}\ and\ \citenamefont
  {Li}(2011)}]{liao2011variational}%
  \BibitemOpen
  \bibfield  {author} {\bibinfo {author} {\bibfnamefont {H.}~\bibnamefont
  {Liao}}\ and\ \bibinfo {author} {\bibfnamefont {T.}~\bibnamefont {Li}},\
  }\bibfield  {title} {\bibinfo {title} {Variational study of the quantum phase
  transition in the bilayer heisenberg model with bosonic rvb wavefunction},\
  }\href@noop {} {\bibfield  {journal} {\bibinfo  {journal} {Journal of
  Physics: Condensed Matter}\ }\textbf {\bibinfo {volume} {23}},\ \bibinfo
  {pages} {475602} (\bibinfo {year} {2011})}\BibitemShut {NoStop}%
\bibitem [{\citenamefont {Hida}(1990)}]{hida1990low}%
  \BibitemOpen
  \bibfield  {author} {\bibinfo {author} {\bibfnamefont {K.}~\bibnamefont
  {Hida}},\ }\bibfield  {title} {\bibinfo {title} {Low temperature properties
  of the double layer quantum heisenberg antiferromagnet-modified spin wave
  method},\ }\href@noop {} {\bibfield  {journal} {\bibinfo  {journal} {Journal
  of the Physical Society of Japan}\ }\textbf {\bibinfo {volume} {59}},\
  \bibinfo {pages} {2230} (\bibinfo {year} {1990})}\BibitemShut {NoStop}%
\bibitem [{\citenamefont {Millis}\ and\ \citenamefont
  {Monien}(1993)}]{Millis93}%
  \BibitemOpen
  \bibfield  {author} {\bibinfo {author} {\bibfnamefont {A.~J.}\ \bibnamefont
  {Millis}}\ and\ \bibinfo {author} {\bibfnamefont {H.}~\bibnamefont
  {Monien}},\ }\bibfield  {title} {\bibinfo {title} {Spin gaps and spin
  dynamics in
  ${\mathrm{la}}_{2\ensuremath{-}x}{\mathrm{sr}}_{x}\mathrm{Cu}{\mathrm{o}}_{4}$
  and
  $\mathrm{Y}{\mathrm{ba}}_{2}{\mathrm{cu}}_{3}{\mathrm{o}}_{7\ensuremath{-}\ensuremath{\delta}}$},\
  }\href {https://doi.org/10.1103/PhysRevLett.70.2810} {\bibfield  {journal}
  {\bibinfo  {journal} {Phys. Rev. Lett.}\ }\textbf {\bibinfo {volume} {70}},\
  \bibinfo {pages} {2810} (\bibinfo {year} {1993})}\BibitemShut {NoStop}%
\bibitem [{\citenamefont {Arovas}\ and\ \citenamefont
  {Auerbach}(1988)}]{arovas88}%
  \BibitemOpen
  \bibfield  {author} {\bibinfo {author} {\bibfnamefont {D.~P.}\ \bibnamefont
  {Arovas}}\ and\ \bibinfo {author} {\bibfnamefont {A.}~\bibnamefont
  {Auerbach}},\ }\bibfield  {title} {\bibinfo {title} {Functional integral
  theories of low-dimensional quantum heisenberg models},\ }\href
  {https://doi.org/10.1103/PhysRevB.38.316} {\bibfield  {journal} {\bibinfo
  {journal} {Phys. Rev. B}\ }\textbf {\bibinfo {volume} {38}},\ \bibinfo
  {pages} {316} (\bibinfo {year} {1988})}\BibitemShut {NoStop}%
\bibitem [{\citenamefont {Auerbach}(2012)}]{auerbach2012interacting}%
  \BibitemOpen
  \bibfield  {author} {\bibinfo {author} {\bibfnamefont {A.}~\bibnamefont
  {Auerbach}},\ }\href@noop {} {\emph {\bibinfo {title} {Interacting electrons
  and quantum magnetism}}}\ (\bibinfo  {publisher} {Springer Science \&
  Business Media},\ \bibinfo {year} {2012})\BibitemShut {NoStop}%
\bibitem [{Note1()}]{Note1}%
  \BibitemOpen
  \bibinfo {note} {In the literature, these are usually called ``bond
  operators''. In this work we will call them ``link operators'' to distinguish
  them from the ``bond operators'' acting on each dimer unit.}\BibitemShut
  {Stop}%
\bibitem [{\citenamefont {Dahlbom}\ \emph
  {et~al.}(2022{\natexlab{a}})\citenamefont {Dahlbom}, \citenamefont {Zhang},
  \citenamefont {Miles}, \citenamefont {Bai}, \citenamefont {Batista},\ and\
  \citenamefont {Barros}}]{Dahlbom22}%
  \BibitemOpen
  \bibfield  {author} {\bibinfo {author} {\bibfnamefont {D.}~\bibnamefont
  {Dahlbom}}, \bibinfo {author} {\bibfnamefont {H.}~\bibnamefont {Zhang}},
  \bibinfo {author} {\bibfnamefont {C.}~\bibnamefont {Miles}}, \bibinfo
  {author} {\bibfnamefont {X.}~\bibnamefont {Bai}}, \bibinfo {author}
  {\bibfnamefont {C.~D.}\ \bibnamefont {Batista}},\ and\ \bibinfo {author}
  {\bibfnamefont {K.}~\bibnamefont {Barros}},\ }\bibfield  {title} {\bibinfo
  {title} {{Geometric integration of classical spin dynamics via a mean-field
  {Schr\"odinger} equation}},\ }\href
  {https://doi.org/10.1103/PhysRevB.106.054423} {\bibfield  {journal} {\bibinfo
   {journal} {Phys. Rev. B}\ }\textbf {\bibinfo {volume} {106}},\ \bibinfo
  {pages} {054423} (\bibinfo {year} {2022}{\natexlab{a}})}\BibitemShut
  {NoStop}%
\bibitem [{\citenamefont {Dahlbom}\ \emph
  {et~al.}(2022{\natexlab{b}})\citenamefont {Dahlbom}, \citenamefont {Miles},
  \citenamefont {Zhang}, \citenamefont {Batista},\ and\ \citenamefont
  {Barros}}]{Dahlbom22b}%
  \BibitemOpen
  \bibfield  {author} {\bibinfo {author} {\bibfnamefont {D.}~\bibnamefont
  {Dahlbom}}, \bibinfo {author} {\bibfnamefont {C.}~\bibnamefont {Miles}},
  \bibinfo {author} {\bibfnamefont {H.}~\bibnamefont {Zhang}}, \bibinfo
  {author} {\bibfnamefont {C.~D.}\ \bibnamefont {Batista}},\ and\ \bibinfo
  {author} {\bibfnamefont {K.}~\bibnamefont {Barros}},\ }\bibfield  {title}
  {\bibinfo {title} {{Langevin dynamics of generalized spins as {\normalfont
  {SU}(${N}$)} coherent states}},\ }\href
  {https://doi.org/10.1103/PhysRevB.106.235154} {\bibfield  {journal} {\bibinfo
   {journal} {Phys. Rev. B}\ }\textbf {\bibinfo {volume} {106}},\ \bibinfo
  {pages} {235154} (\bibinfo {year} {2022}{\natexlab{b}})}\BibitemShut
  {NoStop}%
\bibitem [{Note2()}]{Note2}%
  \BibitemOpen
  \bibinfo {note} {These link-fields are usually called ``bond-fields'' in the
  literature. Here we use the name ``link-fields'' to distinguish them from the
  intra-dimer ``bond-fields''.}\BibitemShut {Stop}%
\bibitem [{\citenamefont {Ghioldi}\ \emph {et~al.}(2018)\citenamefont
  {Ghioldi}, \citenamefont {Gonzalez}, \citenamefont {Zhang}, \citenamefont
  {Kamiya}, \citenamefont {Manuel}, \citenamefont {Trumper},\ and\
  \citenamefont {Batista}}]{ghioldi2018dynamical}%
  \BibitemOpen
  \bibfield  {author} {\bibinfo {author} {\bibfnamefont {E.~A.}\ \bibnamefont
  {Ghioldi}}, \bibinfo {author} {\bibfnamefont {M.~G.}\ \bibnamefont
  {Gonzalez}}, \bibinfo {author} {\bibfnamefont {S.-S.}\ \bibnamefont {Zhang}},
  \bibinfo {author} {\bibfnamefont {Y.}~\bibnamefont {Kamiya}}, \bibinfo
  {author} {\bibfnamefont {L.~O.}\ \bibnamefont {Manuel}}, \bibinfo {author}
  {\bibfnamefont {A.~E.}\ \bibnamefont {Trumper}},\ and\ \bibinfo {author}
  {\bibfnamefont {C.~D.}\ \bibnamefont {Batista}},\ }\bibfield  {title}
  {\bibinfo {title} {Dynamical structure factor of the triangular
  antiferromagnet: Schwinger boson theory beyond mean field},\ }\href@noop {}
  {\bibfield  {journal} {\bibinfo  {journal} {Physical Review B}\ }\textbf
  {\bibinfo {volume} {98}},\ \bibinfo {pages} {184403} (\bibinfo {year}
  {2018})}\BibitemShut {NoStop}%
\bibitem [{\citenamefont {Ghioldi}\ \emph {et~al.}(2022)\citenamefont
  {Ghioldi}, \citenamefont {Zhang}, \citenamefont {Kamiya}, \citenamefont
  {Manuel}, \citenamefont {Trumper},\ and\ \citenamefont
  {Batista}}]{ghioldi2022evidence}%
  \BibitemOpen
  \bibfield  {author} {\bibinfo {author} {\bibfnamefont {E.~A.}\ \bibnamefont
  {Ghioldi}}, \bibinfo {author} {\bibfnamefont {S.-S.}\ \bibnamefont {Zhang}},
  \bibinfo {author} {\bibfnamefont {Y.}~\bibnamefont {Kamiya}}, \bibinfo
  {author} {\bibfnamefont {L.~O.}\ \bibnamefont {Manuel}}, \bibinfo {author}
  {\bibfnamefont {A.~E.}\ \bibnamefont {Trumper}},\ and\ \bibinfo {author}
  {\bibfnamefont {C.}~\bibnamefont {Batista}},\ }\bibfield  {title} {\bibinfo
  {title} {Evidence of two-spinon bound states in the magnetic spectrum of
  {B}a$_3${C}o{S}b$_2${O}$_9$},\ }\href@noop {} {\bibfield  {journal} {\bibinfo
   {journal} {Physical Review B}\ }\textbf {\bibinfo {volume} {106}},\ \bibinfo
  {pages} {064418} (\bibinfo {year} {2022})}\BibitemShut {NoStop}%
\bibitem [{\citenamefont {Zhang}\ \emph {et~al.}(2022)\citenamefont {Zhang},
  \citenamefont {Ghioldi}, \citenamefont {Manuel}, \citenamefont {Trumper},\
  and\ \citenamefont {Batista}}]{zhang2022schwinger}%
  \BibitemOpen
  \bibfield  {author} {\bibinfo {author} {\bibfnamefont {S.-S.}\ \bibnamefont
  {Zhang}}, \bibinfo {author} {\bibfnamefont {E.}~\bibnamefont {Ghioldi}},
  \bibinfo {author} {\bibfnamefont {L.~O.}\ \bibnamefont {Manuel}}, \bibinfo
  {author} {\bibfnamefont {A.~E.}\ \bibnamefont {Trumper}},\ and\ \bibinfo
  {author} {\bibfnamefont {C.~D.}\ \bibnamefont {Batista}},\ }\bibfield
  {title} {\bibinfo {title} {Schwinger boson theory of ordered magnets},\
  }\href@noop {} {\bibfield  {journal} {\bibinfo  {journal} {Physical Review
  B}\ }\textbf {\bibinfo {volume} {105}},\ \bibinfo {pages} {224404} (\bibinfo
  {year} {2022})}\BibitemShut {NoStop}%
\bibitem [{\citenamefont {Millis}\ and\ \citenamefont
  {Monien}(1994)}]{Millis94}%
  \BibitemOpen
  \bibfield  {author} {\bibinfo {author} {\bibfnamefont {A.~J.}\ \bibnamefont
  {Millis}}\ and\ \bibinfo {author} {\bibfnamefont {H.}~\bibnamefont
  {Monien}},\ }\bibfield  {title} {\bibinfo {title} {Spin gaps and bilayer
  coupling in
  ${\mathrm{yba}}_{2}$${\mathrm{cu}}_{3}$${\mathrm{o}}_{7\mathrm{\ensuremath{-}}\mathrm{\ensuremath{\delta}}}$
  and ${\mathrm{yba}}_{2}$${\mathrm{cu}}_{4}$${\mathrm{o}}_{8}$},\ }\href
  {https://doi.org/10.1103/PhysRevB.50.16606} {\bibfield  {journal} {\bibinfo
  {journal} {Phys. Rev. B}\ }\textbf {\bibinfo {volume} {50}},\ \bibinfo
  {pages} {16606} (\bibinfo {year} {1994})}\BibitemShut {NoStop}%
\bibitem [{Note3()}]{Note3}%
  \BibitemOpen
  \bibinfo {note} {In the SP approximation, the two-particle continuum
  corresponds to the excitation of a singlet boson and a triplet boson (see
  Eq.~\protect \eqref {eq:antisymm_spin}). However, this is a spurious
  continuum, as the singlet boson is not a physical mode. The physical
  continuum appears at higher energies and corresponds to a three-triplon
  excitations, which are only obtained after properly accounting for quantum
  fluctuations (this is a generic feature of the large-$N$ theories). In
  contrast, the two-particle continuum in the symmetric channel of the DSSF is
  physical, as it consists of two triplon excitations.}\BibitemShut {Stop}%
\bibitem [{\citenamefont {Podolsky}\ \emph {et~al.}(2011)\citenamefont
  {Podolsky}, \citenamefont {Auerbach},\ and\ \citenamefont
  {Arovas}}]{podolsky2011visibility}%
  \BibitemOpen
  \bibfield  {author} {\bibinfo {author} {\bibfnamefont {D.}~\bibnamefont
  {Podolsky}}, \bibinfo {author} {\bibfnamefont {A.}~\bibnamefont {Auerbach}},\
  and\ \bibinfo {author} {\bibfnamefont {D.~P.}\ \bibnamefont {Arovas}},\
  }\bibfield  {title} {\bibinfo {title} {Visibility of the amplitude (higgs)
  mode in condensed matter},\ }\href@noop {} {\bibfield  {journal} {\bibinfo
  {journal} {Physical Review B}\ }\textbf {\bibinfo {volume} {84}},\ \bibinfo
  {pages} {174522} (\bibinfo {year} {2011})}\BibitemShut {NoStop}%
\bibitem [{\citenamefont {Podolsky}\ and\ \citenamefont
  {Sachdev}(2012)}]{podolsky2012spectral}%
  \BibitemOpen
  \bibfield  {author} {\bibinfo {author} {\bibfnamefont {D.}~\bibnamefont
  {Podolsky}}\ and\ \bibinfo {author} {\bibfnamefont {S.}~\bibnamefont
  {Sachdev}},\ }\bibfield  {title} {\bibinfo {title} {Spectral functions of the
  higgs mode near two-dimensional quantum critical points},\ }\href@noop {}
  {\bibfield  {journal} {\bibinfo  {journal} {Physical Review B}\ }\textbf
  {\bibinfo {volume} {86}},\ \bibinfo {pages} {054508} (\bibinfo {year}
  {2012})}\BibitemShut {NoStop}%
\bibitem [{\citenamefont {Gazit}\ \emph {et~al.}(2013)\citenamefont {Gazit},
  \citenamefont {Podolsky},\ and\ \citenamefont {Auerbach}}]{gazit2013fate}%
  \BibitemOpen
  \bibfield  {author} {\bibinfo {author} {\bibfnamefont {S.}~\bibnamefont
  {Gazit}}, \bibinfo {author} {\bibfnamefont {D.}~\bibnamefont {Podolsky}},\
  and\ \bibinfo {author} {\bibfnamefont {A.}~\bibnamefont {Auerbach}},\
  }\bibfield  {title} {\bibinfo {title} {Fate of the higgs mode near quantum
  criticality},\ }\href@noop {} {\bibfield  {journal} {\bibinfo  {journal}
  {Physical Review Letters}\ }\textbf {\bibinfo {volume} {110}},\ \bibinfo
  {pages} {140401} (\bibinfo {year} {2013})}\BibitemShut {NoStop}%
\bibitem [{\citenamefont {Pollet}\ and\ \citenamefont
  {Prokof'Ev}(2012)}]{pollet2012higgs}%
  \BibitemOpen
  \bibfield  {author} {\bibinfo {author} {\bibfnamefont {L.}~\bibnamefont
  {Pollet}}\ and\ \bibinfo {author} {\bibfnamefont {N.}~\bibnamefont
  {Prokof'Ev}},\ }\bibfield  {title} {\bibinfo {title} {Higgs mode in a
  two-dimensional superfluid},\ }\href@noop {} {\bibfield  {journal} {\bibinfo
  {journal} {Physical Review Letters}\ }\textbf {\bibinfo {volume} {109}},\
  \bibinfo {pages} {010401} (\bibinfo {year} {2012})}\BibitemShut {NoStop}%
\bibitem [{\citenamefont {Perelomov}(1972)}]{Perelomov72}%
  \BibitemOpen
  \bibfield  {author} {\bibinfo {author} {\bibfnamefont {A.~M.}\ \bibnamefont
  {Perelomov}},\ }\bibfield  {title} {\bibinfo {title} {Coherent states for
  arbitrary lie group},\ }\href {https://doi.org/10.1007/BF01645091} {\bibfield
   {journal} {\bibinfo  {journal} {Communications in Mathematical Physics}\
  }\textbf {\bibinfo {volume} {26}},\ \bibinfo {pages} {222} (\bibinfo {year}
  {1972})}\BibitemShut {NoStop}%
\bibitem [{\citenamefont {Do}\ \emph {et~al.}(2021)\citenamefont {Do},
  \citenamefont {Zhang}, \citenamefont {Williams}, \citenamefont {Hong},
  \citenamefont {Garlea}, \citenamefont {Rodriguez-Rivera}, \citenamefont
  {Jang}, \citenamefont {Cheong}, \citenamefont {Park}, \citenamefont
  {Batista},\ and\ \citenamefont {Christianson}}]{Do2021}%
  \BibitemOpen
  \bibfield  {author} {\bibinfo {author} {\bibfnamefont {S.-H.}\ \bibnamefont
  {Do}}, \bibinfo {author} {\bibfnamefont {H.}~\bibnamefont {Zhang}}, \bibinfo
  {author} {\bibfnamefont {T.~J.}\ \bibnamefont {Williams}}, \bibinfo {author}
  {\bibfnamefont {T.}~\bibnamefont {Hong}}, \bibinfo {author} {\bibfnamefont
  {V.~O.}\ \bibnamefont {Garlea}}, \bibinfo {author} {\bibfnamefont {J.~A.}\
  \bibnamefont {Rodriguez-Rivera}}, \bibinfo {author} {\bibfnamefont {T.-H.}\
  \bibnamefont {Jang}}, \bibinfo {author} {\bibfnamefont {S.-W.}\ \bibnamefont
  {Cheong}}, \bibinfo {author} {\bibfnamefont {J.-H.}\ \bibnamefont {Park}},
  \bibinfo {author} {\bibfnamefont {C.~D.}\ \bibnamefont {Batista}},\ and\
  \bibinfo {author} {\bibfnamefont {A.~D.}\ \bibnamefont {Christianson}},\
  }\bibfield  {title} {\bibinfo {title} {Decay and renormalization of a
  longitudinal mode in a quasi-two-dimensional antiferromagnet},\ }\href
  {https://doi.org/10.1038/s41467-021-25591-7} {\bibfield  {journal} {\bibinfo
  {journal} {Nature Communications}\ }\textbf {\bibinfo {volume} {12}},\
  \bibinfo {pages} {5331} (\bibinfo {year} {2021})}\BibitemShut {NoStop}%
\bibitem [{\citenamefont {Bai}\ \emph {et~al.}(2023)\citenamefont {Bai},
  \citenamefont {Zhang}, \citenamefont {Zhang}, \citenamefont {Dun},
  \citenamefont {Phelan}, \citenamefont {Garlea}, \citenamefont {Mourigal},\
  and\ \citenamefont {Batista}}]{Bai2023}%
  \BibitemOpen
  \bibfield  {author} {\bibinfo {author} {\bibfnamefont {X.}~\bibnamefont
  {Bai}}, \bibinfo {author} {\bibfnamefont {S.-S.}\ \bibnamefont {Zhang}},
  \bibinfo {author} {\bibfnamefont {H.}~\bibnamefont {Zhang}}, \bibinfo
  {author} {\bibfnamefont {Z.}~\bibnamefont {Dun}}, \bibinfo {author}
  {\bibfnamefont {W.~A.}\ \bibnamefont {Phelan}}, \bibinfo {author}
  {\bibfnamefont {V.~O.}\ \bibnamefont {Garlea}}, \bibinfo {author}
  {\bibfnamefont {M.}~\bibnamefont {Mourigal}},\ and\ \bibinfo {author}
  {\bibfnamefont {C.~D.}\ \bibnamefont {Batista}},\ }\bibfield  {title}
  {\bibinfo {title} {Instabilities of heavy magnons in an anisotropic magnet},\
  }\href {https://doi.org/10.1038/s41467-023-39940-1} {\bibfield  {journal}
  {\bibinfo  {journal} {Nature Communications}\ }\textbf {\bibinfo {volume}
  {14}},\ \bibinfo {pages} {4199} (\bibinfo {year} {2023})}\BibitemShut
  {NoStop}%
\bibitem [{\citenamefont {Matsushita}\ \emph {et~al.}(1999)\citenamefont
  {Matsushita}, \citenamefont {P.~Gelfand},\ and\ \citenamefont
  {Ishii}}]{matsushita1999bond}%
  \BibitemOpen
  \bibfield  {author} {\bibinfo {author} {\bibfnamefont {Y.}~\bibnamefont
  {Matsushita}}, \bibinfo {author} {\bibfnamefont {M.}~\bibnamefont
  {P.~Gelfand}},\ and\ \bibinfo {author} {\bibfnamefont {C.}~\bibnamefont
  {Ishii}},\ }\bibfield  {title} {\bibinfo {title} {Bond-operator mean field
  theory for the bilayer heisenberg model},\ }\href@noop {} {\bibfield
  {journal} {\bibinfo  {journal} {Journal of the Physical Society of Japan}\
  }\textbf {\bibinfo {volume} {68}},\ \bibinfo {pages} {247} (\bibinfo {year}
  {1999})}\BibitemShut {NoStop}%
\end{thebibliography}

%

\end{document}